\documentclass[preprint,prd,aps,floatfix,preprintnumbers,superscriptaddress,bibnotes,nofootinbib]{revtex4-1}

\usepackage{epsfig}
\usepackage{amsmath}
\usepackage{latexsym}
\usepackage[psamsfonts]{amssymb}
\usepackage{graphicx}
\usepackage{ulem}
\usepackage{longtable}
\usepackage{epstopdf}
\usepackage{bm}% bold math
\newcommand{\be}{\begin{equation}}
\newcommand{\ee}{\end{equation}}
\newcommand{\bea}{\begin{eqnarray}}
\newcommand{\eea}{\end{eqnarray}}
\newcommand{\bi}{\begin{itemize}}
\newcommand{\ei}{\end{itemize}}

\begin{document}

%-- title ---

\title{
 Glueball spectroscopy in lattice QCD using gradient flow
}

%-- author list ---

\author{Keita Sakai${}^{1}$}\email{sakai@nucl.phys.tohoku.ac.jp}
\author{Shoichi Sasaki${}^{1}$}\email{ssasaki@nucl.phys.tohoku.ac.jp}

\affiliation{${}^{1}$Department of Physics, Tohoku University, Sendai 980-8578, Japan}

\date{\today}
%-- abstract ---
\begin{abstract}
Removing ultraviolet noise from the gauge fields is necessary for glueball spectroscopy in lattice QCD. 
It is known that the Yang-Mills gradient flow method is an alternative approach instead of 
link smearing or link fuzzing in various aspects. In this work we study the application of the 
gradient flow technique to the construction of the extended glueball operators. 
We examine a simple application of the gradient flow method, which has some problems in 
glueball mass calculations at large flow time because of its nature of diffusion in space-time. 
To avoid this problem, the spatial links are evolved by the ``spatial gradient flow'', that 
is defined to restrict the diffusion to spatial directions only. 
We test the spatial gradient flow in calculations of glueball two-point functions and Wilson loops 
as a new smearing method, 
and then discuss its efficiency in comparison with the original gradient flow method 
and the conventional method.
Furthermore, to demonstrate the feasibility of our proposed method, we determine the masses of 
the three lowest-lying glueball states, corresponding to the $0^{++}$, $2^{++}$, and $0^{-+}$ glueballs, 
in the continuum limit in pure Yang-Mills theory.

\end{abstract}


\maketitle

%--- main text -------------------------------------------------------
 
%%%%%%%%%%%%%%  SEC 1  %%%%%%%%%%%%%%%%%%%%%%%%
\section{Introduction}
\label{sec:INTRO}

The existence of composite states consisting solely of gluons, called glueballs, 
is one of the important predictions of QCD. Since none of them have been identified 
in experiments as a glueball state, the lattice QCD results play an essential  
role in studying properties of the glueball states including their masses.
However, ultraviolet noise from the gauge fields makes it difficult to calculate
the glueball spectrum in lattice QCD.
Therefore, noise reduction techniques such as the single-link smearing or the double-link fuzzing procedure
plays an increasingly important role in the construction of the extended glueball operators. 

Various smearing techniques such as APE smearing~\cite{APE:1987ehd}, HYP-smearing~\cite{Hasenfratz:2001hp}, 
and stout smearing~\cite{Morningstar:2003gk} are developed for many purposes, while the 
fuzzing approach is proposed for a specific purpose that requires a significant improvement of having
a better overlap with the glueball ground states~\cite{Teper:1987wt}.  
In several previous works on the lattice glueball mass 
calculations~\cite{{Morningstar:1999rf},{Chen:2005mg},{Meyer:2004gx},{Athenodorou:2020ani}}, 
some sophisticated combinations of both the single-link smearing and the double-link fuzzing schemes are 
conventionally adopted (denoted as the conventional approach). 

Recently, it was found that the Yang-Mills gradient flow method~\cite{Luscher:2010iy} 
is an alternative approach instead of the smearing in various aspects (e.g., the computation of 
topological charge~\cite{Alexandrou:2017hqw}). Indeed, the gradient flow equation can be regarded as a continuous version of
the recursive update procedure in the stout-link smearing 
with the small smearing parameter~\cite{{Alexandrou:2017hqw},{Luscher:2010iy}}.
Therefore, in this study, we investigate the application of the gradient flow 
technique to the glueball calculation and also demonstrate its effectiveness 
in comparison to the conventional approach. 

This paper is organized as follows. In Sec.~\ref{sec:METHOD_GF}, after a brief
introduction of the original Yang-Mills gradient flow, we describe our proposal
of ``spatial gradient flow" as a new smearing method. In Sec.~\ref{sec:GB_2PT},
we give a short outline of how to construct glueball two-point functions 
based on spacelike Wilson loops. In Sec.~\ref{sec:SET_UP},
we first briefly summarize the numerical ensembles used in this study. 
Then we present results of the static quark potential computed
with the spatial gradient flow on every ensemble.  
Section~\ref{sec:FEATURES} gives the features of
the spatial gradient flow in glueball spectroscopy. 
The results of glueball masses obtained by the spatial gradient flow
are summarized in Sec.~\ref{sec:RESULTS}. 
Finally, we close with summary in Sec.~\ref{sec:SUMMARY}.

%\clearpage
%%%%%%%%%%%%%%  SEC 2  %%%%%%%%%%%%%%%%%%%%%%%%
\section{Calculation method I: Gradient flow method}
\label{sec:METHOD_GF}

In this section, we first provide a brief review of the original Yang-Mills gradient flow, which
makes the Wilson flow diffused in the four-dimensional space-time, in Sec.~\ref{sec:FORM_ORGFLOW}.
As reported in Ref.~\cite{Sakai:2022mgd}, a simple application of the gradient flow 
technique to the glueball calculation has some problem in measuring the glueball mass 
from the two-point function. We thus propose the ``spatial gradient flow" as a new smearing method, which
is described in Sec.~\ref{sec:FORM_SPATIALFLOW}.

%%%%%%%%%%5
\subsection{Original gradient flow}
\label{sec:FORM_ORGFLOW}

The Yang-Mills gradient flow on the lattice is a kind of diffusion equation, where the link variables
$U_\mu(x)$ evolve smoothly as a function of fictitious time $\tau$ (denoted as flow time)~\cite{Luscher:2010iy}. 
The associated flow 
$V_\mu(x,\tau)$ of the link variables (hereafter called the Wilson flow)
driven by the gradient of the action with respect the link variables~\cite{Luscher:2010iy}. 
The simplest choice of the action of the link variables $U_{\mu}(x)$ 
is the standard Wilson plaquette action,
%
% Eq.1
%
\begin{widetext}
\begin{equation}
S_{W}[U]=\frac{2}{g_0^2}\sum_{x, \mu>\nu}{\rm Tr}\left\{1-{\rm Re}\left[U_{\mu}(x)
    U_{\nu}(x+\hat{\mu})U^\dagger_{\mu}(x+\hat{\nu},\tau)U^\dagger_{\nu}(x)\right]\right\},
    \label{eq:WilsonAction}
\end{equation}
\end{widetext}
where $g_0$ is the bare coupling.
The flowed link variables $V_\mu(x,\tau)$ are defined by the following equation 
with the initial conditions $V_\mu(x,0)=U_\mu(x)$,
%
% Eq.2
%
\begin{equation}
   \frac{\partial}{\partial \tau}V_\mu(x,\tau)\cdot V^{-1}_\mu(x,\tau) =-g_0^2\partial_{x,\mu} S_W[V],
    \label{eq:gradient_flow}
\end{equation}
where ${S_W}[V]$ denotes the standard Wilson plaquette action 
in terms of the flowed link variables (see Appendix~\ref{sec:FLOW_STOUT} for the definition of the link 
derivative operator $\partial_{x,\mu}$ and the explicit expression of $\partial_{x,\mu} S_W$).

According to Eq.~\eqref{eq:gradient_flow}, the link variables are diffused 
in the four-dimensional {\it space-time}, so that the Wilson flow is approximately spread out 
in a Gaussian distribution with the diffusion radius (or length) of $R_d=\sqrt{8\tau}$~\cite{Luscher:2010iy}. 
Although such smearing procedure works well with the longer flow time, 
too much smearing will destroy or hide the true temporal correlation 
of the glueball two-point function due to the overlap of two glueball operators given by the Wilson flow
as discussed in Ref.~\cite{Sakai:2022mgd}. 
Therefore, the longer flow is not applicable for the glueball spectroscopy 
to avoid {\it over smearing}, that was observed in Ref.~\cite{Chowdhury:2015hta}.

\subsection{Spatial gradient flow as a new smearing method}
\label{sec:FORM_SPATIALFLOW}

As described in Sec~\ref{sec:FORM_ORGFLOW}, the previous attempt to apply the gradient flow
to the glueball spectroscopy is not fully satisfactory~\cite{Chowdhury:2015hta}. 
We propose the ``spatial gradient flow'' as a new smearing method
in order to overcome the limited usage of the Wilson flow due to over smearing. 
The spatial gradient flow is defined to restrict the diffusion to spatial directions only,
so that the spatial links $U_i(x)$ are evolved into the spatial Wilson flow 
$\mathcal{V}_i(x,\tau)$ as the initial conditions of 
$\mathcal{V}_i(x,0)=U_i(x)$
in the following gradient flow equation:
%
% Eq.3
%
\begin{equation}
   \frac{\partial}{\partial \tau}\mathcal{V}_i(x,\tau)\cdot \mathcal{V}^{-1}_i(x,\tau) =-g^2_0\partial_{x,i} S_{sW}[\mathcal{V}_i(x,\tau)].
    \label{eq:spatial_flow}
\end{equation}
Here, $S_{sW}$ denotes the spatial part of the standard Wilson plaquette action,
%
% Eq.4
%
\begin{widetext}
\begin{equation}
    S_{sW}[\mathcal{V}_i(x,\tau)]=\frac{2}{g_0^2}\sum_{x, i>j}{\rm Tr}\left\{1-{\rm Re}\left[\mathcal{V}_{i}(x,\tau)
    \mathcal{V}_{j}(x+\hat{i},\tau)\mathcal{V}^\dagger_{i}(x+\hat{j},\tau)\mathcal{V}^\dagger_{j}(x,\tau)\right]\right\},
    \label{eq:spatial_WilsonAction}
\end{equation}
\end{widetext}
where the plaquette values are composed only of the spatial links.
The indices ${i}$ and ${j}$ run only over spatial directions.
Since the spatial Wilson flow is diffused only {\it in three-dimensional space}, 
its diffusion radius is given by ${\mathcal{R}_d=\sqrt{6\tau}}$.
We will later show that this new smearing works well even for the glueball spectroscopy {\it without over smearing}. 

%\clearpage
%%%%%%%%%%%%%%  SEC 3  %%%%%%%%%%%%%%%%%%%%%%%%
\section{Calculation method II: Glueball two-point function}
\label{sec:GB_2PT}

We are interested in three lowest-lying glueball states, which carry specific  
quantum numbers, $J^{PC}=0^{++}$ (scalar), $0^{-+}$ (pseudoscalar), or $2^{++}$ (tensor), in this study. 
In this section, we briefly describe how to construct two-point correlation functions of the 
desired glueball state having spin $J$, parity $P$, and charge-conjugation parity $C$.

First of all, on the lattice, rotational symmetry is reduced to the octahedral point group $O$,
which is a finite subgroup of the rotation group $SO(3)$. There are 24 rotational operations
associated with all proper rotations in the group $O$. The irreducible representations (irreps) $R$
of $O$ are the counterparts of spin $J$ for the continuum rotation group $SO(3)$. 
There are five irreps, which are classified by two one-dimensional representations (denoted as $A_1$ and $A_2$), 
one two-dimensional representation (denoted as $E$), and two three-dimensional representations 
(denoted as $T_1$ and $T_2$)~\cite{Johnson:1982yq}. 

The inclusion of inversion ($X$ is mapped to $-X$) results in the symmetry group known as
$O_h$, which has 48(=24$\times$2) symmetry operations.
The irreducible representations of $O_h$ are obtained from those of $O$ by appending the index $g$ (gerade) or
$u$ (ungerade), which indicates even or odd parity~\cite{Berg:1982kp}. 
For convenience, we use the indices $+$ and $-$, instead of $g$ and $u$. 
Therefore, ten different irreducible representations of $O_h$ are denoted as $R^P$, hereafter. 
The glueball states are also eigenstates of charge conjugation. 
Therefore, the quantum number of the lattice glueball state 
is specified by $R^{PC}$.  The quantum number $R^{PC}$ is expected to have the following correspondence: 
$0^{++} \leftrightarrow A_1^{++}$, $0^{-+} \leftrightarrow A_1^{-+}$
and $2^{++} \leftrightarrow E^{++}\oplus T_2^{++}$ in the continuum limit~\cite{{Johnson:1982yq},{Berg:1982kp}}.

%
% FIG.1
%
\begin{figure}[h]
      \includegraphics[width=0.5\textwidth, bb=0 0 1024 768]{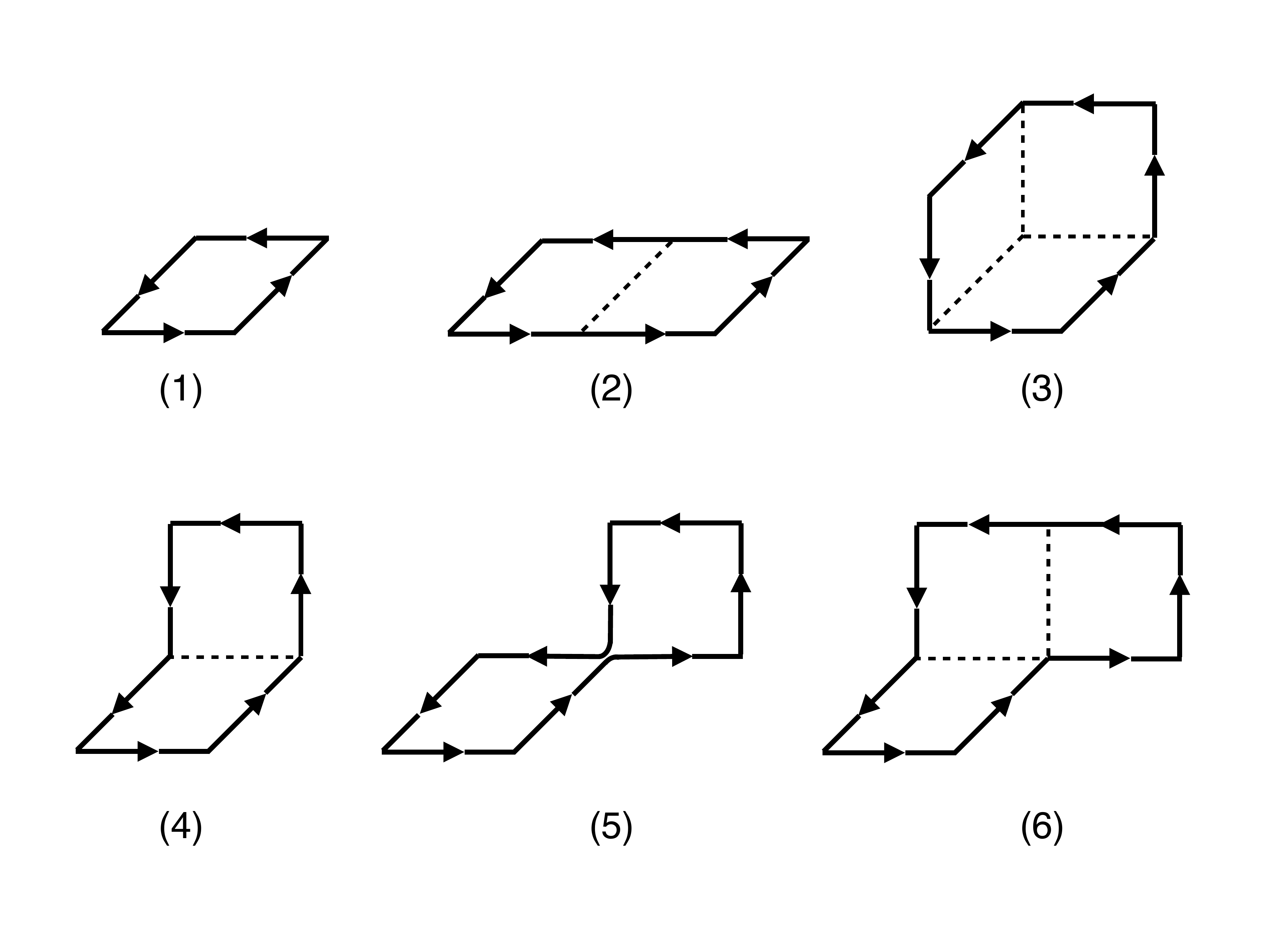}
  \caption{Six prototypes of spacelike Wilson loops used to construct the glueball operators:
  (1) plaquette operator (denoted as ${\cal O}_{\rm plaq}$), (2) rectangle operator (denoted as ${\cal O}_{\rm rect}$), (3) twist operator (denoted as ${\cal O}_{\rm twist}$), (4) chair operator (denoted as ${\cal O}_{\rm chair}$), (5) ``fish-shaped'' operator (denoted as ${\cal O}_{\rm fish}$), and (6) ``hand-shaped'' operator (denoted as ${\cal O}_{\rm hand}$). 
  \label{fig:WL_SHAPES}
  }
  \end{figure}
%

%
% TABLE.1
%
\begin{table*}[ht]                                                                                       
  \caption{
  Classification of spacelike Wilson loops used in this study.
  Each operator ${\cal O}_k$ is classified with the shape (labeled as $k$) characterized by the ordered closed path
  (denoted as ${\cal C}_k$)
  and the numbers of the links (denoted as $l$) involved in the Wilson loop. In the table, the paths for prototype of Wilson loops 
  depicted in Fig.~\ref{fig:WL_SHAPES} are given with an $l$-tuple composed of the direction of the coordinate axes as 
  $\pm X$, $\pm Y$, and $\pm Z$. The minus sign indicates the path along in the negative direction. 
  The check mark symbol ($\checkmark$) in the table indicates that the operator contains the corresponding irreducible representation~\cite{Berg:1982kp}.
  \label{tab:WL_OP}}
\begin{ruledtabular}                                                                              
\begin{tabular}{| c  c  c  c  c  c  c  |}
\hline
Label& No. of links & Prototype path&\multicolumn{4}{c|}{Target irreps} \cr
$k$ & $l$ & ${\cal C}_k$ & $A_1^{++}$ & $A_1^{-+}$ & $E^{++}$  &$T_2^{++}$ \cr
\hline
Plaq & 4  & $[X, Y, -X, -Y]$ &  \checkmark&  & \checkmark & \cr
Rect & 6  & $[X, X, Y, -X, -X, -Y]$&  \checkmark&  & \checkmark & \cr
Twist & 6  & $[X, Y, Z, -X, -Y, -Z]$ &  \checkmark& & & \checkmark\cr 
Chair & 6  & $[X, Y, Z, -X, -Z, -Y]$ &  \checkmark& &\checkmark& \checkmark\cr
Fish  & 8  & $[X, Y, X, Z, -X, -Z, -X, -Y]$ &  \checkmark& \checkmark& \checkmark& \checkmark\cr
Hand  & 8 & $[X, Y, X, Z, -X, -X, -Z, -Y]$ &  \checkmark& \checkmark& \checkmark& \checkmark\cr 
\hline
\end{tabular}
\end{ruledtabular}                                                                                
\end{table*}

\subsection{Glueball operators}
We use six prototypes of spacelike Wilson loops to construct the glueball operators 
for four specific channels of $R^{PC}=A_1^{++}$, $A_{1}^{-+}$, $E^{++}$ and $T_2^{++}$ 
as depicted in Fig.~\ref{fig:WL_SHAPES}.
First four operators are all types of spacelike Wilson loops with four and six links, 
while the remaining two operators are chosen from spacelike Wilson loops of eight links. 
Each operator ${\cal O}_k$ is classified with the numbers of the links (denoted as $l$) involved in the Wilson loop, 
the shape characterized by the ordered closed path ${\cal C}_k$ and the associated orientation 
as listed in Table~\ref{tab:WL_OP}. In the case of $SU(N)$ with $N\ge 3$, the real part of 
Wilson loops has a charge conjugation parity $C=+1$, while the imaginary part has $C$-parity $C=-1$~\cite{Berg:1982kp}. 
In this study, we restrict ourselves to consider the real part of Wilson loops, since the three lowest-lying glueball 
states carry $C=+1$.

We take the following procedure to get the irreducible contents of the representation $R^P$ 
with fixed $C$-parity from the operators ${\cal O}_k$ according to Ref.~\cite{Berg:1982kp}. 
First, 48 symmetry operations $\hat{S}_i$, which are defined in Table~\ref{tab:48_OP}, 
are applied to each prototype of Wilson loops ${\cal O}_k[{\cal C}_k]$ 
so as to obtain 48 copies with different orientations of Wilson loops ${\cal O}_k[\hat{S}_i {\cal C}_k]$.
A linear combination of them with weights equal to the irreducible characters $\chi_{\Gamma}(S_i)$ of $S_i$ for the $\Gamma$ irreps provides the operator projected onto the $\Gamma$ irrep as
%
% Eq.5
%
\begin{equation}
    {\cal O}^{\Gamma}_k=\hat{P}_{\Gamma}{\cal O}_k[{\cal C}_k] =\frac{1}{48}\sum_{i=1}^{48}\chi^{\ast}_{\Gamma}(S_i)
    {\cal O}_k[\hat{S}_i {\cal C}_k], 
    \label{eq:projection}
\end{equation}
where the characters $\chi_{\Gamma}(S_i)$ of the irreps $\Gamma=A_1^+$, $A_1^-$, $E^+$ and $T_2^+$ are listed in Table~\ref{tab:CharacterTable}. 

We next construct two-point correlation functions of glueball states for given irreps $\Gamma$ as 
%
% Eq.6
%
\begin{equation}
C^{\Gamma}_k(t)=\sum_{t^\prime}\langle 0| \tilde{\cal O}^{\Gamma}_k(t+t^\prime)\tilde{\cal O}^{\Gamma}_k(t^\prime)^\dagger|0\rangle, 
\end{equation}
where the tilde over ${\cal O}^{\Gamma}_k$ implies the vacuum-subtracted operator defined 
as $\tilde{\cal O}^{\Gamma}_k(t)={\cal O}^{\Gamma}_k(t)-\langle 0|{\cal O}^{\Gamma}_k(t)|0\rangle$.
We may also consider an $N\times N$ correlation matrix using a set of different shape operators for given irreps $\Gamma$~\cite{Michael:1988jr} as
%
% Eq.7
%
\begin{equation}
C^{\Gamma}_{kk^\prime}(t)=\sum_{t^\prime}\langle 0| \tilde{\cal O}^{\Gamma}_k(t+t^\prime)\tilde{\cal O}^{\Gamma}_{k^\prime}(t^\prime)^\dagger|0\rangle, 
\end{equation}
which allow us to perform the variational method~\cite{{Michael:1985ne},{Luscher:1990ck}}.

%
% TABLE.2
%
\begin{table*}[ht]                                                                                       
  \caption{Each of 48 elements $S_i$ of the group $O_{h}$, which are presented by the coordinate transformations, 
  is divided into ten different conjugacy classes $\{ E, 3{C_4}^2, 6C_4, 8C_3, 6{C_2}^\prime, I, 3\sigma_h, 6IC_4, 8IC_3, 6\sigma_d\}\in O_h$.
  \label{tab:48_OP}
  }
      \begin{ruledtabular}                                                                              
\begin{tabular}{| c  c  c | c  c  c |}
\hline
Class & $i$ & Operation & Class & $i$ & Operation  \cr
\hline
$E$  & 1 &  $(X,Y,Z) \rightarrow (X,Y,Z)$ & $I$  & 25 &  $(X,Y,Z) \rightarrow (-X,-Y,-Z)$\cr 
\hline
$3{C_4}^2$ & 2 &  $(X,Y,Z) \rightarrow (-X,-Y,Z)$ & $3\sigma_h$ & 26 &  $(X,Y,Z) \rightarrow (X,Y,-Z)$\cr
                  & 3 &  $(X,Y,Z) \rightarrow (-X,Y,-Z)$ &                    & 27 &  $(X,Y,Z) \rightarrow (X,-Y,Z)$\cr
                  & 4 &  $(X,Y,Z) \rightarrow (X,-Y,-Z)$ &                    & 28 &  $(X,Y,Z) \rightarrow (-X,Y,Z)$\cr
\hline
$6C_4$ & 5 &  $(X,Y,Z) \rightarrow (-Y,X,Z)$ & $6IC_4$ & 29 &  $(X,Y,Z) \rightarrow (Y,-X,-Z)$\cr
            & 6 &  $(X,Y,Z) \rightarrow (Y,-X,Z)$ &               & 30 &  $(X,Y,Z) \rightarrow (-Y,X,-Z)$\cr
            & 7 &  $(X,Y,Z) \rightarrow (Z,Y,-X)$  &              & 31 &  $(X,Y,Z) \rightarrow (-Z,-Y,X)$\cr
            & 8 &  $(X,Y,Z) \rightarrow (-Z,Y,X)$ &               & 32 &  $(X,Y,Z) \rightarrow (Z,-Y,-X)$\cr
            & 9 &  $(X,Y,Z) \rightarrow (X,-Z,Y)$ &               & 33 &  $(X,Y,Z) \rightarrow (-X,Z,-Y)$ \cr
            & 10 &  $(X,Y,Z) \rightarrow (X,Z,-Y)$&              & 34 &  $(X,Y,Z) \rightarrow (-X,-Z,Y)$\cr
\hline
$8C_3$ & 11 &  $(X,Y,Z) \rightarrow (Y,Z,X)$ & $8IC_3$ & 35 &  $(X,Y,Z) \rightarrow (-Y,-Z,-X)$\cr
             & 12 &  $(X,Y,Z) \rightarrow (Z,X,Y)$ &              & 36 &  $(X,Y,Z) \rightarrow (-Z,-X,-Y)$\cr
             & 13 &  $(X,Y,Z) \rightarrow (Y,-Z,-X)$ &             & 37 &  $(X,Y,Z) \rightarrow (-Y,Z,X)$ \cr
             & 14 &  $(X,Y,Z) \rightarrow (-Z,-X,Y)$ &             & 38 &  $(X,Y,Z) \rightarrow (Z,X,-Y)$\cr
             & 15 &  $(X,Y,Z) \rightarrow (-Y,Z,-X)$ &              & 39 &  $(X,Y,Z) \rightarrow (Y,-Z,X)$ \cr
             & 16 &  $(X,Y,Z) \rightarrow (Z,-X,-Y)$ &              & 40 &  $(X,Y,Z) \rightarrow (-Z,X,Y)$ \cr
             & 17 &  $(X,Y,Z) \rightarrow (-Y,-Z,X)$ &              & 41 &  $(X,Y,Z) \rightarrow (Y,Z,-X)$ \cr
             & 18 &  $(X,Y,Z) \rightarrow (-Z,X,-Y)$ &              & 42 &  $(X,Y,Z) \rightarrow (Z,-X,Y)$\cr
\hline
$6{C_2}^\prime$ & 19  & $(X,Y,Z) \rightarrow (Y,X,-Z)$ & $6\sigma_d$ & 43  & $(X,Y,Z) \rightarrow (-Y,-X,Z)$\cr
                          &  20  & $(X,Y,Z) \rightarrow (Z,-Y,X)$ &                     & 44  & $(X,Y,Z) \rightarrow (-Z,Y,-X)$\cr
                          & 21  & $(X,Y,Z) \rightarrow (-X,Z,Y)$ &                     & 45  & $(X,Y,Z) \rightarrow (X,-Z,-Y)$\cr
                          & 22  & $(X,Y,Z) \rightarrow (-Y,-X,-Z)$ &                   & 46  & $(X,Y,Z) \rightarrow (Y,X,Z)$\cr
                          & 23  & $(X,Y,Z) \rightarrow (-Z,-Y,-X)$ &                   & 47  & $(X,Y,Z) \rightarrow (Z,Y,X)$\cr
                          & 24  & $(X,Y,Z) \rightarrow (-X,-Z,-Y)$ &                   & 48  & $(X,Y,Z) \rightarrow (X,Z,Y)$ \cr
\hline
\end{tabular}
\end{ruledtabular}                                                                                
\end{table*}

%
% TABLE.3
%
\begin{table*}[ht]                                                                                       
  \caption{
  Table of characters $\chi_{\Gamma}(S_i)$ for four irreps, $A_1^+$, $A_1^{-}$, $E^+$, and $T_2^+$ of the group $O_h$. 
  The elements $S_i$ of the group $O_h$ belong to ten different conjugacy classes $\{ E, 3{C_4}^2, 6C_4, 8C_3, 6{C_2}^\prime, I, 3\sigma_h, 6IC_4, 8IC_3, 6\sigma_d\}$~\cite{Berg:1982kp}.
  \label{tab:CharacterTable}
  }
      \begin{ruledtabular}                                                                              
\begin{tabular}{| c  c  c  c  c  c  c  c  c  c  c  |}
\hline
Irreps & $E$  & $3{C_4}^2$ & $6C_4$ & $8C_3$ & $6{C_2}^\prime$ & $I$  & $3\sigma_h$
& $6IC_4$ & $8IC_3$ & $6\sigma_d$ \cr
\hline
$A_1^{+}$  &  $+1$ &  $+1$ & $+1$  & $+1$ & $+1$ &  $+1$ & $+1$ & $+1$ & $+1$ & $+1$ \cr
$A_1^{-}$   & $+1$  &  $+1$ & $+1$  & $+1$ & $+1$ & $-1$ & $-1$ & $-1$ & $-1$ & $-1$ \cr
$E^{+}$      &  $+2$ &  $+2$ & $0$  & $-1$ & $0$ &  $+2$ & $+2$ &  $0$ & $-1$ & $0$\cr
$T_2^{+}$  &  $+3$ & $-1$ & $-1$ &  $0$ &  $+1$ &  $+3$ & $-1$ & $-1$ &  $0$ & $+1$\cr
\hline
\end{tabular}
\end{ruledtabular}                                                                                
\end{table*}

%\clearpage
%%%%%%%%%%%%%%  SEC 4  %%%%%%%%%%%%%%%%%%%%%%%%
\section{Lattice setup}
\label{sec:SET_UP}

\subsection{Gauge ensembles}
\label{sec:GAUGE_ENSEMBLE}

We perform the pure Yang-Mills lattice simulations
using the standard Wilson plaquette action
with a fixed physical volume ($La\approx 1.6$ fm)
at four different gauge couplings ($\beta=6/g_0^2=6.2$, 6.4, 6.71 and 6.93).
The gauge configurations in each simulation are 
separated by $n_{\rm update}$ sweeps after $n_{\rm therm}$ sweeps for thermalization
as summarized in Table~\ref{tab:Sims}.
Each sweep consists of one heat bath~\cite{Cabibbo:1982zn} combined with four 
over-relaxation~\cite{Creutz:1987xi} steps.
The number of configurations analyzed is ${\cal O}$(500--4000).
All lattice spacings are set by the Sommer scale ($r_0=0.5$ fm)~\cite{{Sommer:1993ce},{Necco:2001xg}}.

For both original and spatial gradient flows, the forth-order Runge-Kutta scheme is used with
an integration step size of $\epsilon=0.025$. 
The flow time $\tau$ is given by $n_{\rm flow}\times \epsilon$ where $n_{\rm flow}$ 
denotes the number of flow iterations.
Our simulation code for the gradient flow had been already checked in Table II of Ref.~\cite{Kamata:2016any}, 
where the values of the gradient flow reference scale are directly compared with the results given 
in the original work of L\"uscher~\cite{Luscher:2010iy}.

%
% TABLE.4
%
\begin{table*}[ht]                                                                                       
  \caption{
  Summary of the gauge ensembles: gauge coupling, lattice size ($L^3\times T$), plaquette value, 
  lattice spacing ($a$), spatial extent ($La$), the Sommer scale ($r_0$), 
  the number of the gauge field configurations ($N_{\rm conf}$),  
  the number of thermalization sweeps ($n_{\rm therm}$)  
  and the number of update sweeps ($n_{\rm update}$).
  All lattice spacings are set by the Sommer scale ($r_0=0.5$ fm)~\cite{{Sommer:1993ce},{Necco:2001xg}}.
  \label{tab:Sims}
  }
      \begin{ruledtabular}                                                                              
\begin{tabular}{| c c c c c c c c c         |}
\hline
$\beta=6/g_0^2$ & $L^3\times T$ & plaquette & $a$ [fm] & $\sim La$ [fm] & $r_0/a$ (Ref.~\cite{Necco:2001xg}) & $N_{\rm conf}$ & $n_{\rm therm}$ & $n_{\rm update}$ \cr 
\hline
6.20& $24^3 \times 24$ &  
0.613644(3)&0.0677(3) & 1.62 & 7.38(3) & 4000 & 5000 &200 \cr
6.40& $32^3 \times 32$ & 
0.630646(2)&0.0513(3) & 1.64 & 9.74(5) & 3000 & 5000 &200 \cr
6.71& $48^3 \times 48$ &   
0.653298(2)&0.0345(2) & 1.66 & 14.49(10) & 1000 & 25000 &200 \cr
6.93& $64^3 \times 64$ &   
0.667376(1)&0.0256(2) & 1.64 & 19.48(12) & 500 & 64000 &600 \cr
\hline
\end{tabular}
\end{ruledtabular}                                                                                
\end{table*}

\subsection{Scale setting and static quark potential}
\label{sec:POTENTIAL}

The static potential $V(r)$ between 
heavy quark and antiquark pairs, which are 
separated by relative distance $r$, is calculated from the 
Wilson-loop expectation value $\langle W(r,t)\rangle$ with spatial 
extent $r$ and temporal extent $t$ as
%
% Eq.8
%
\be
\langle W(r,t)\rangle =C(r) e^{-V(r)t} + \cdot\cdot\cdot, 
\ee
where the ellipsis denotes some contribution from the excited states. 

To determine the static quark potential $V(r)$, let us consider the following quantity:
%
% Eq.9
%
\be
V(r,t)=\ln\left\{\frac{\langle W(r,t)\rangle}{\langle W(r,t+1)\rangle}
\right\}.
\label{eq:Wloop-Pot}
\ee
Since the $t$-dependence of $V(r,t)$ is expected to disappear
for sufficiently large $t$, the static potential $V(r)$
can be determined from a plateau seen in $V(r,t)$ as
$t$ increases for fixed $r$.

\begin{figure}
\centering
\includegraphics*[width=0.5\textwidth]{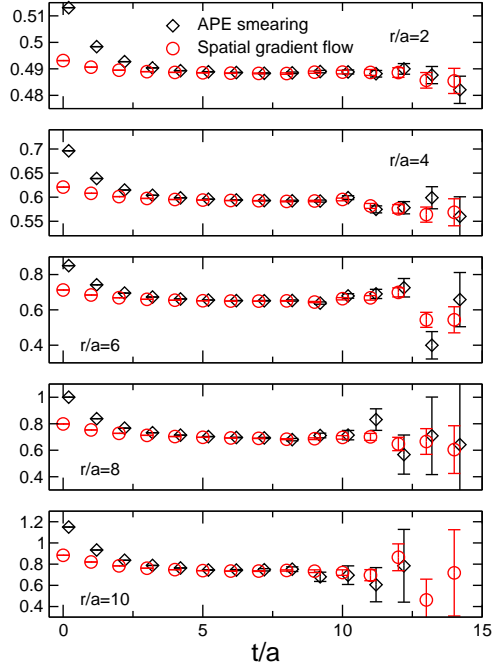} 
\caption{The $t$-dependence of $V(r,t)$ for several values of fixed $r/a$ (
$r/a=2, 4, 6, 8, 10$ from top panel to bottom panel).
The Wilson loops $W(r,t)$ are constructed by the smeared spatial links, which 
are computed with either APE smearing (diamonds) or spatial gradient flow (circles). 
The spatial gradient flow provides longer plateau behaviors than those of APE smearing.
}
\label{fig:tdep-Wloop}
\end{figure}

Figure~\ref{fig:tdep-Wloop} shows the $t$-dependence of $V(r,t)$ calculated
at $\beta=6.4$ for fixed $r/a$ ($r/a=2, 4, 6, 8, 10$ from top panel to bottom panel)
as typical examples. 
The Wilson loops $W(r,t)$ are constructed by the smeared spatial links, which 
are computed with either APE smearing or spatial gradient flow. 
The diamond symbols represent the results calculated using APE smearing 
with $\alpha_{\mathrm{APE}}=0.5$ and 5 steps, while the circle symbols
represent the results calculated using spatial gradient flow with $n_{\mathrm{flow}}=50$. 

At glance, spatial gradient flow provides the better behavior, where
the plateau starts at earlier $t$ and extends to larger $t$, 
comparing with APE smearing. 
It indicates that the systematic uncertainties stemming
from the excited-state contamination are better under control 
to determine the static potential using the spatial gradient flow method.
%                                                                             
%  FIG.2                                                                           
% 

Hereafter, we adopt the spatial gradient flow method to evaluate 
the value of $V(r)$ from the Wilson-loop expectation value, and then 
aim to determine the Sommer scale from the resulting static potential 
at each $\beta$. In this study, the Wilson loops are restricted to on-axis loops only. 
To extract the value of $V(r)$ from $\langle W(r,t)\rangle$ 
at fixed $r$, we use the double-exponential fit, where
the a correlation among $\langle W(r,t)\rangle$ at different value of $t$ 
is taken into account by using a covariance matrix, for our final analysis.
To apply a tree-level improvement on the static quark potential, we consider
%
% Eq.11
%
\be
V_I(R)=V(r)
\ee
with $(4\pi R)^{-1}=G(r, 0, 0)$ where $G(\mathbf{r})$ 
is the (scalar) lattice propagator in three dimensions~\cite{Necco:2001xg}.

\begin{figure}
\centering
\includegraphics*[width=0.5\textwidth]{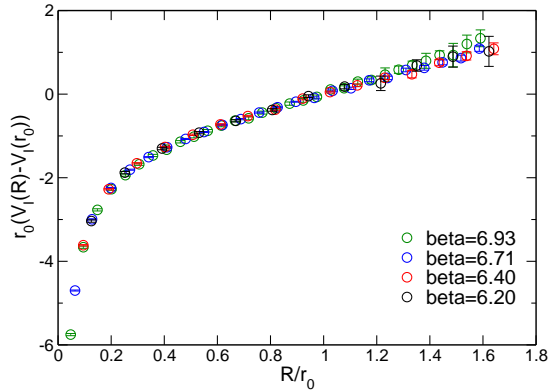} 
\caption{The lattice spacing dependence of $V_I(R)$. The vertical and horizontal axes
are normalized by the Sommer scale $r_0$ given in Ref.~\cite{Necco:2001xg}.
For clarity of the figure, a constant shift has been applied by subtraction
of the value at $R=r_0$.
}
\label{fig:Static-Potential}
\end{figure}

In Fig.~\ref{fig:Static-Potential}, all data points of $V_I(R)$, which are computed at four different lattice spacings, are plotted as a function of $R$. The vertical and horizontal axes are normalized by the Sommer scale $r_0$ given in Ref.~\cite{Necco:2001xg}.
For clarity of the figure, the self-energy contribution is subtracted by the value at $R=r_0$. 
Our results of $V_I(R)$ obtained by spatial gradient flow exhibit good scaling behavior 
with the literature values of $r_0/a$~\cite{Necco:2001xg}. We finally determine $r_0$ from our
results of the static quark potentials computed at four gauge couplings.
For this purpose, we first adopt the Cornell potential parametrization
by fitting the data of $V_I(R)$ as
%
% Eq.12
%
\be
V_I(R)=V_0-\frac{A}{R}+\sigma R
\ee
with the Coulombic coefficient $A$, the string tension $\sigma$, and a constant $V_0$. 
Finally, the parameters are $A$ and $\sigma$ can be used to determine the Sommer scale $r_0$ as
%
% Eq.13
%
\be
\label{Eq:r_0}
r_0=\sqrt{\frac{1.65-A}{\sigma}}
\ee
for each gauge coupling $\beta$.  
In Table~\ref{tab:Cornel_para}, we summarize the fit results of the Cornell potential parameters ($V_0$, $A$, and $\sqrt{\sigma}$) and the Sommer parameter $r_0$ obtained from all four ensembles, in lattice units. Although all measured values of the Sommer parameter are barely consistent with the literature values of $r_0/a$~\cite{Necco:2001xg},
while our estimates of $r_0/a$ are systematically underestimated. 

The origin for the underestimation of $r_0/a$ is twofold. According to Eq.~(\ref{Eq:r_0}), one is the overestimation of $A$, while the other is the overestimation of $\sigma$. 
In the string regime ($R>1~{\rm fm}$)~\cite{Luscher:2002qv}, the effective string theory predicts
that the coefficient $A$ is given by the universal L\"uscher constant $A=\pi/12$~\cite{Luscher:1980ac}, which is smaller than our estimates of $A$. 
Thus, we alternatively choose fits with $A$ fixed at $\pi/12$, though 
our data falls outside of this range. Nevertheless, the obtained results for the string tension $\sigma$
becomes slightly larger than the values tabulated in Table~\ref{tab:Cornel_para}, 
so that the resulting values of $r_0/a$ get smaller and go slightly further away from the literature values. 
We thus consider that the systematic underestimation of our values 
of $r_0$ is mainly caused by a slight overestimation of the string tension since the excited-state contaminations 
are not fully eliminated in our analysis of $V(r)$ especially for large $r$. 
We simply use the double-exponential fit to 
determine $V(r)$ from $\langle W(r,t)\rangle$ instead of the variational method that was adopted in Ref.~\cite{Necco:2001xg}.

Figure~\ref{fig:Static-Potential} shows the lattice spacing dependence of $V_I(R)$. The vertical and horizontal axes
are normalized by the Sommer scale $r_0$ given in Ref.~\cite{Necco:2001xg}. For clarity of the figure, a constant shift has been applied by subtraction of the value at $R=r_0$. Indeed, the scaling behavior, where the data points of $V_I(R)$ measured at different lattice spacings collapse on a single curve, is clearly seen in Fig.~\ref{fig:Static-Potential}.
We hereafter use the literature values of $r_0/a$~\cite{Necco:2001xg} for whole analysis instead of our measured values.

%
%  TABLE.4
%
\begin{table*}[ht]                                                                             
  \caption{
  Summary of the Cornell potential parameters ($V_0$, $A$, and $\sqrt{\sigma}$) and the Sommer parameter $r_0$ 
  in lattice units for all four ensembles.
  \label{tab:Cornel_para}
  }
\begin{ruledtabular}                                                                              
\begin{tabular}{| l c c c c  c c |}
\hline
$\beta$ & $aV_0$ & $A$ & $a\sqrt{\sigma}$ & $r_0/a$ 
& [$r_{\rm min}/a, r_{\rm max}/a$]& $\chi^2/{\rm dof}$ \cr 
\hline
6.2  &  0.623(7)  & 0.257(9)   & 0.1647(38) & 7.16(15) & [2, 9] & 1.02\cr
6.4  &  0.606(9)  & 0.284(26) & 0.1220(32) & 9.58(17) &[4,13] & 0.81\cr
6.71&  0.569(13) & 0.331(65) & 0.0819(39) &14.02(33) & [8,16] &1.15 \cr
6.93&  0.541(8)  & 0.339(43) & 0.0608(29) & 18.84(61) & [8,21] & 0.98\cr
\hline
\end{tabular}
\end{ruledtabular}                                                                                
\end{table*}

%                                                                             
%  FIG.3                                                                           
%                                                                             
%  

%\end{document}
%\clearpage
%%%%%%%%%%%%%%  SEC 5  %%%%%%%%%%%%%%%%%%%%%%%%
\section{Features of the spatial gradient flow}
\label{sec:FEATURES}

\subsection{Comparison with the original gradient flow}
\label{sec:COMP_ORGFLOW}

\begin{figure*}[h]
      \includegraphics[width=0.48\textwidth, bb=0 0 792 612]{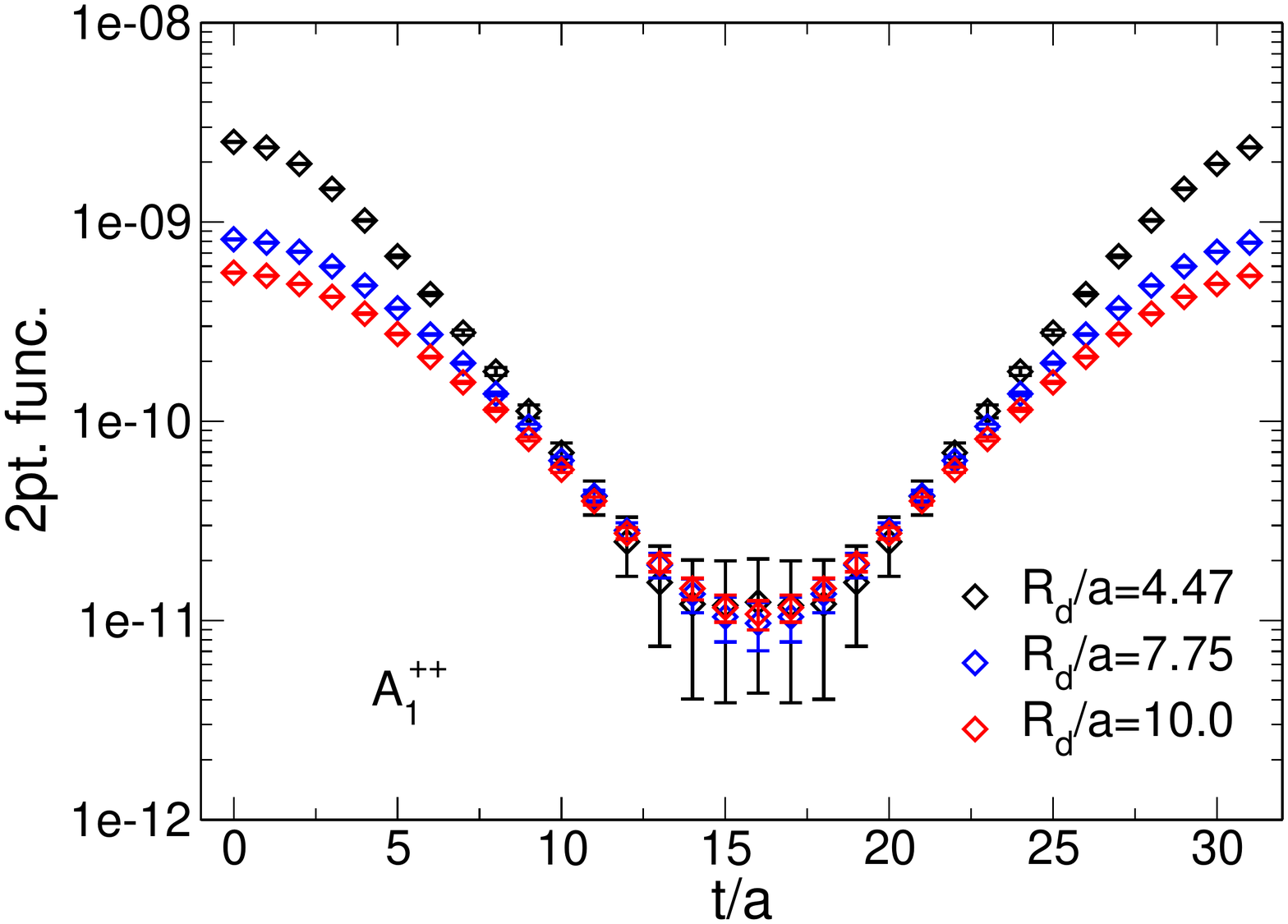}
      \includegraphics[width=0.48\textwidth, bb=0 0 792 612]{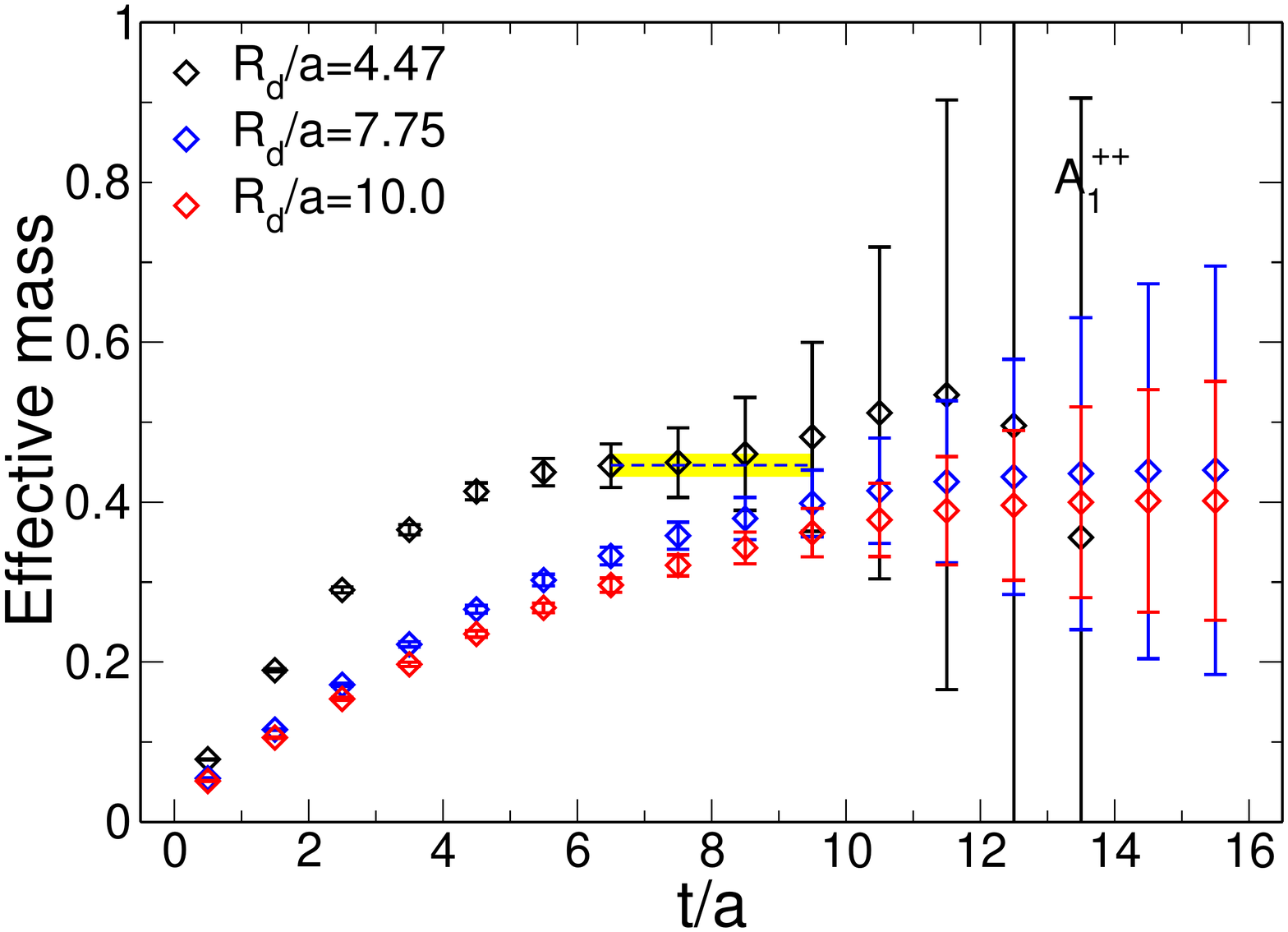}
  \caption{
  Examples of the $A_1^{++}$ glueball results obtained from {\it the original gradient flow}: two-point functions (left) and their effective mass plots (right) as functions of the time slice $t$ for three values of flow time $\tau$.\label{fig:ORG_FLOW}}
\end{figure*}

We first recapitulate the problem of a simple application of {\it the original gradient flow}
to calculate the glueball two-point functions.  In this subsection, we focus on the results 
of the $A_1^{++}$ glueball state calculated on a ${32^4}$ lattice at ${\beta=6.4}$ 
with the ``plaquette" glueball operator ${\cal O}_{\rm plaq}$ as a typical example.
In Fig.~\ref{fig:ORG_FLOW}, we show the results of two-point functions (left panel) and their 
effective mass plots (right panel) using the original gradient flow with three values of flow time $\tau$,
which are represented by the values of the diffusion radius ${R_d}=\sqrt{8\tau}$ in lattice units.
%
% FIG.4
%

\begin{figure*}[h]
      \includegraphics[width=0.48\textwidth, bb=0 0 792 612]{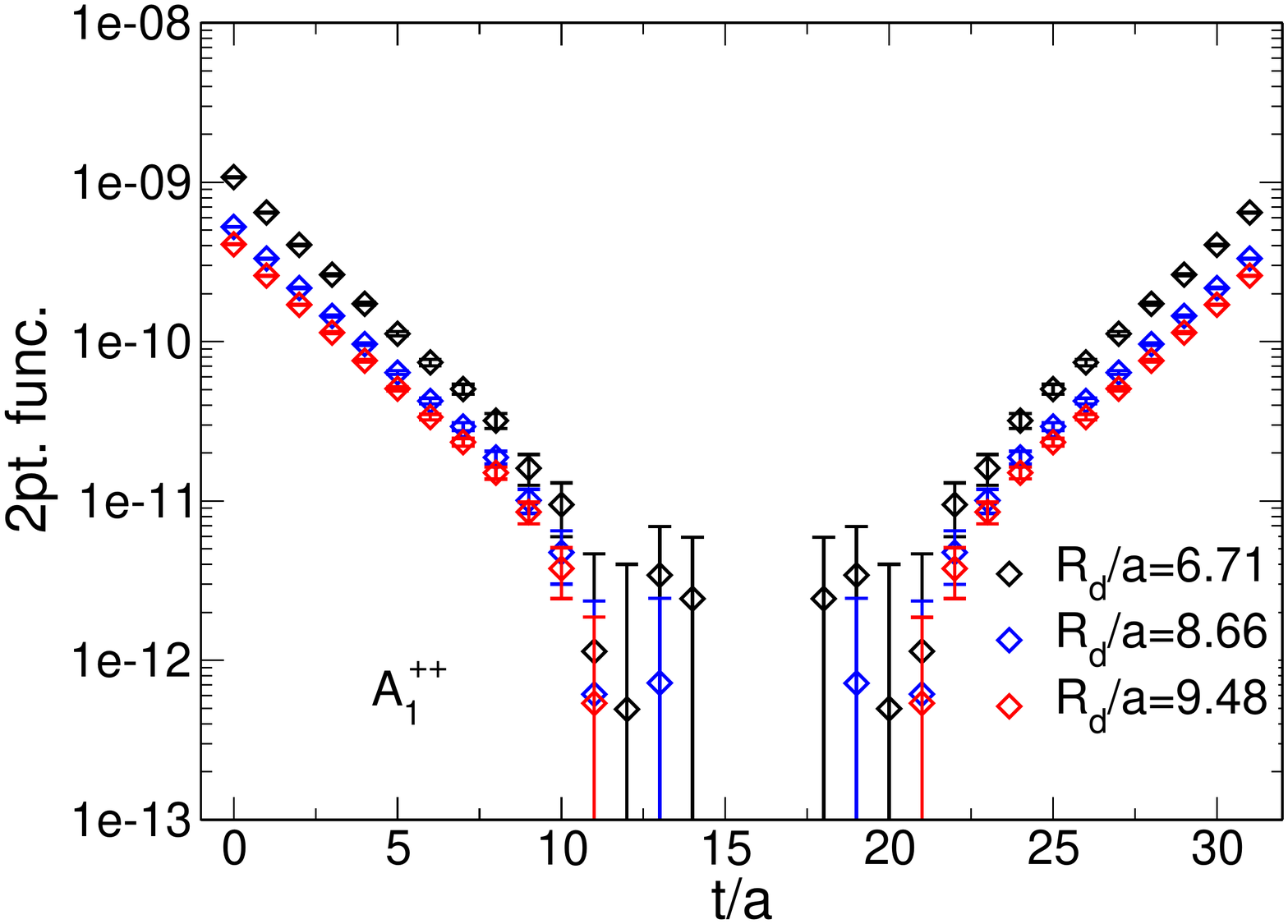}
      \includegraphics[width=0.48\textwidth, bb=0 0 792 612]{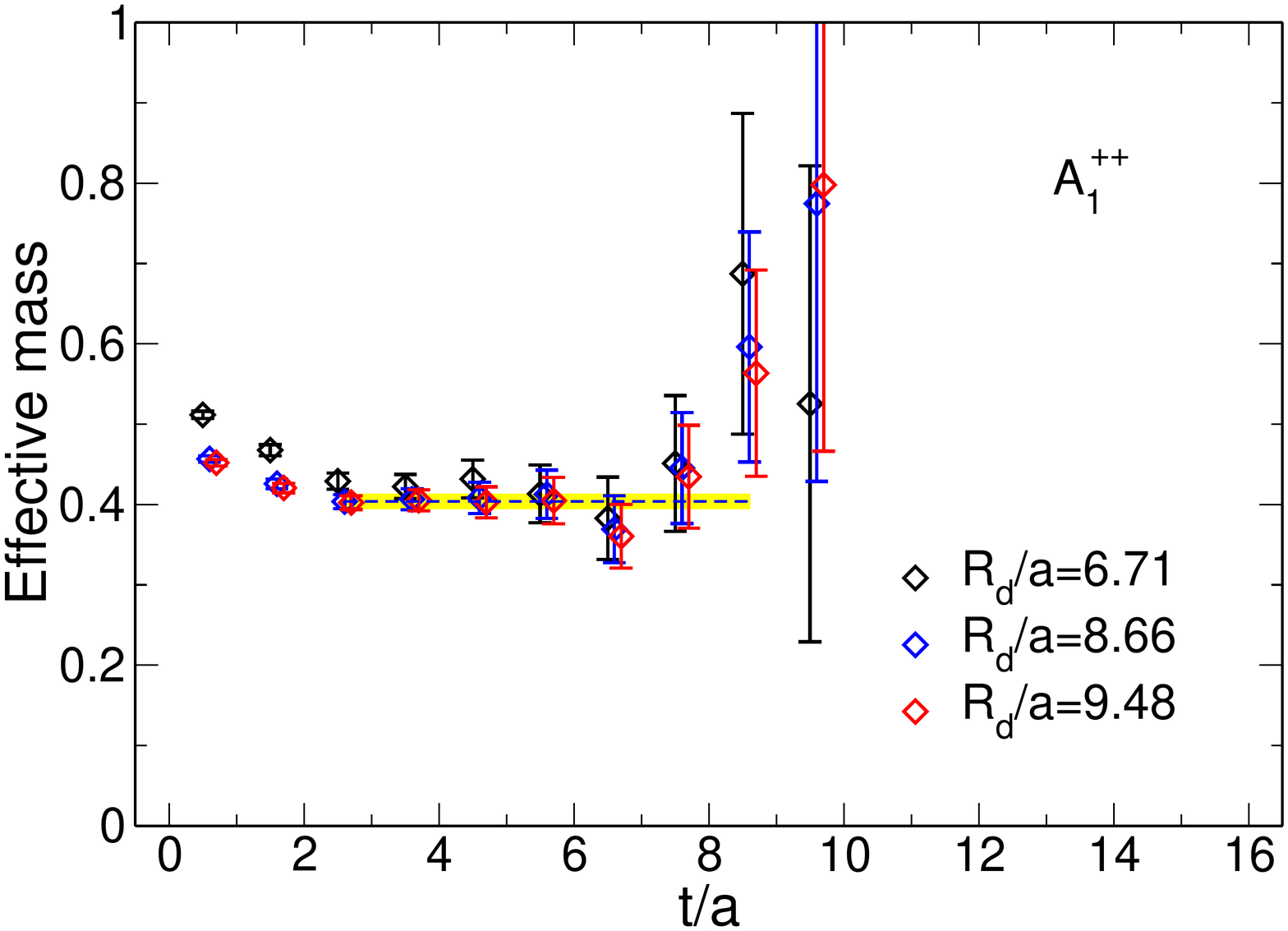}
  \caption{Examples of the $A_1^{++}$ glueball results obtained from {\it the spatial gradient flow}: two-point functions (left) and their effective mass plots (right) as functions of the time slice $t$ for three values of flow time $\tau$.\label{fig:SP_FLOW}}
\end{figure*}

As shown in the left panel of Fig.~\ref{fig:ORG_FLOW}, the statistical errors on the glueball two-point function are dramatically reduced up to the large time slice region as the flow time increases. 
However, the temporal correlation in the region of $t < R_d$ become suffered from the
overlap of two glueball operators which are smeared in space-time 
according to a Gaussian spread. In fact that if the two-point function ${C(t)}$ forms a Gaussian shape, 
$C(t)\propto C_{\rm guass}(t)= e^{-t^2/(2R_d^2)}$, with a Gaussian width corresponding to the diffusion radius ${R_d}$, 
its effective mass gives rise to a peculiar $t$-dependence as 
\begin{equation}
  M_{\rm eff}(t^\prime)=\ln\left\{\frac{C(t)}{C(t+1)}\right\}\approx \frac{t^\prime}{R_d^2},
  \label{eq:linear_emass}
\end{equation}
whose value linearly increases from zero with
a coefficient of $1/R_d^2$ as a function of the time slice $t^\prime=t+\tfrac{1}{2}$.
This feature can be observed in the right panel of Fig.~\ref{fig:ORG_FLOW}, where
each effective mass~\footnote{
To take into account ``the wrap-around effect'' due to 
the periodic boundary condition, 
the effective masses $M_{\rm eff}(t^\prime)$ are given by a solution of 
$
\frac{C(t)}{C(t+1)}=\frac{\cosh[M_{\rm eff}(t^\prime)(t-T/2)]}{\cosh[M_{\rm eff}(t^\prime)(t+1-T/2)]}
$ in the right panels of Fig.~\ref{fig:ORG_FLOW} and
Fig.~\ref{fig:SP_FLOW}.} approximately starts from zero and linearly raise up to around $t\approx R_d$ with increasing of the time slice $t$.
Furthermore, as expected in Eq.~(\ref{eq:linear_emass}), it is easily observed that the slope of the linear dependence 
decreases with the larger flow time. When the shorter flow time such as the case 
of ${R_d/a=4.47}$ is chosen to avoid over smearing, the effective mass shows a plateau behavior in the region of $t > R_d$.
For the longer flow time such as the cases of ${R_d/a=7.75}$ and 10.0, the plateau formation becomes uncertain 
because $R_d$ approaches the vicinity of the temporal midpoint ($t/a=16$), where the signals of the effective mass get noisier.
As a result, the plateau behavior, which highly depends on the choice of the flow time, is too uncertain to extract the ground-state mass of the glueball with high accuracy.

%However, the effective mass tends to approach zero 
%toward the temporal midpoint 
%($t/a=16$) due to the wrap-around effect, so that the %plateau region highly depends on the choice of the flow %time.
%As a result, the plateau behavior is too uncertain to %extract the ground-state mass of the glueball with high %accuracy.

We next show the results obtained from {\it the spatial gradient flow} in Fig.~\ref{fig:SP_FLOW}.
First of all, in the left panel of Fig.~\ref{fig:SP_FLOW}, the exponential falloffs
are clearly seen for all three values of the flow time and their slopes in the asymptotic region 
are independent of the choice of the flow time. The latter point can be confirmed in the right panel of
Fig.~\ref{fig:SP_FLOW}, where their effective mass plots are displayed. For sufficiently large flow time
(${\cal R}_d/a > 6.71$), the plateau behavior in the effective mass plot does not change 
with variation in flow time.
This is a great advantage compared to the original gradient flow. Furthermore, the plateau behavior starts at a smaller time slice, where the true temporal correlation of the glueball two-point function
is kept unaffected during the smearing procedure contrast to the original gradient flow. 
It is another advantage for extracting the ground-state mass of the glueball with high accuracy,
though the large statistical fluctuations still remain in the large $t$ region.

%
% FIG.5
%

Finally, we calculate the ground-state mass of the $A_1^{++}$ glueball by fitting
the glueball two-point function with a single exponential form
for both gradient flow cases. The choice of ${R_d/a=4.47}$ for the original gradient flow is taken 
to avoid over smearing, while the data with ${{\cal R}_d/a=8.66}$ is used
for the spatial gradient flow as a typical example. 
The $A_1^{++}$ glueball masses are respectively evaluated from two types of the gradient flow as below:
\begin{equation}
aM_{A_1^{++}}=\begin{cases}
0.446(14) & \mbox{(original gradient flow)}  \\
0.404(9) & \mbox{(spatial gradient flow)}.
\end{cases}
\end{equation}

In the right panel of Fig.~\ref{fig:ORG_FLOW} and Fig.~\ref{fig:SP_FLOW},
each of the central values and errors is displayed as a blue dotted line and yellow shaded bands within 
the fit range. 
The statistical error on the original gradient flow result
is slightly larger than that of the spatial gradient flow, while
the central value of the former is slightly overestimated in comparison to
the latter. Recall that the central value of the original gradient flow result
tends to be lower when the flow time is taken longer regardless of over smearing. 
Needless to say, the original gradient flow requires the optimal choice of the flow time,
while the spatial gradient flow result becomes stable for the large flow time. 

As will be discussed in Appendix~\ref{sec:FLOW_STOUT}, the spatial gradient flow
is slightly more efficient than the gradient flow, which is diffused in the four-dimensional space-time, in terms of reduction of relative uncertainties. 
As reported in an earlier work~\cite{Teper:1991un}, although the cooling method that {\it can
smoothen the whole four-dimensional space-time} was also used for calculating 
the string tension and glueball masses, the similar conclusion is
made that the results were not better than the conventional
approach that can smoothen only the three-dimensional space.  

%\clearpage	
\subsection{Equivalence to the stout smearing}
\label{sec:EQUIVE}

We will later show numerical equivalence between the spatial gradient flow and the stout smearing in the glueball calculations. 
As emphasized in Ref.~\cite{Morningstar:2003gk}, the stout smearing is a relatively new type 
of smearing technique, which can keep the differentiability with respect to the link 
variables during the smearing procedure.
This property is maintained by the use of the exponential function 
in the stout-link smearing algorithm to remain within the $SU(3)$ group. 
For the gradient flow, the numerical integrations of Eqs.~(\ref{eq:gradient_flow}) and (\ref{eq:spatial_flow})
with respect to the flow time are perform with the Runge-Kutta scheme to obtain the Wilson flow
as a solution of Eqs.~(\ref{eq:gradient_flow}) and (\ref{eq:spatial_flow}). 
This procedure requires the exponentiation of the ``Lie-algebra fields'' for the integration. 
In this sense, neither of the two methods uses the projection
into $SU(3)$ for the flowed or smeared link variables. 

The gradient flow equation can be regarded as a continuous version of
the recursive update procedure in the stout-link smearing 
as pointed out in the original paper~\cite{Luscher:2010iy}.
Moreover, the authors of Ref.~\cite{Alexandrou:2017hqw} relate
the smoothing parameter for other smearing schemes to the gradient flow time $\tau$
under the assumption that the lattice spacing and the smearing parameters are small enough.
For the case of the stout smearing, the corresponding flow time $\tau$ is 
given by the matching relation of $\tau=\rho n_{\rm st}$ 
with the number of stout smearing steps $n_{\rm st}$ for the isotropic 
four-dimensional case of the stout smearing parameters (${\rho}_{\mu \nu}=\rho$)~\cite{Alexandrou:2017hqw}.
We will later rederive the above matching relation in more rigorous manner
in Appendix~\ref{sec:FLOW_STOUT}.

\begin{figure*}[h]
      \includegraphics[width=0.48\textwidth, bb=0 0 792 612]{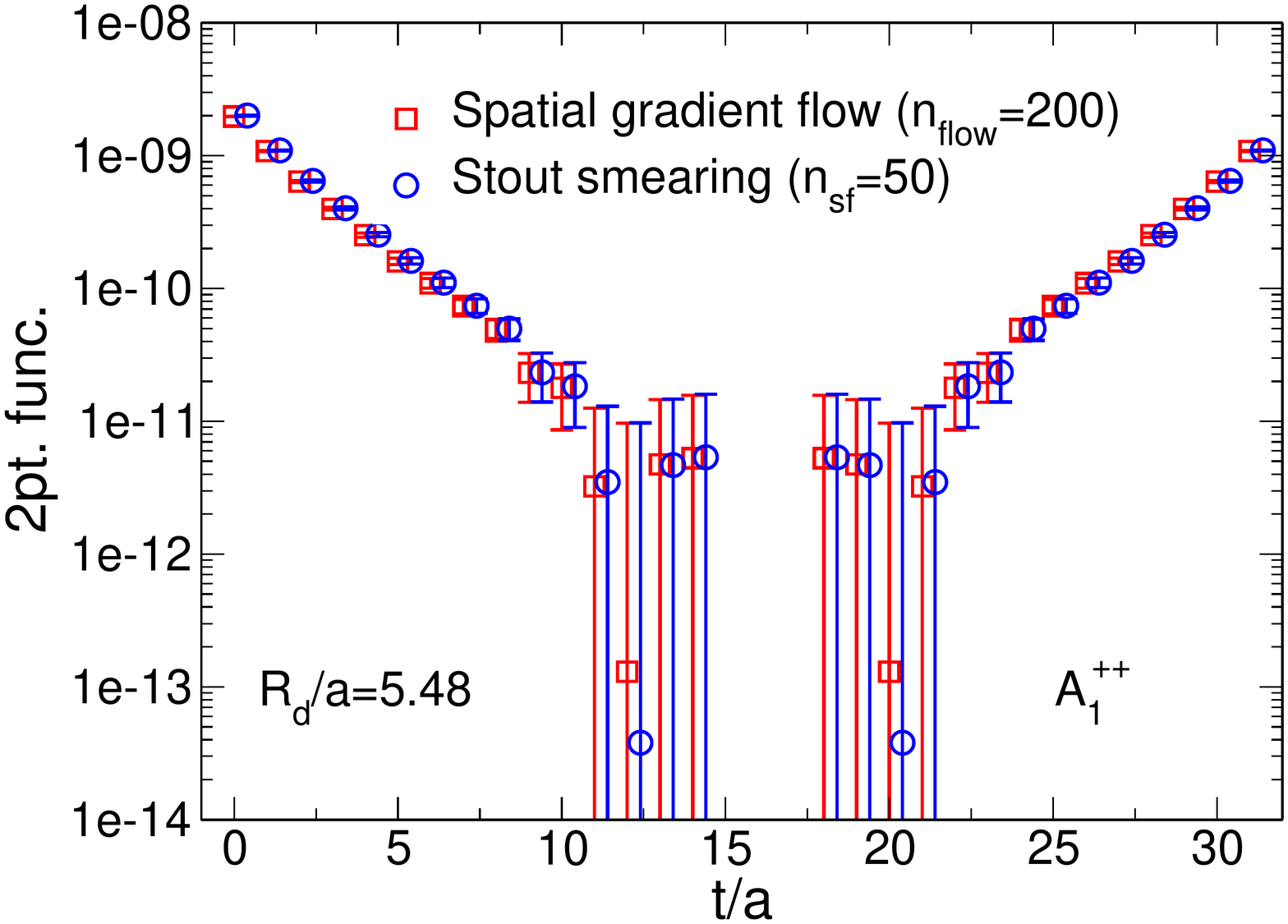}
      \includegraphics[width=0.48\textwidth, bb=0 0 792 612]{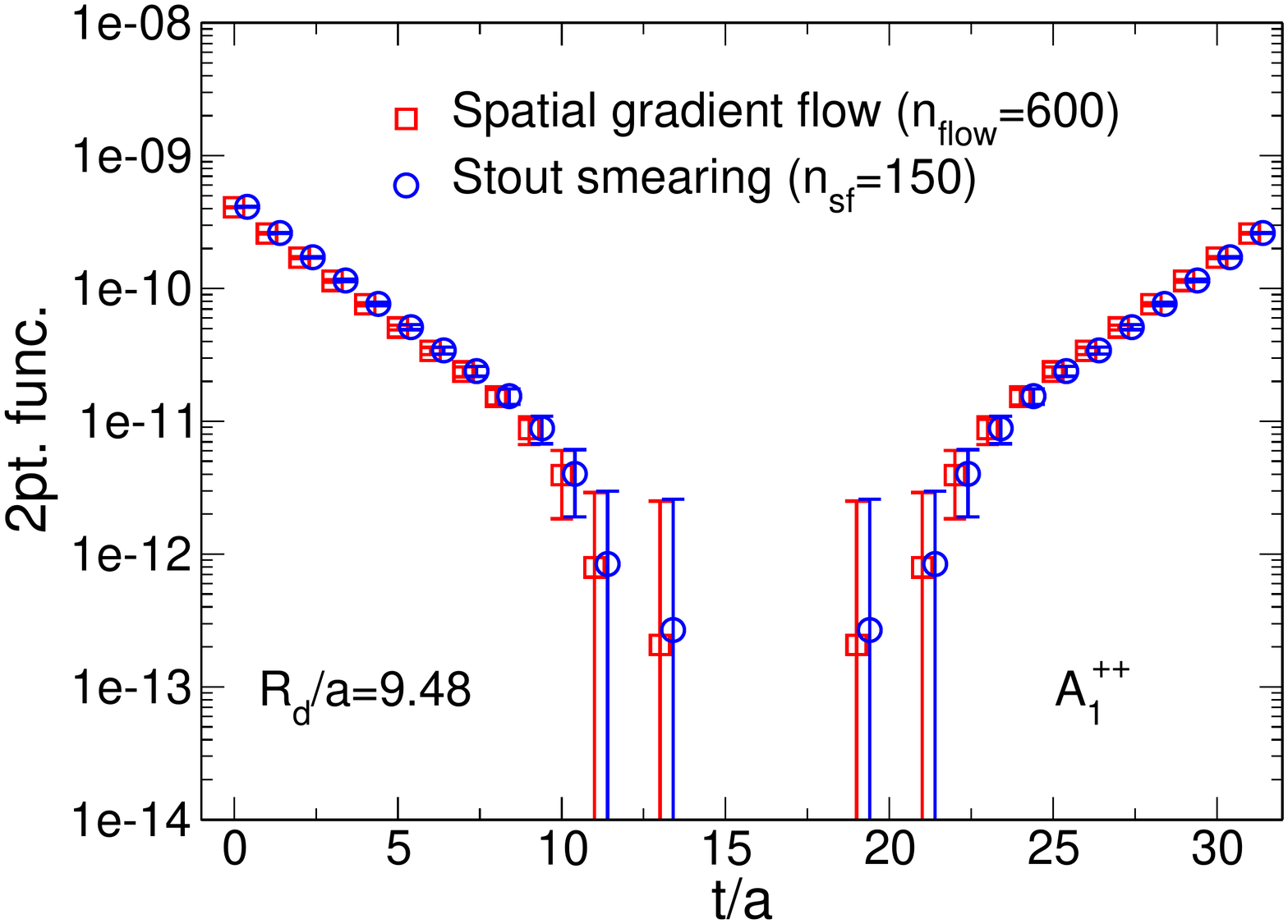}
  \caption{Comparisons of the two-point functions for the $A_1^{++}$ glueball using  
  the spatial gradient flow and the stout smearing.
  The left panel is for the lower-diffusion case (${\cal R}_d/a=5.48$), while the right panel is for the higher-diffusion case (${\cal R}_d/a=9.48$).\label{fig:2PT_SPFLOW_STOUT}}
\end{figure*}

In Fig.~\ref{fig:EM_SPFLOW_STOUT}, we show the effective masses of the $A_1^{++}$ 
glueball state obtained
from the spatial gradient flow and the stout smearing at the same flow time 
$\tau$ that is determined by the matching relation, $\tau=\rho n_{\rm st}$, between the two methods.
In this work, for the stout smearing, the spatially isotropic three-dimensional parameter 
set is chosen to be $\rho_{ij}=\rho=0.1$ and $\rho_{4\mu}=\rho_{\mu 4}=0$.
The numerical equivalence between the two methods is clearly observed 
in both the lower and higher-diffusion cases as shown in Fig.~\ref{fig:2PT_SPFLOW_STOUT}.
It is worth remarking that the values of $n_{\rm st}$ adopted in Fig.~\ref{fig:EM_SPFLOW_STOUT}, 
is much larger than a typical value of less than ten in the usual usage.  Although the usage of the stout smearing 
with a small value of $n_{\rm st}$ is not effective for the glueball calculations, the almost identical 
result to the one made by the spatial gradient flow with the diffusion 
radius ($\sqrt{6\tau}$) can be obtained by the case if the same amount of the diffusion 
radius ($\sqrt{6\rho n_{\rm st}}$) is adopted in the stout smearing.

%
% FIG.6
%
%

%
% FIG.7
%
\begin{figure*}[h]
      \includegraphics[width=0.48\textwidth]{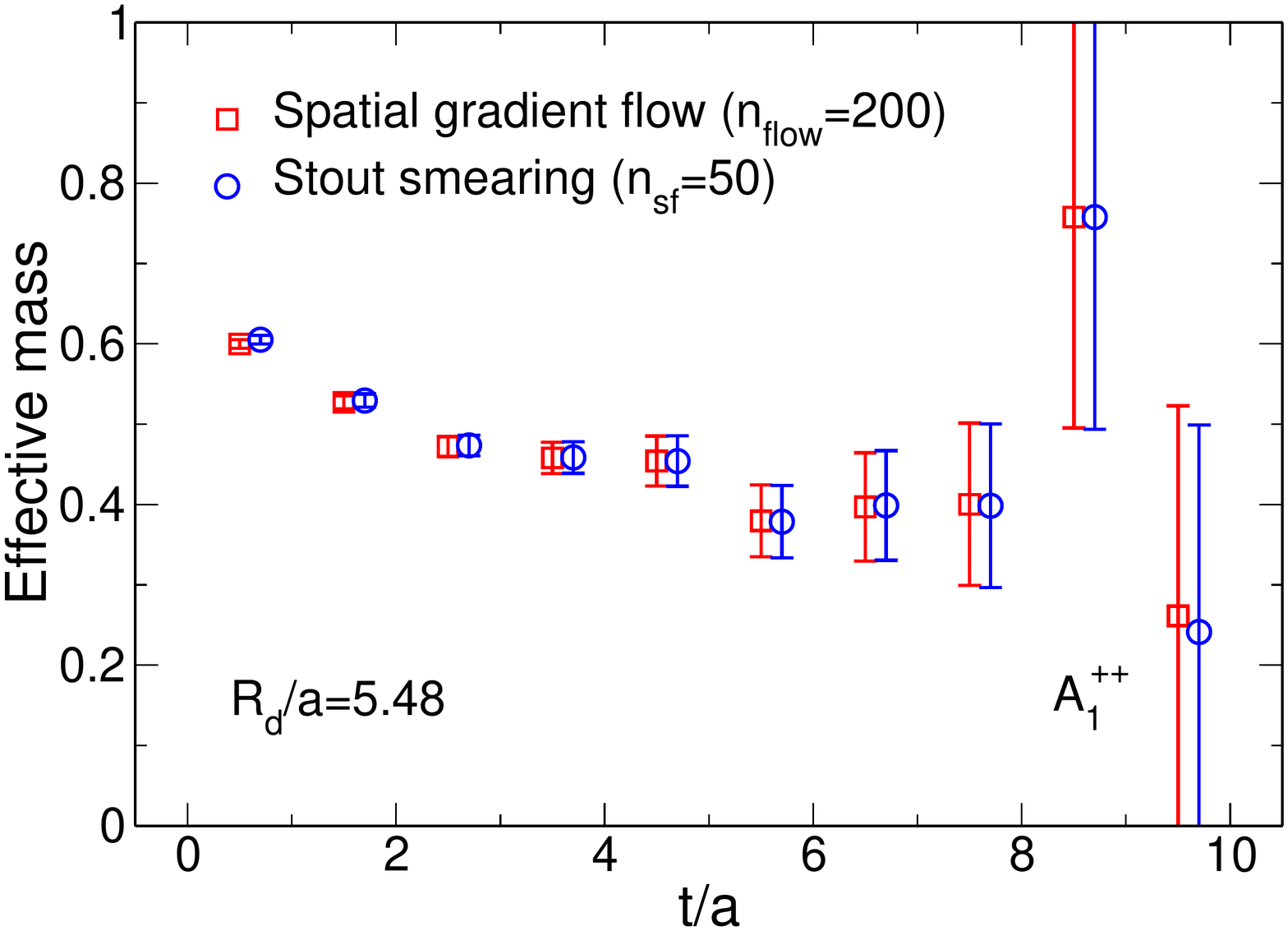}
      \includegraphics[width=0.48\textwidth]{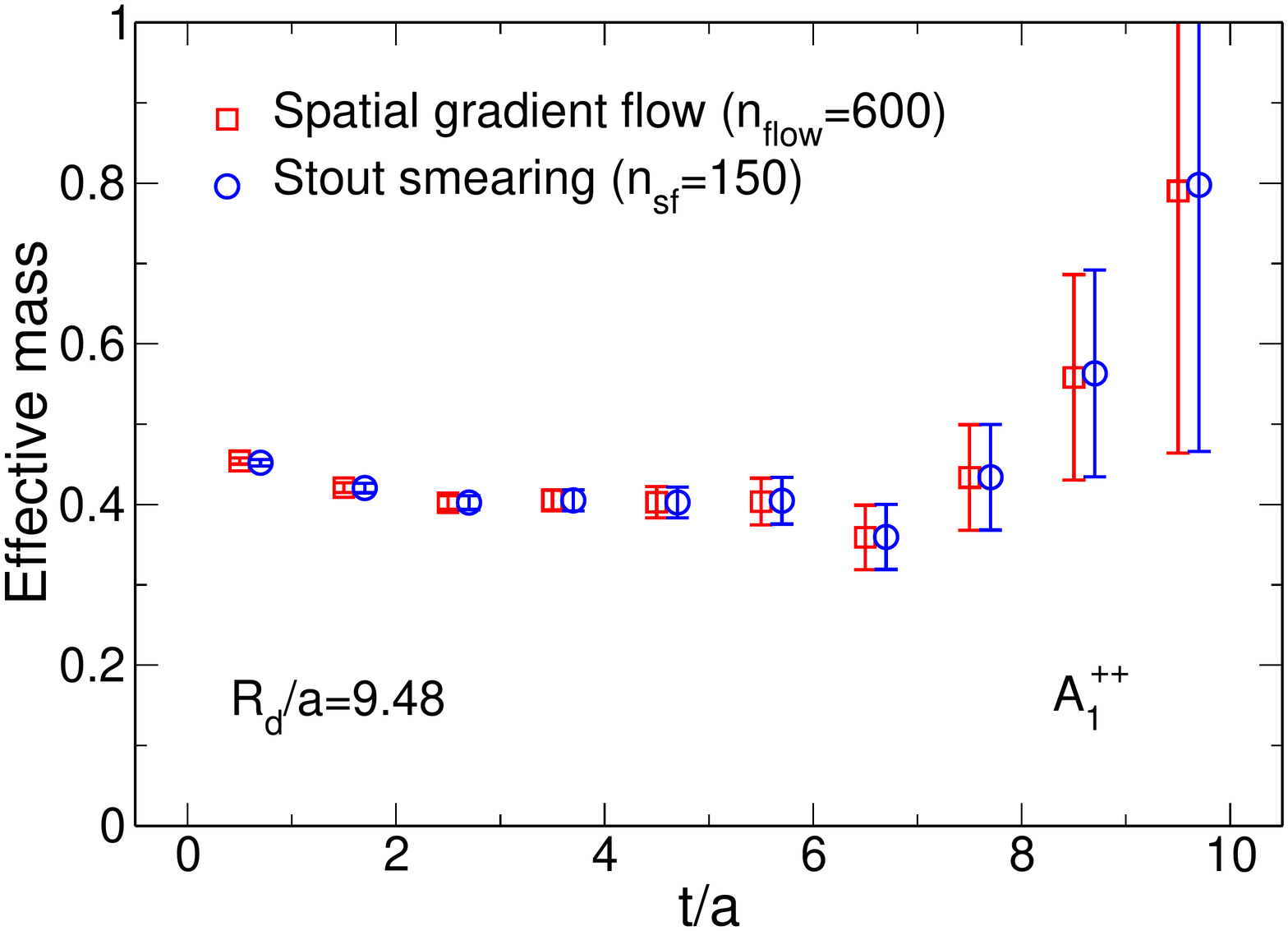}
  \caption{Comparisons of the effective mass plots for the $A_1^{++}$ glueball using the spatial 
  gradient flow and the stout smearing. The left panel is for the lower-diffusion case (${\cal R}_d/a=5.48$), 
  while the right panel is for the higher-diffusion case (${\cal R}_d/a=9.48$).\label{fig:EM_SPFLOW_STOUT}}
\end{figure*}
%

%\clearpage
%%%%%%%%%%%%%%  SEC 6  %%%%%%%%%%%%%%%%%%%%%%%%
\section{Results}
\label{sec:RESULTS}

In this section, we present the results of glueball masses in the four channels ($A_1^{++}$, $A_{1}^{-+}$, $E^{++}$
and $T_2^{++}$) using the spatial gradient flow. To perform the continuum extrapolation,  
we calculate the glueball masses at four different gauge couplings with a fixed physical volume 
($La \approx 1.6$ fm). In this section, we rotate the temporal direction using hypercubic symmetry of 
each gauge configuration and then increase the total number of glueball mass measurements 
by a factor of four as listed in Table~\ref{tab:GB_Paras}. The maximum number of flow iterations corresponds to
the diffusion radius ${\cal R}_d \approx 0.5-0.6$ fm at each ensemble.

%
% TABLE.6
%
\begin{table*}[ht]
  \caption{
  Summary of glueball simulation parameters: gauge coupling, lattice size ($L^3\times T$), 
  the number of the accumulated gauge configurations ($N_{\rm conf}$), 
  the number of measurements per configuration ($N_{\rm meas}$), the number of total measurements ($N_{\rm total}=N_{\rm conf}\times N_{\rm meas}$) and the number of flow iterations ($n_{\rm flow})$.
  \label{tab:GB_Paras}}
  \begin{ruledtabular}  
\begin{tabular}{| l  c c c c l |}
\hline
$\beta$ & $L^3\times T$ & $N_{\rm conf}$ & $N_{\rm meas}$ & $N_{\rm total}$ & $n_{\rm flow}$\cr 
\hline
6.2  & $24^3\times 24$ & 4000 & 4 & 16000 & from 50 to 500 (every 50) \cr
6.4	& $32^3\times 24$ & 3000 & 4 & 12000 & from 50 to 800 (every 50)\cr
6.71& $48^3\times 48$ & 1000 & 4 & 4000  & from 50 to 1400 (every 50)\cr
6.93& $64^3\times 64$ & 500 & 4 & 2000  & from 50 to 2600 (every 50)\cr
\hline
\end{tabular}
\end{ruledtabular}   
\end{table*}

\subsection{Less shape-dependence}
\label{sec:NODEP_SHAPE}

\begin{figure*}
	\includegraphics[width=0.48\textwidth, bb=0 0 792 612]{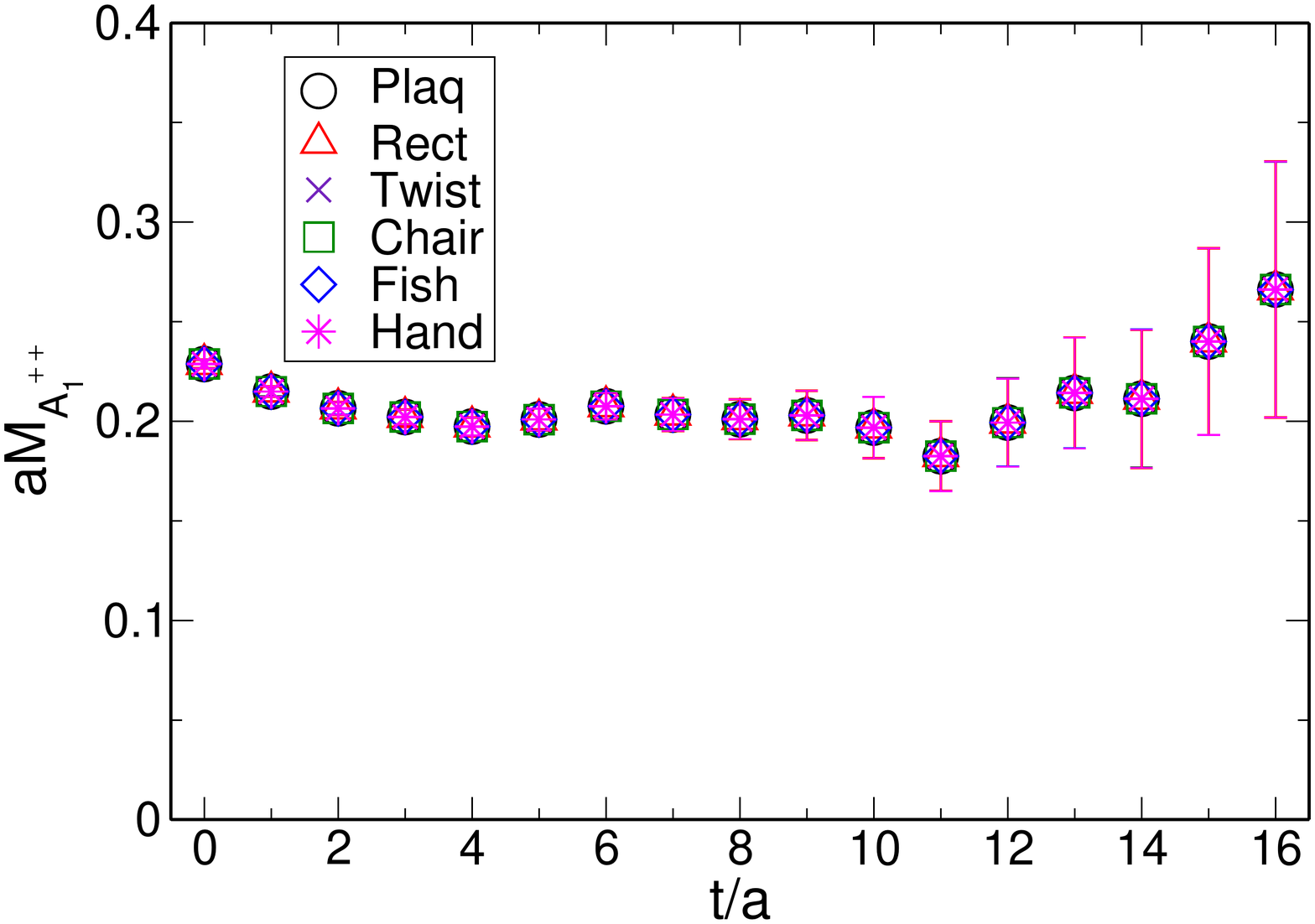}
	\includegraphics[width=0.48\textwidth, bb=0 0 792 612]{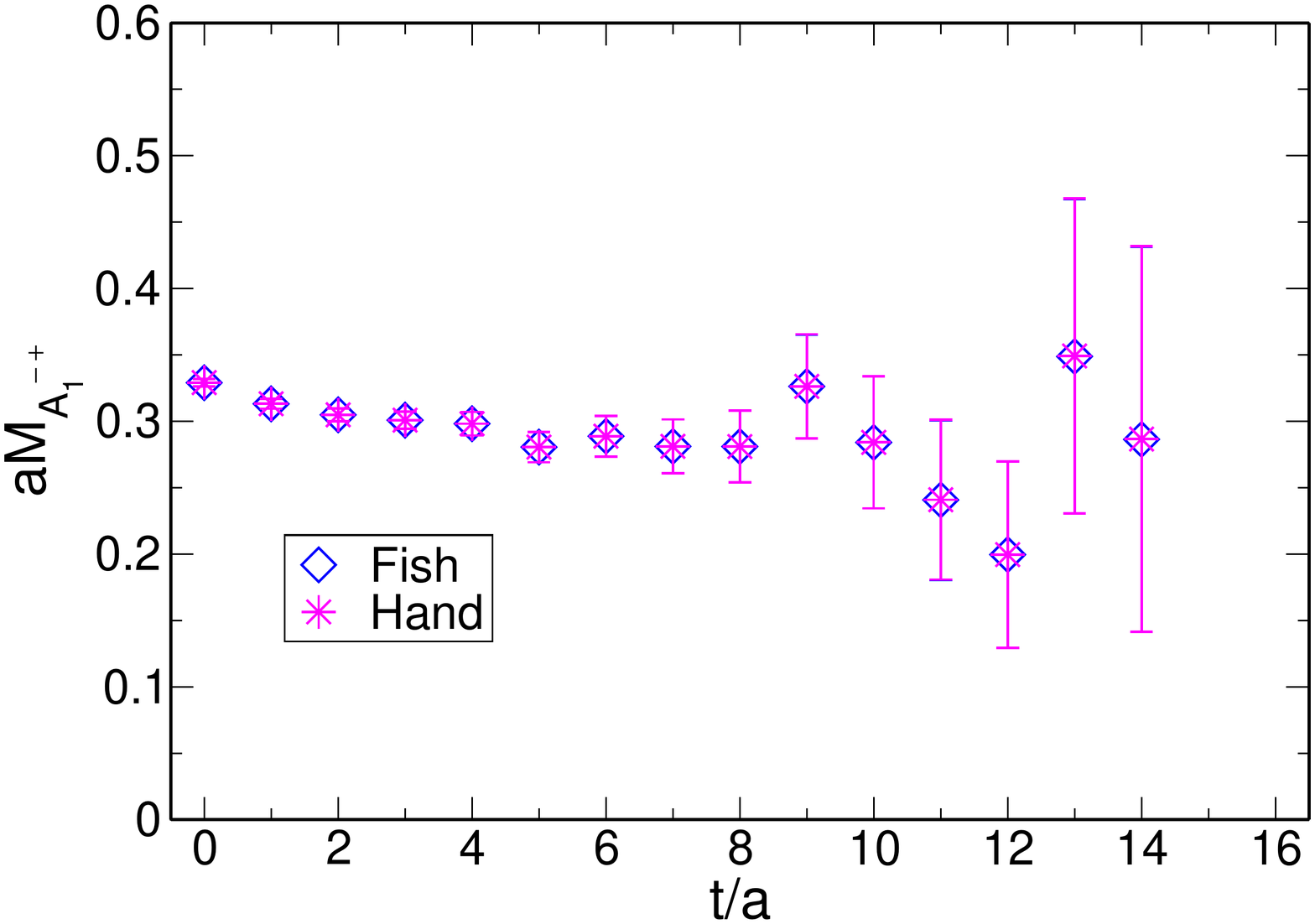}
	\includegraphics[width=0.48\textwidth, bb=0 0 792 612]{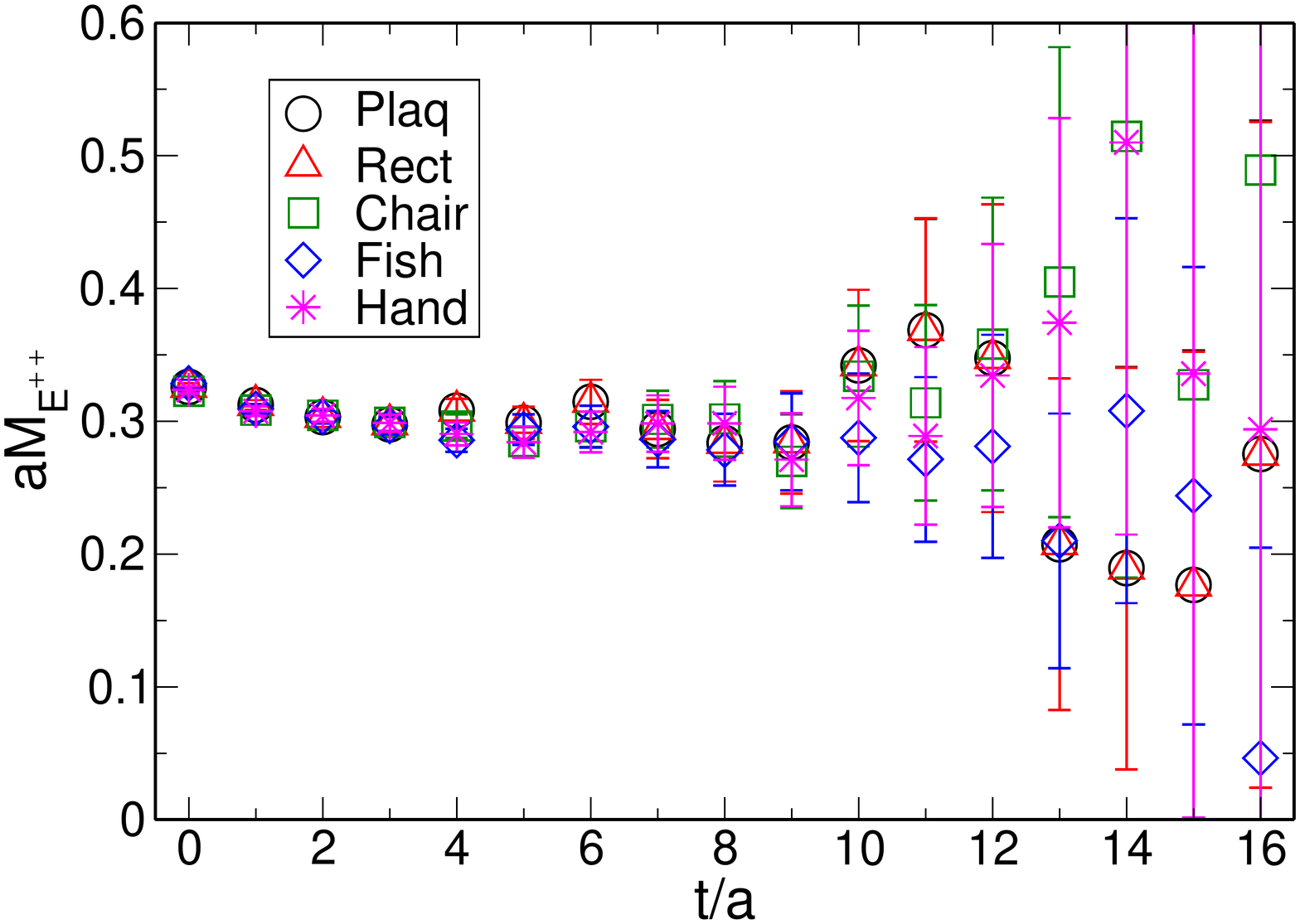}
	\includegraphics[width=0.48\textwidth, bb=0 0 792 612]{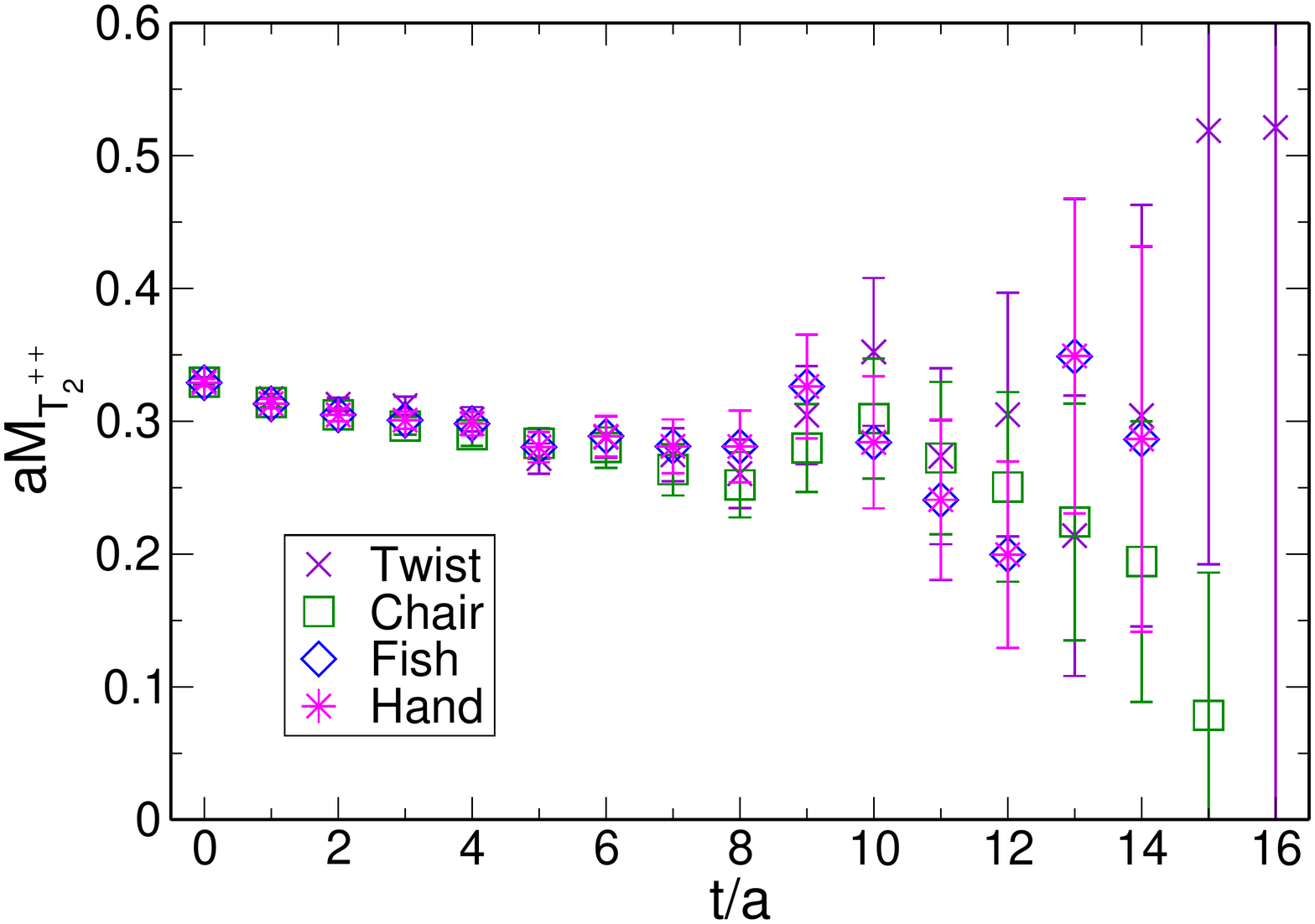}
	\caption{Operator shape-dependence of effective masses with $A_1^{++}$ (top-left), $A_1^{-+}$ (top-right), $E^{++}$ (bottom-left) and $T_{2}^{++}$ (bottom-right) irreps. at $\beta=6.93$ with $n_{\rm flow}=2000$ (${\cal R}_d\approx0.44$ fm).
	\label{fig:NODEP_SHAPE}}
\end{figure*}

As described in Sec.~\ref{sec:EQUIVE}, in this study, we adopt six types of Wilson loop shapes (plaquette, rectangle, 
twist, chair, fish, hand)~\cite{Michael:1988jr} to construct the glueball operators.
In the largest case, the $A_1^{++}$ glueball state can be created with all six operators, and even in the smallest case, 
at least two operators can be used to compute the $A_1^{-+}$ glueball state as summarized in Table~\ref{tab:WL_OP}.
In this sense, the variational analysis~\cite{{Michael:1985ne},{Luscher:1990ck}} based on the different shapes
is in principle applicable, according to Ref.~\cite{Michael:1988jr}.
However, it is found that the shape-dependence of the resulting two-point functions 
disappears due to a strong isotropic nature in the spatial
gradient flow method as shown in Fig.~\ref{fig:NODEP_SHAPE}. 
These figures are displayed as typical examples.

%
% FIG.8
%

The top panels of Fig.~\ref{fig:NODEP_SHAPE} show the effective mass plots for the $A_1^{++}$ (left) and
$A_1^{-+}$ (right) channels at $\beta=6.93$ with $n_{\rm flow}=2000$ (${\cal R}_d\approx0.44$ fm). 
All data points of different symbols including their error bars overlap each other. 
As shown in the bottom panels of Fig.~\ref{fig:NODEP_SHAPE}, some of different operators are almost identical ({\it e.g.} plaquette and rectangle operators for the $E^{++}$ channel, and also fish and hand operators for the $T_2^{++}$ channel), though
all data points of the effective mass are almost similar at a smaller time slice even for the tensor cases.

It is worth remarking that both the spatial gradient flow 
and stout smearing methods share {\it this strong isotropic nature} in the extended glueball operators after the 
large flow time or the high-diffusion case. Therefore, the variational analysis~\cite{{Michael:1985ne},{Luscher:1990ck}} 
based on the different shapes is not applicable at the fixed flow time or the fixed smearing step. 
However, instead of the different shapes, we can use the different diffuseness 
of the extended operator, which is given at the different flow time or the 
different smearing step, to carry out the variational analysis.

\subsection{Variational analysis}
\label{sec:VM_ANAL}

As described in the previous subsection, we perform the variational analysis~\cite{{Michael:1985ne},{Luscher:1990ck}}
with a set of basis operators, which are made of the flowed link variables at the different flow time for a fixed shape $k$. 
In this study, we choose the ``fish" shaped operator ${\cal O}_{\rm fish}$ which contains all irreps of our target states
($A_1^{++}$, $A_1^{-+}$, $E^{++}$, $T_2^{++}$). 

For the variational analysis, we construct the $N\times N$ correlation matrix of two-point functions of glueball states 
for given irreps $\Gamma$ as
%
% Eq.
%
\begin{equation}
C^{\Gamma}_{\alpha \beta}(t)=\sum_{t^\prime}\langle 0| \tilde{\cal O}^{\Gamma}_\alpha(t+t^\prime)
\tilde{\cal O}^{\Gamma}_{\beta}(t^\prime)^\dagger|0\rangle, 
\end{equation}
where the labels of $\alpha$, $\beta$, which run from 1 to $N$, identify the different flow iterations. 
The tilde over ${\cal O}^{\Gamma}_\alpha$ indicates the vacuum-subtracted operator 
as $\tilde{\cal O}^{\Gamma}_\alpha(t)={\cal O}^{\Gamma}_\alpha(t)-\langle 0|{\cal O}^{\Gamma}_\alpha(t)|0\rangle$. 
We next solve the generalized eigenvalue problem,
\begin{equation}
C^{\Gamma}_{\alpha \beta}(t)\omega_{n,\beta}=\lambda_{n, \Gamma}(t,t_0)
\omega_{n,\beta}
\end{equation}
to obtain the $n$th eigenvalue $\lambda_{n, \Gamma}(t,t_0)$, where $t_0$ 
is a reference time slice, and its eigenvector $\omega_{n,\beta}$. 
If only the $N$ lowest states are propagating in the region where $t\ge t_0$,
the $n$th eigenvalue $\lambda_{n, \Gamma}(t,t_0)$ for $n\le N$ is given by a single exponential
form with the rest mass of the $n$th glueball state as 
\begin{equation}
\lambda_{n, \Gamma}(t,t_0)=e^{-(t-t_0)M_{n,\Gamma}},
\end{equation}
which corresponds to the eigenvalue of the transfer matrix
between two time slices $t$ and $t_0$. 
Details of how to practically compute the eigenvalues $\lambda_{n, \Gamma}(t,t_0)$
are described in Appendix B of Ref.~\cite{Sasaki:2006jn}.
An effective mass is defined as
\begin{equation}
M^{\rm eff.}_{n, \Gamma}(t)=\ln \frac{\lambda_{n,\Gamma}(t,t_0)}{\lambda_{n,\Gamma}(t+1,t_0)}, 
\end{equation}
where $\lambda_{n,\Gamma}(t,t_0)$ is the $n$th eigenvalue of the $N\times N$ correlation matrix 
for $\Gamma=A_1^{++}$, $A_1^{-+}$, $E^{++}$, $T_2^{++}$. 
In this study, we choose $N=6$ and the reference time slice as $t_0/a=0$, where the resulting mass is less sensitive 
to variation of $t_0$.

\begin{figure*}
	\includegraphics[width=0.48\textwidth, bb=0 0 792 612]{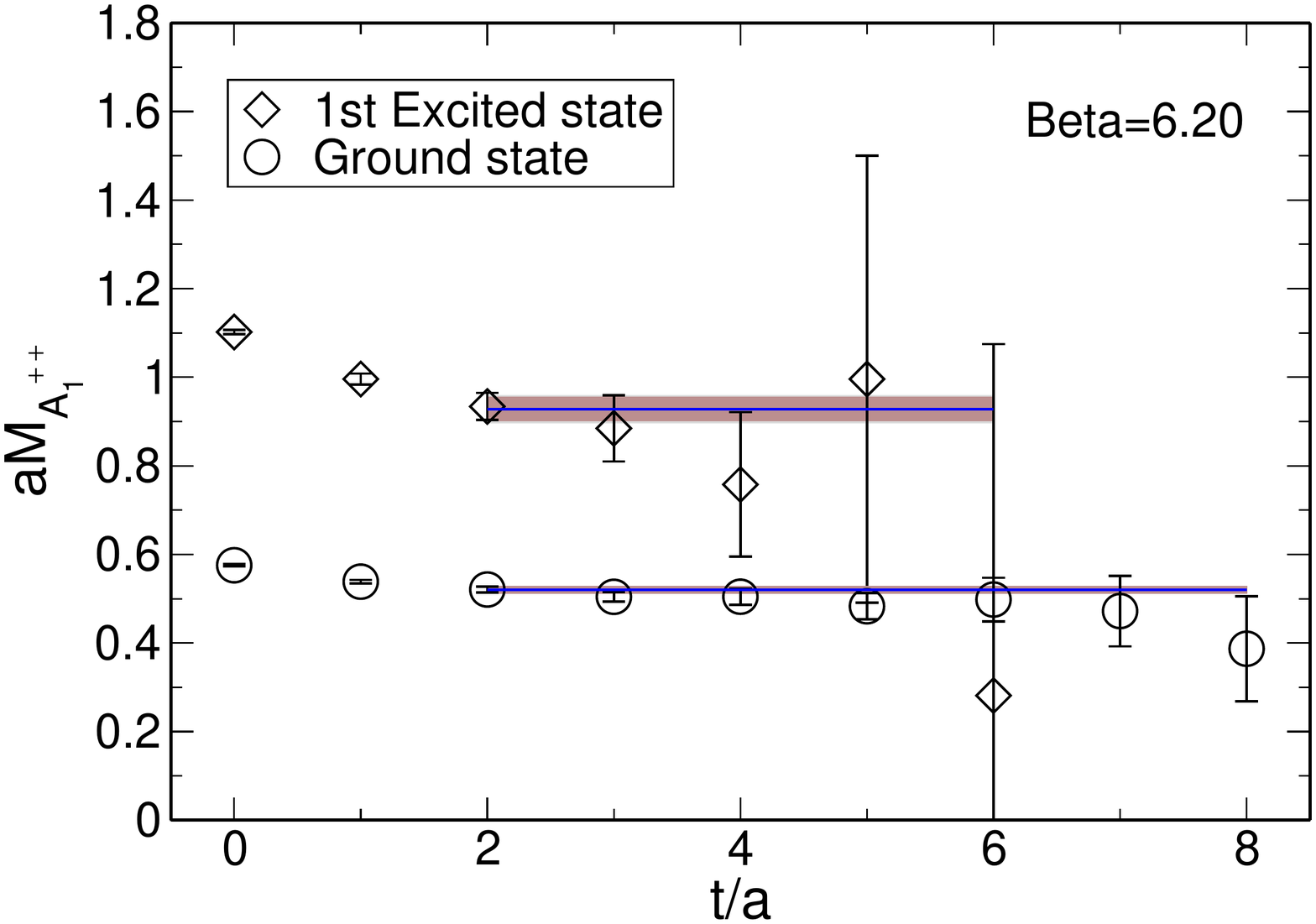}
	\includegraphics[width=0.48\textwidth, bb=0 0 792 612]{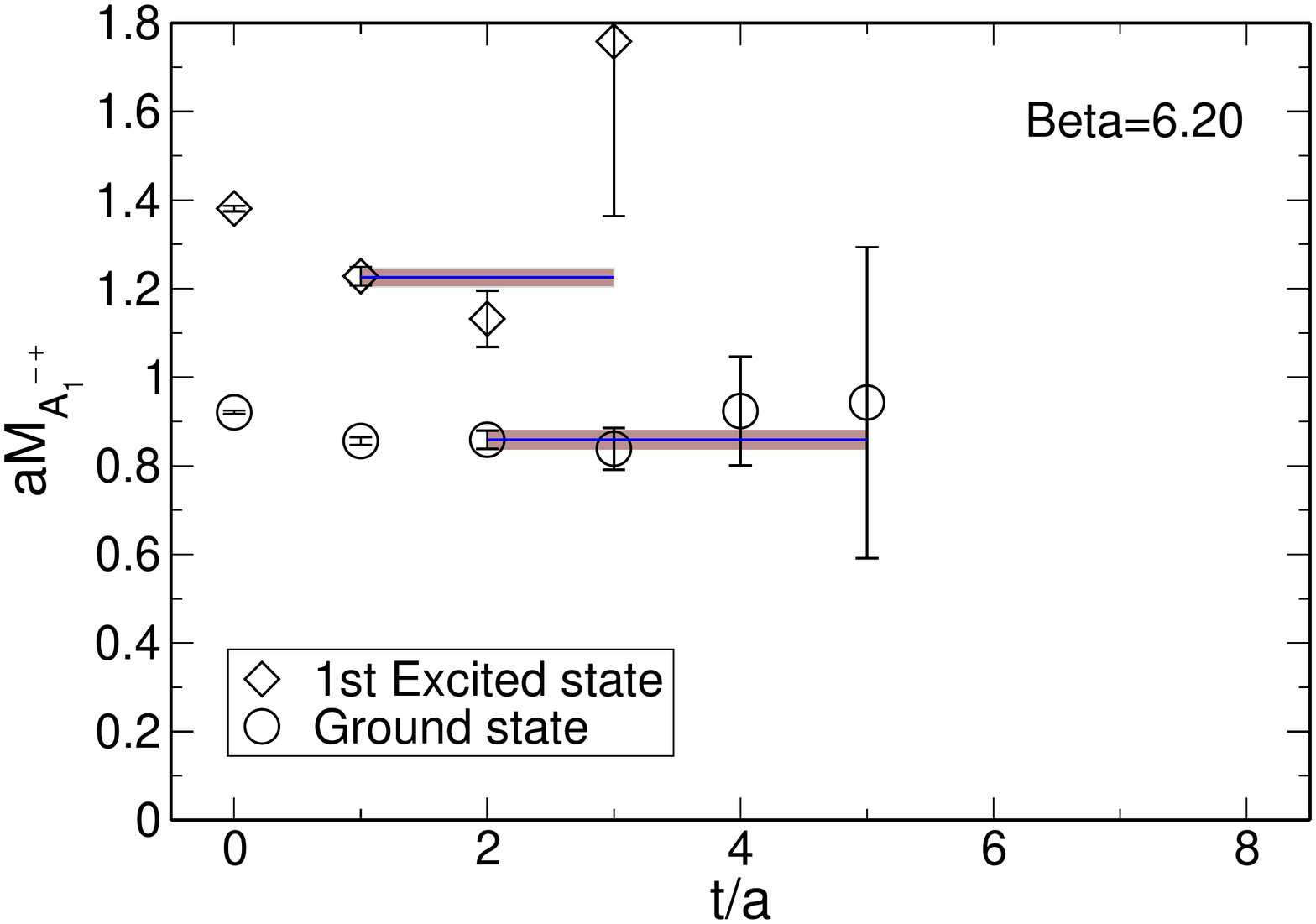}
	\includegraphics[width=0.48\textwidth, bb=0 0 792 612]{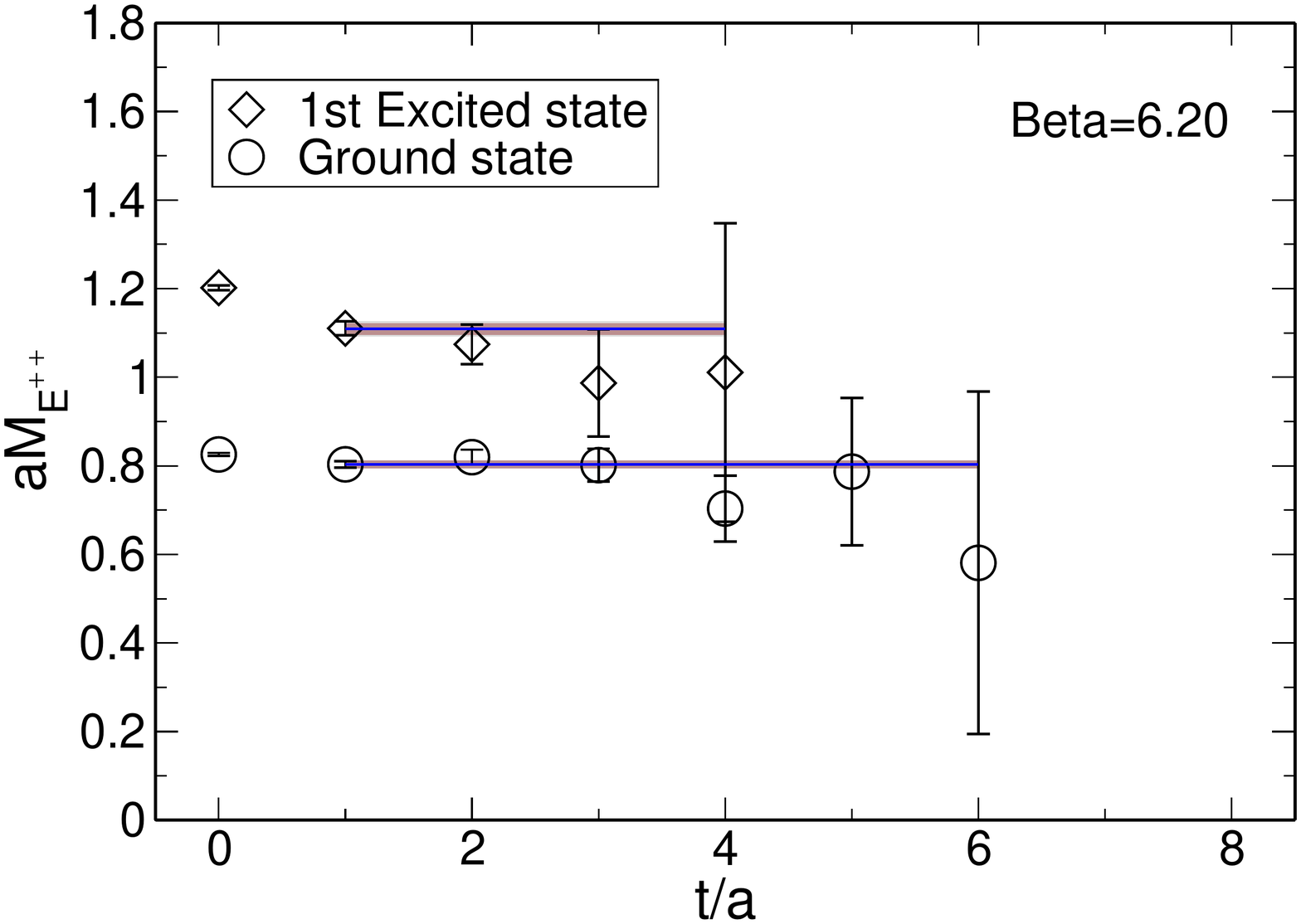}
	\includegraphics[width=0.48\textwidth, bb=0 0 792 612]{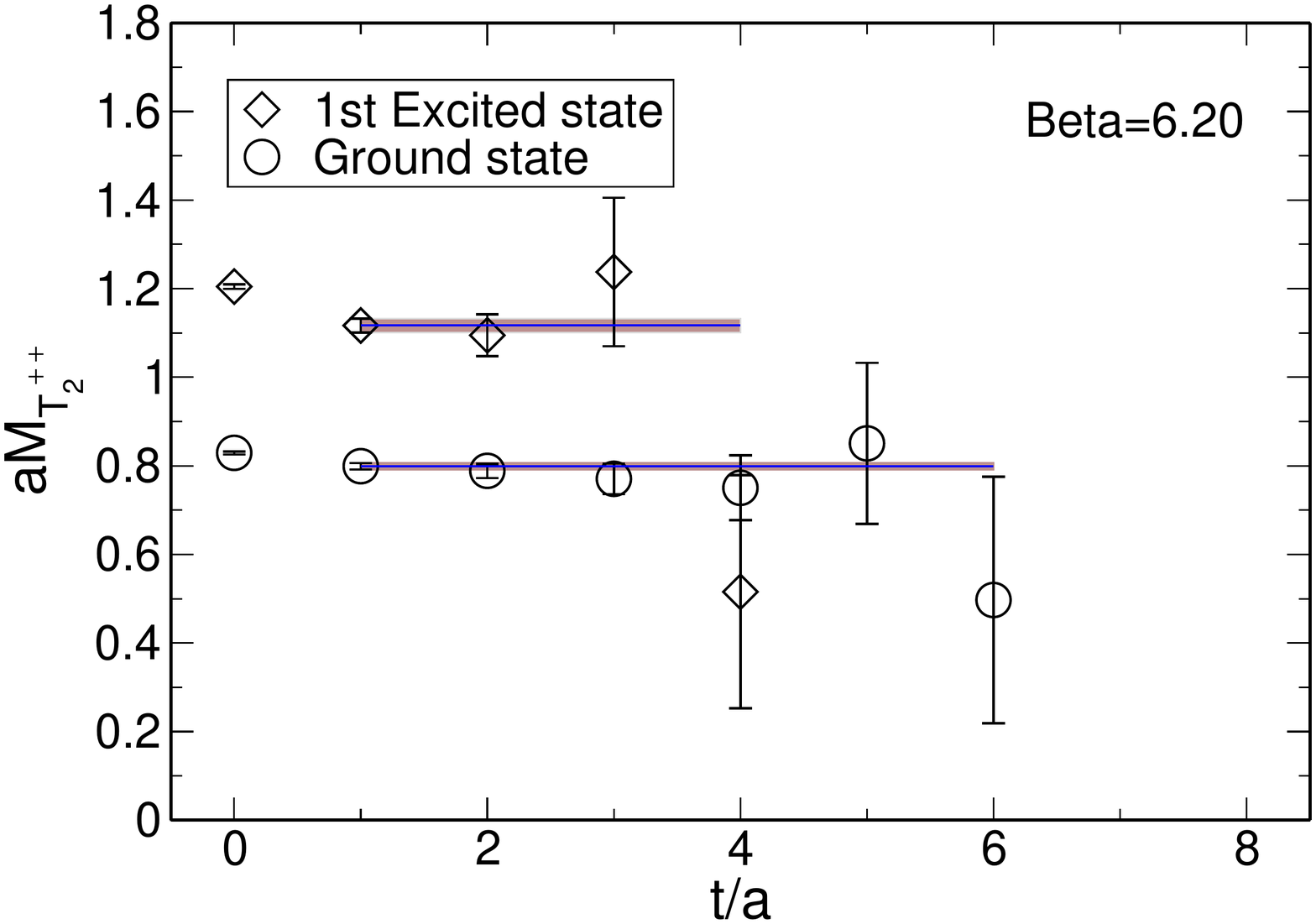}
	\caption{Effective mass plots for the ground state and the first excited state in 	
	 $A_1^{++}$ (top-left), $A_1^{-+}$ (top-right), $E^{++}$ (bottom-left), and $T_{2}^{++}$ (bottom-right) channels 
	 at $\beta=6.20$.
	\label{fig:VM_B620}}
\end{figure*}

\begin{figure*}
	\includegraphics[width=0.48\textwidth, bb=0 0 792 612]{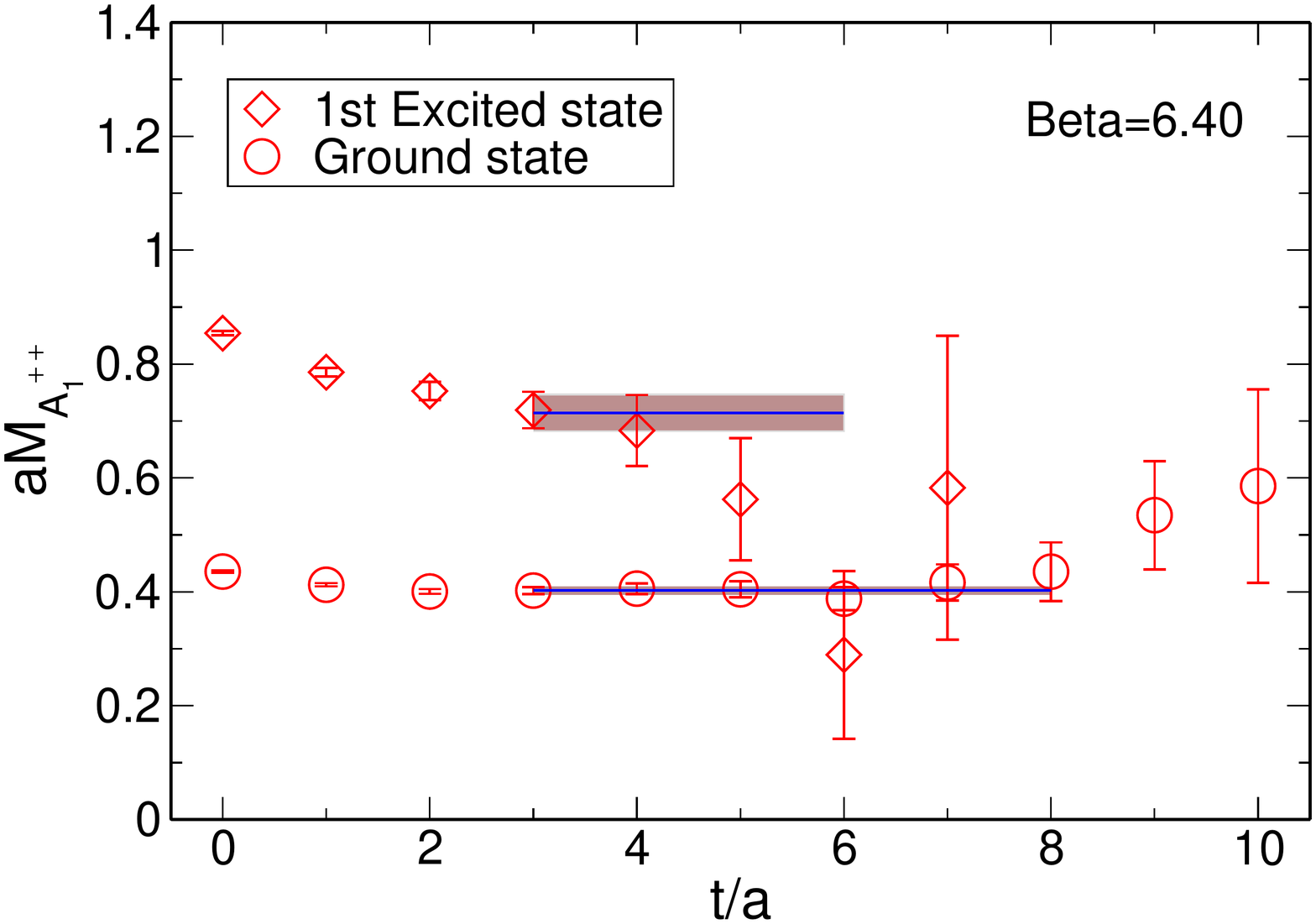}
	\includegraphics[width=0.48\textwidth, bb=0 0 792 612]{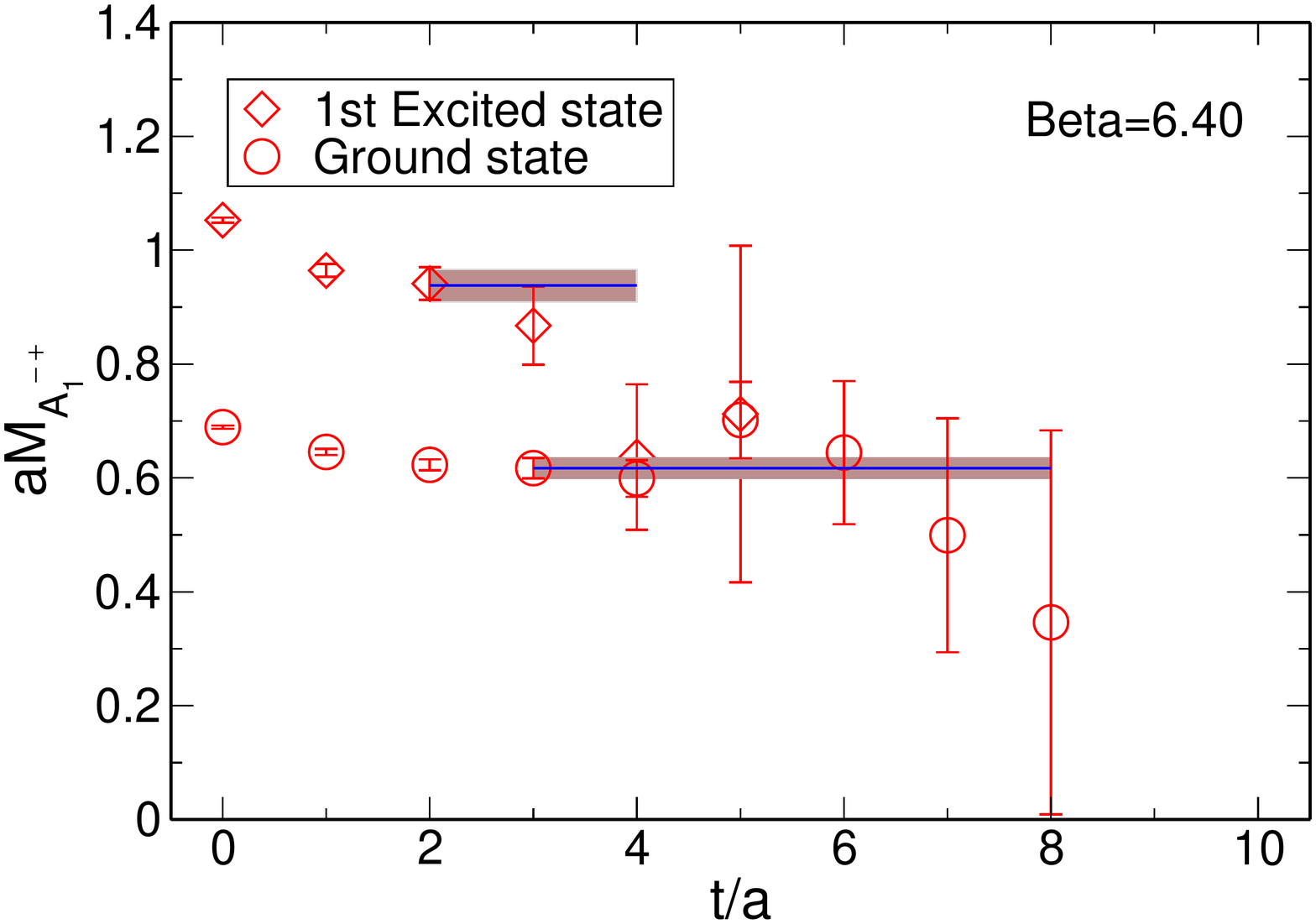}
	\includegraphics[width=0.48\textwidth, bb=0 0 792 612]{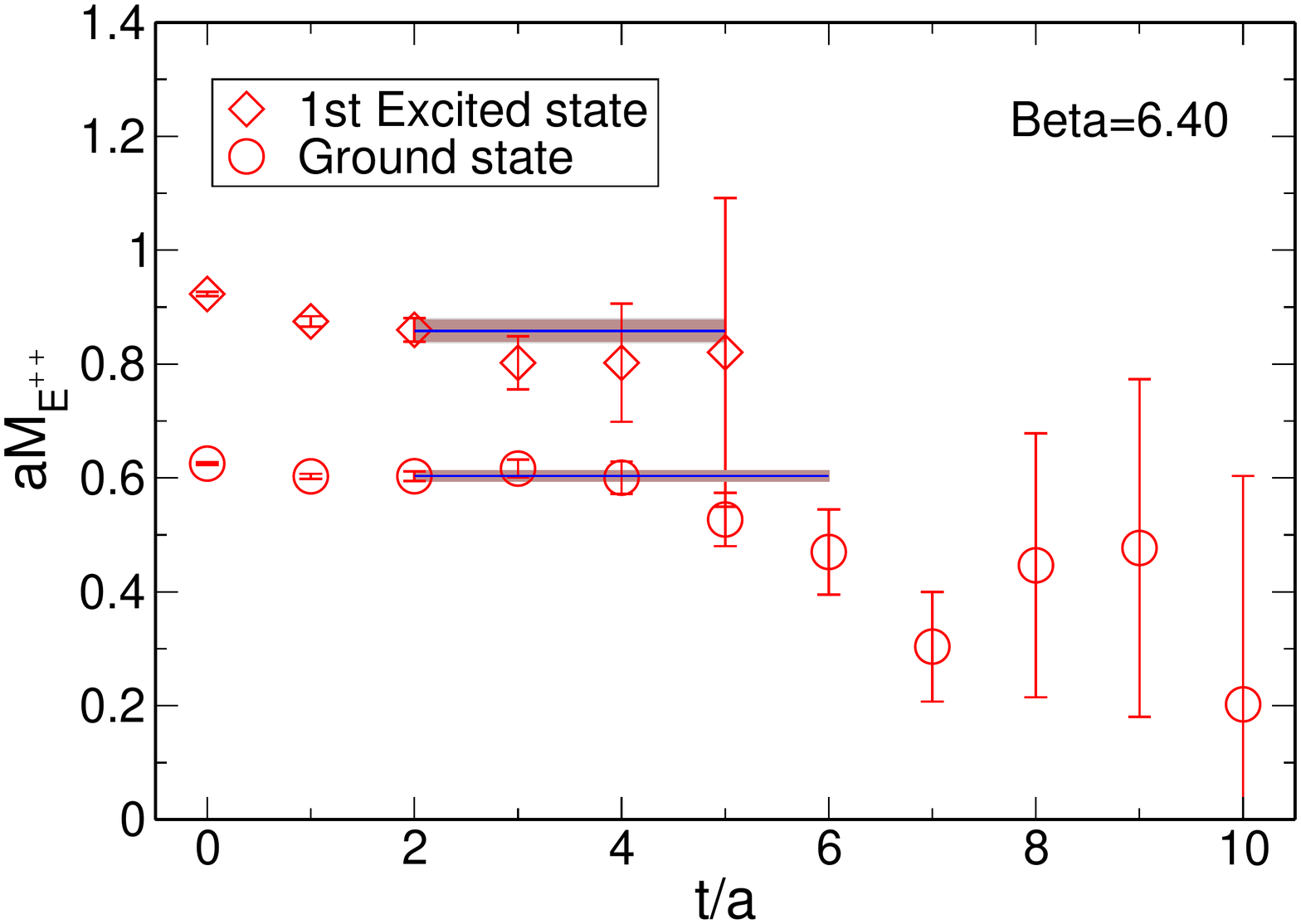}
	\includegraphics[width=0.48\textwidth, bb=0 0 792 612]{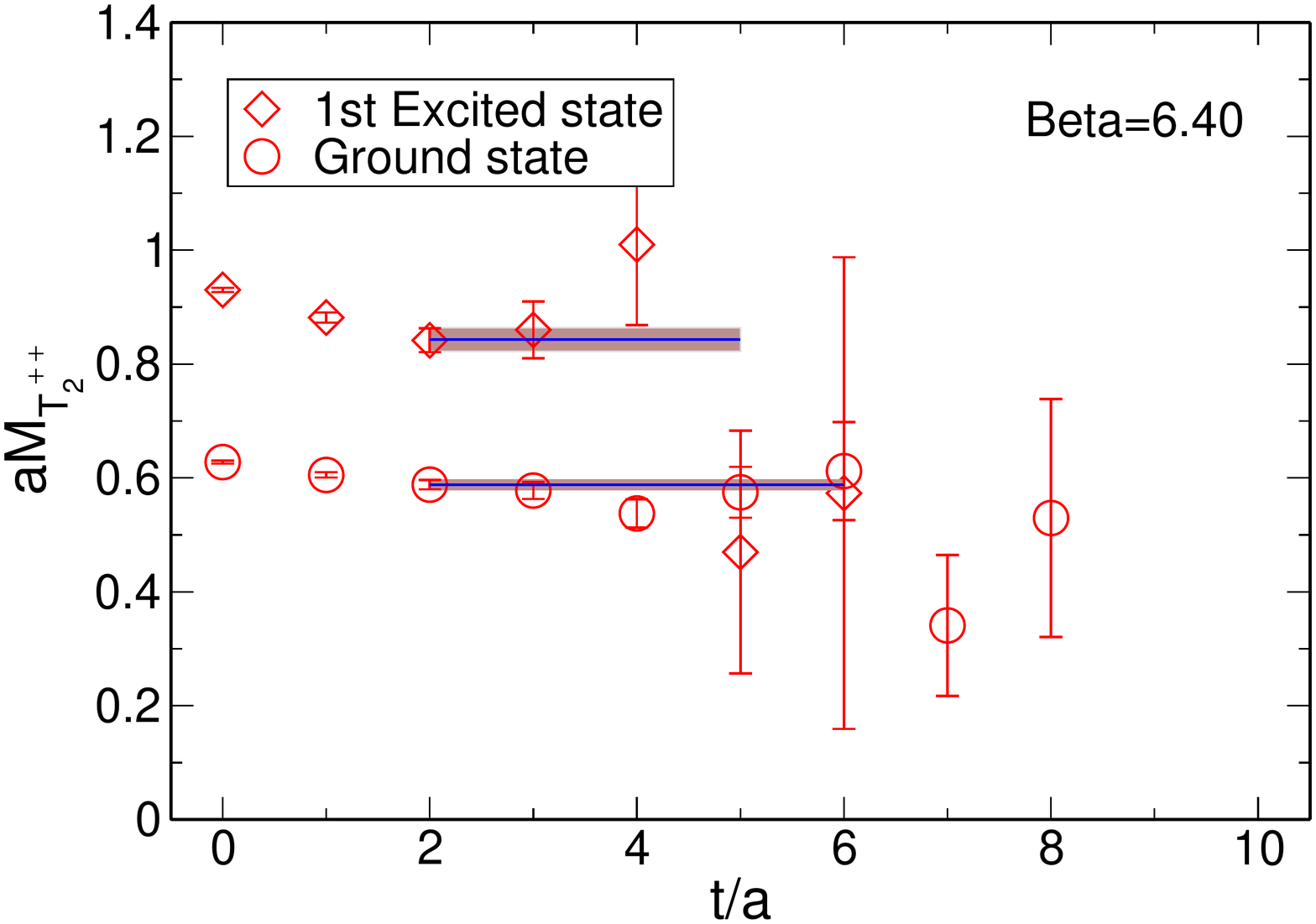}
	\caption{Effective mass plots for the ground state and the first excited state in 	
	 $A_1^{++}$ (top-left), $A_1^{-+}$ (top-right), $E^{++}$ (bottom-left), and $T_{2}^{++}$ (bottom-right) channels 
	 at $\beta=6.40$.
	\label{fig:VM_B640}}
\end{figure*}

\begin{figure*}
	\includegraphics[width=0.48\textwidth, bb=0 0 792 612]{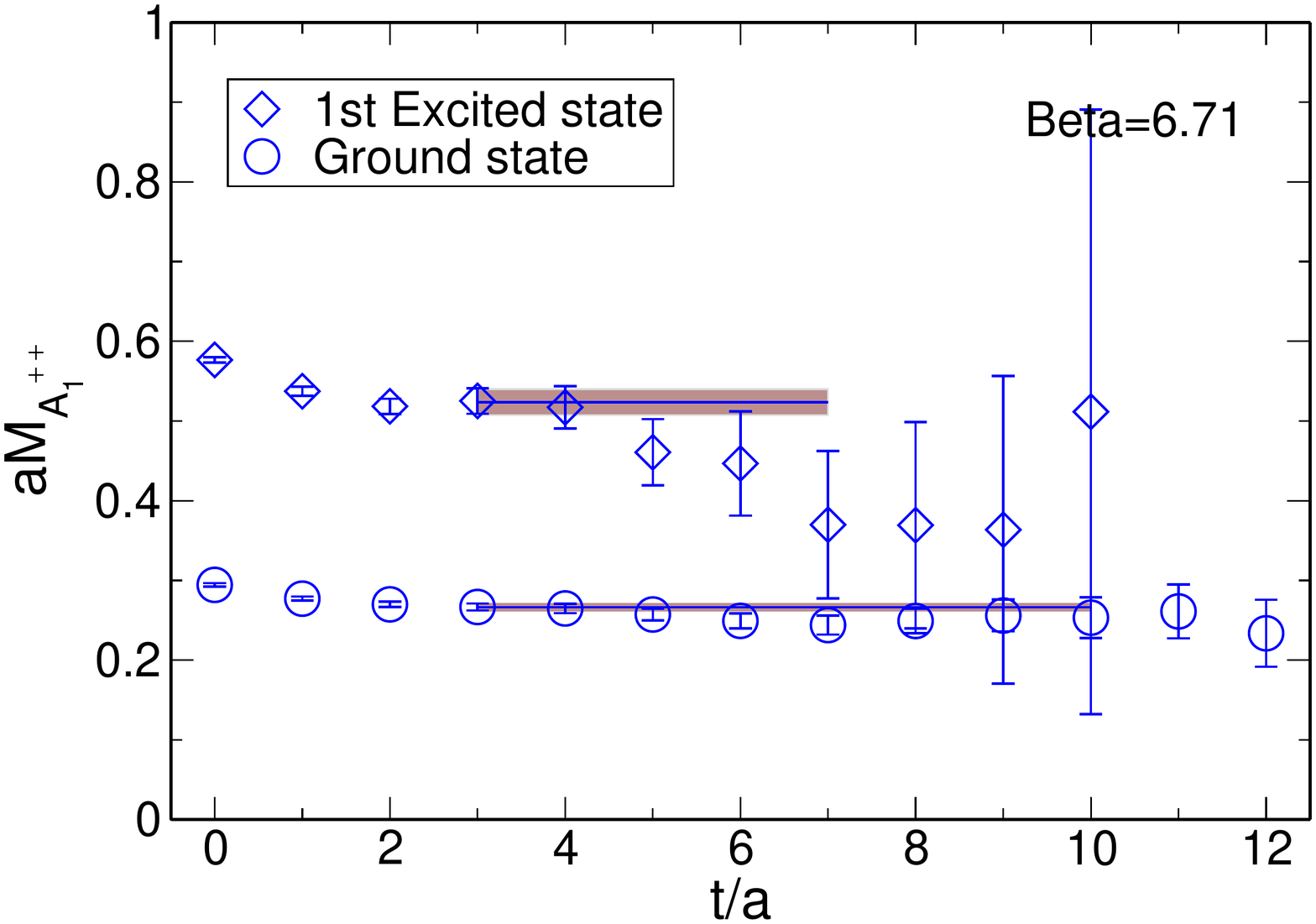}
	\includegraphics[width=0.48\textwidth, bb=0 0 792 612]{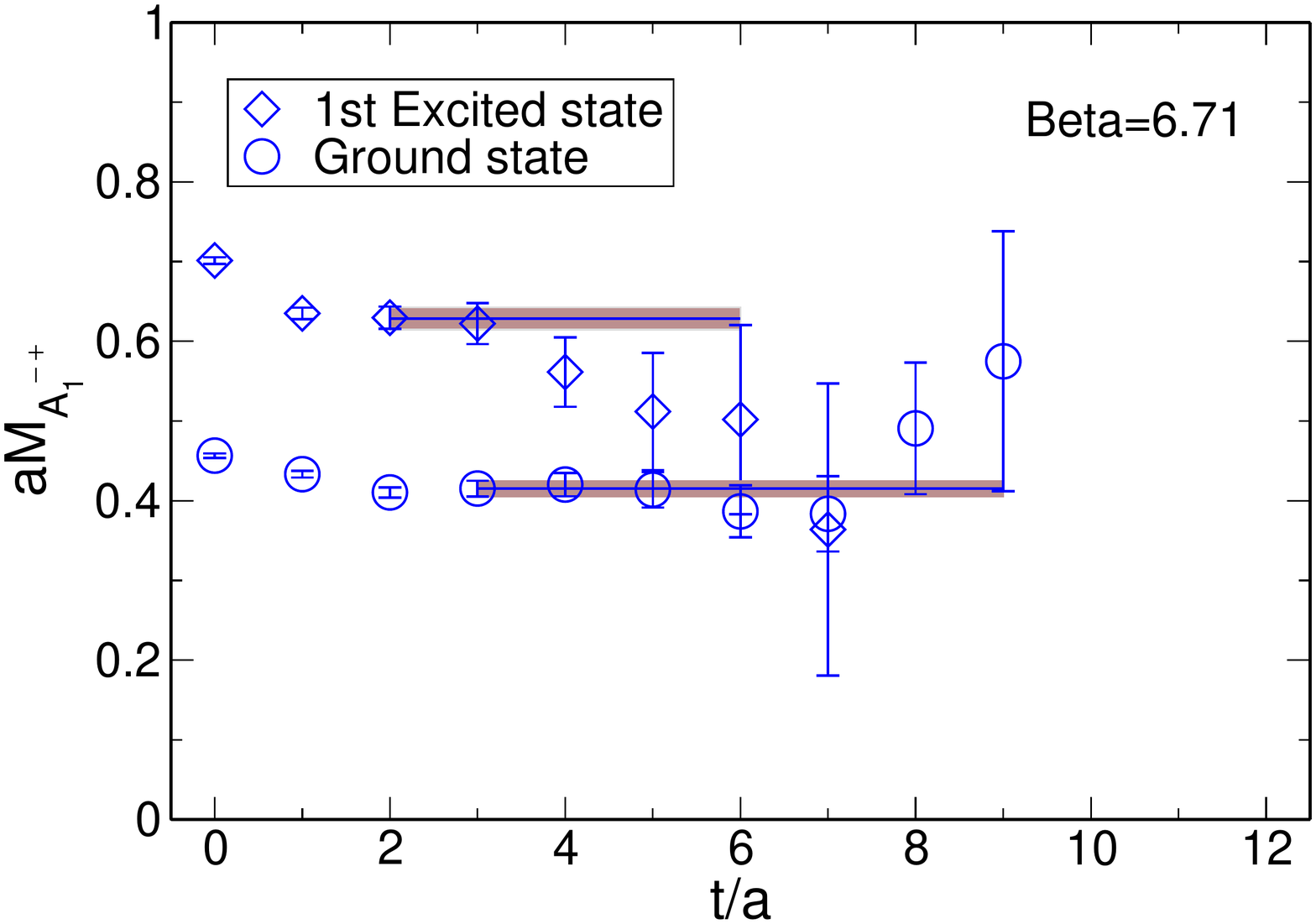}
	\includegraphics[width=0.48\textwidth, bb=0 0 792 612]{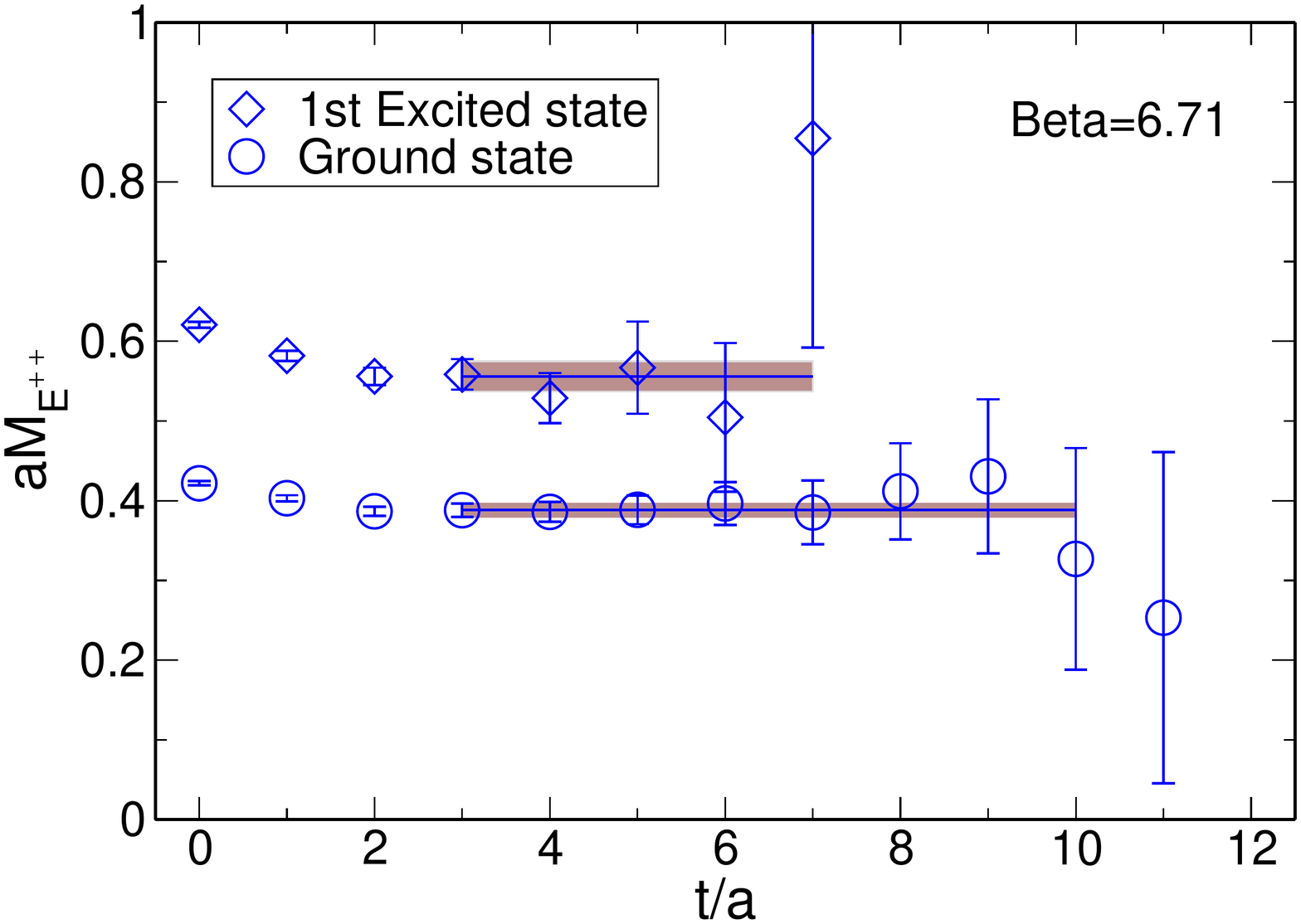}
	\includegraphics[width=0.48\textwidth, bb=0 0 792 612]{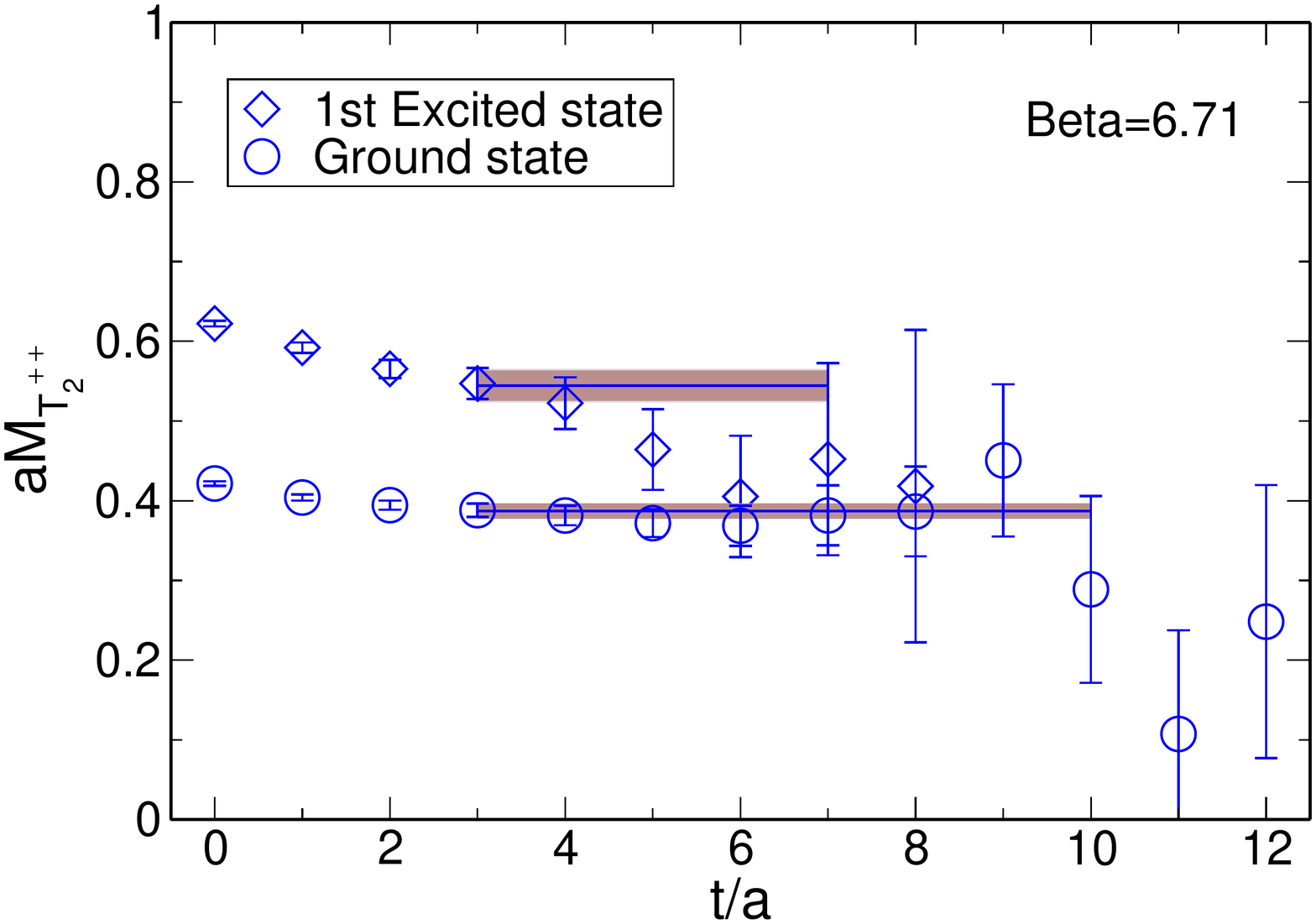}
	\caption{Effective mass plots for the ground state and the first excited state in 	
	 $A_1^{++}$ (top-left), $A_1^{-+}$ (top-right), $E^{++}$ (bottom-left), and $T_{2}^{++}$ (bottom-right) channels 
	 at $\beta=6.71$.
	\label{fig:VM_B671}}
\end{figure*}

\begin{figure*}
	\includegraphics[width=0.48\textwidth, bb=0 0 792 612]{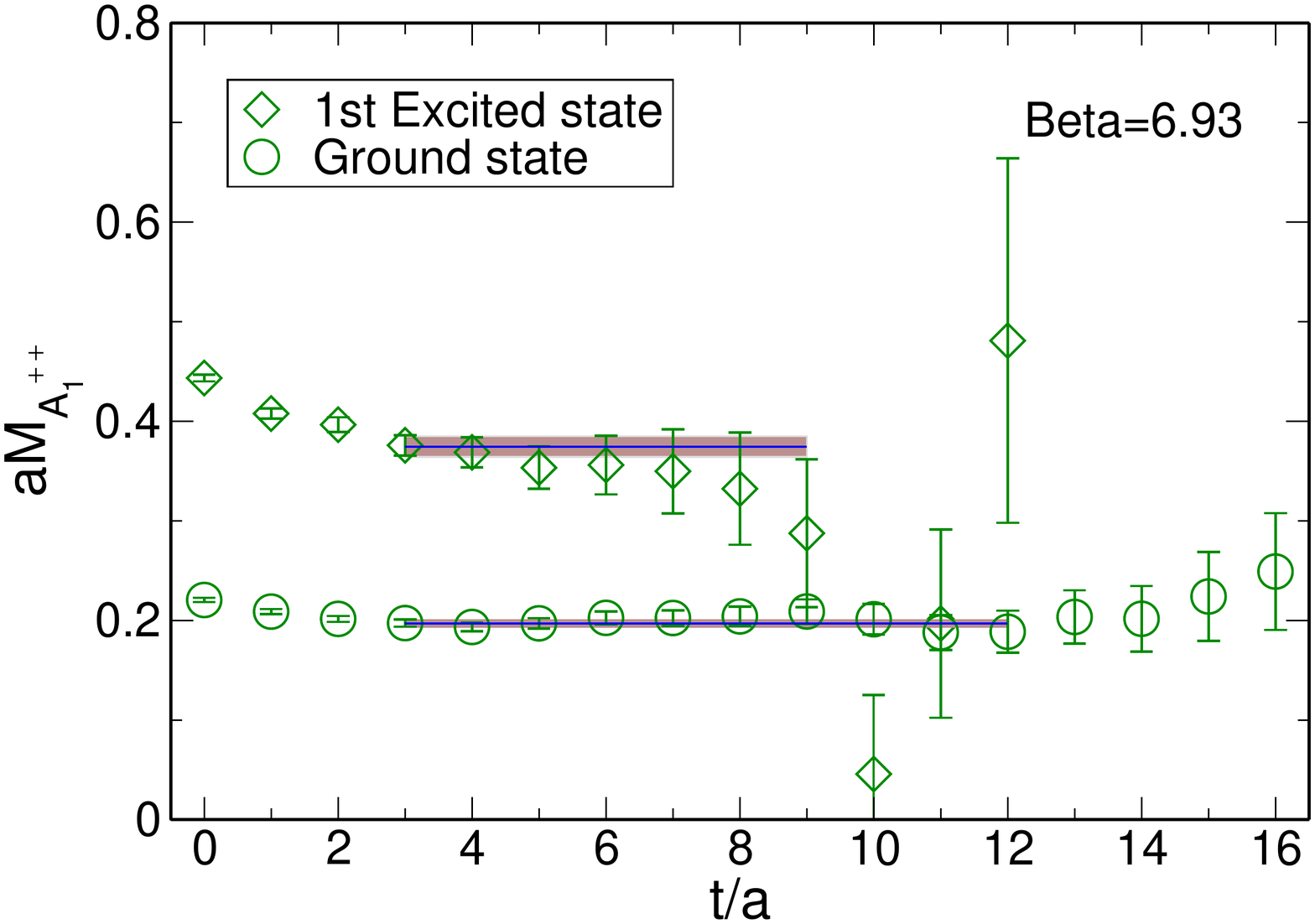}
	\includegraphics[width=0.48\textwidth, bb=0 0 792 612]{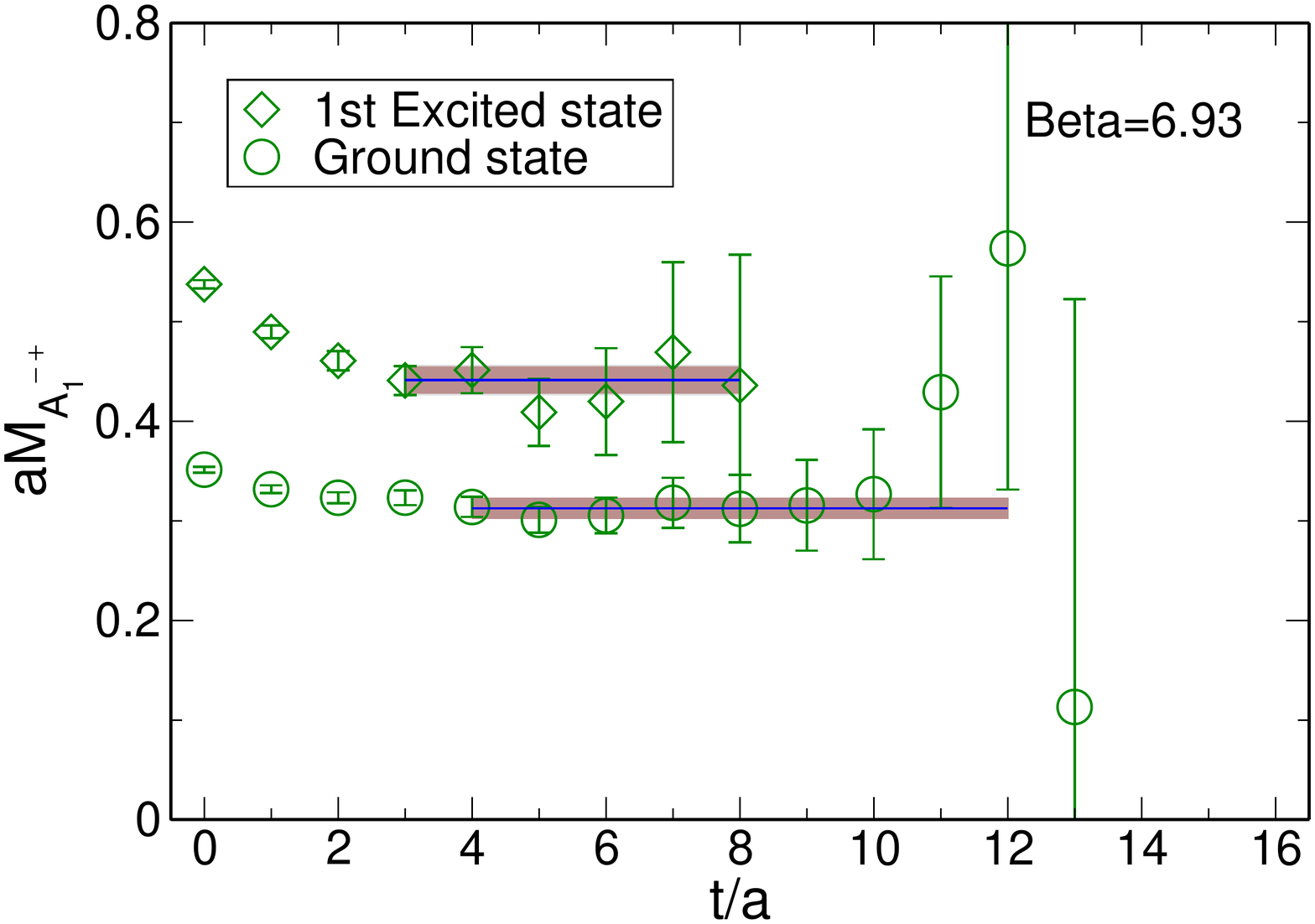}
	\includegraphics[width=0.48\textwidth, bb=0 0 792 612]{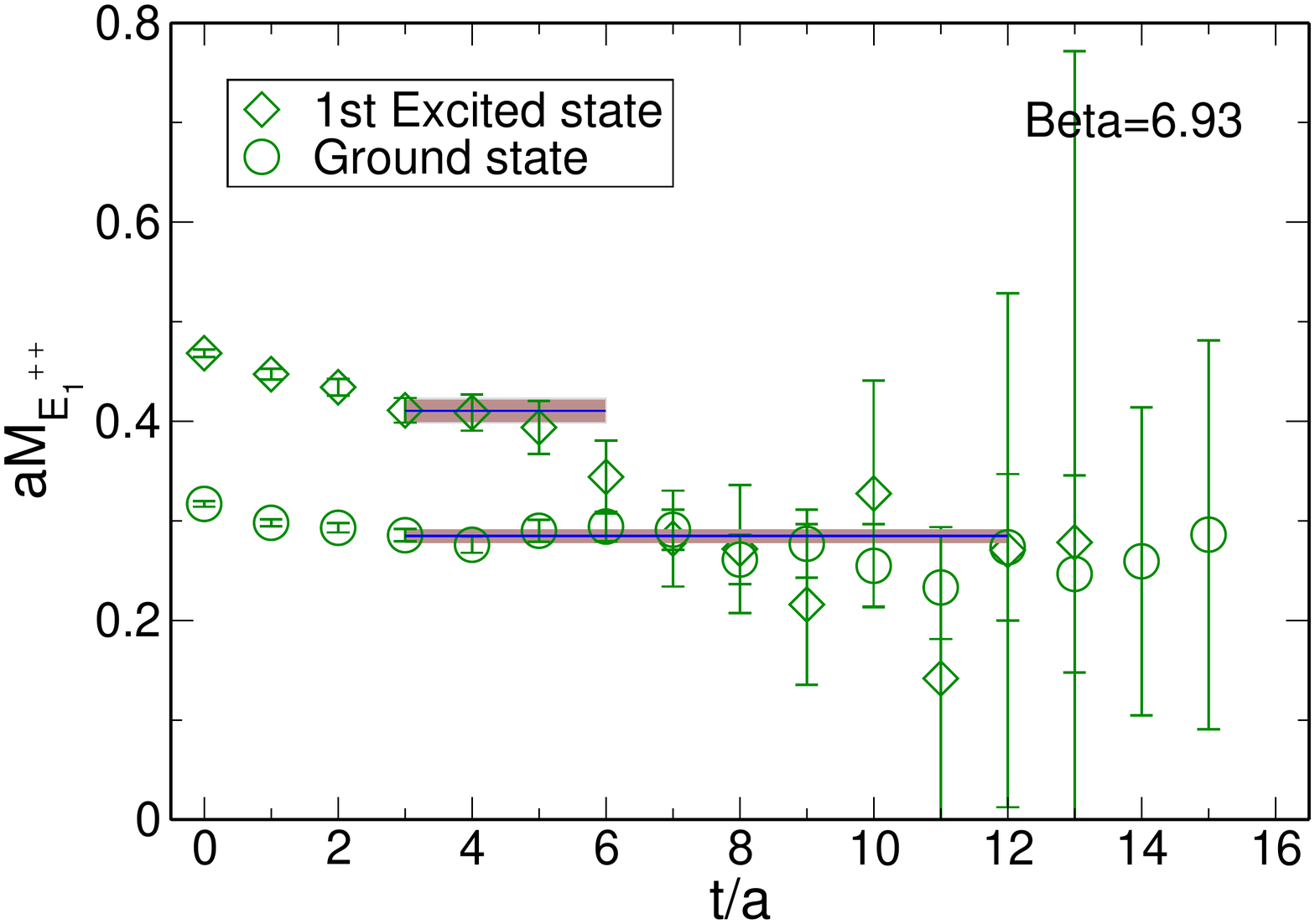}
	\includegraphics[width=0.48\textwidth, bb=0 0 792 612]{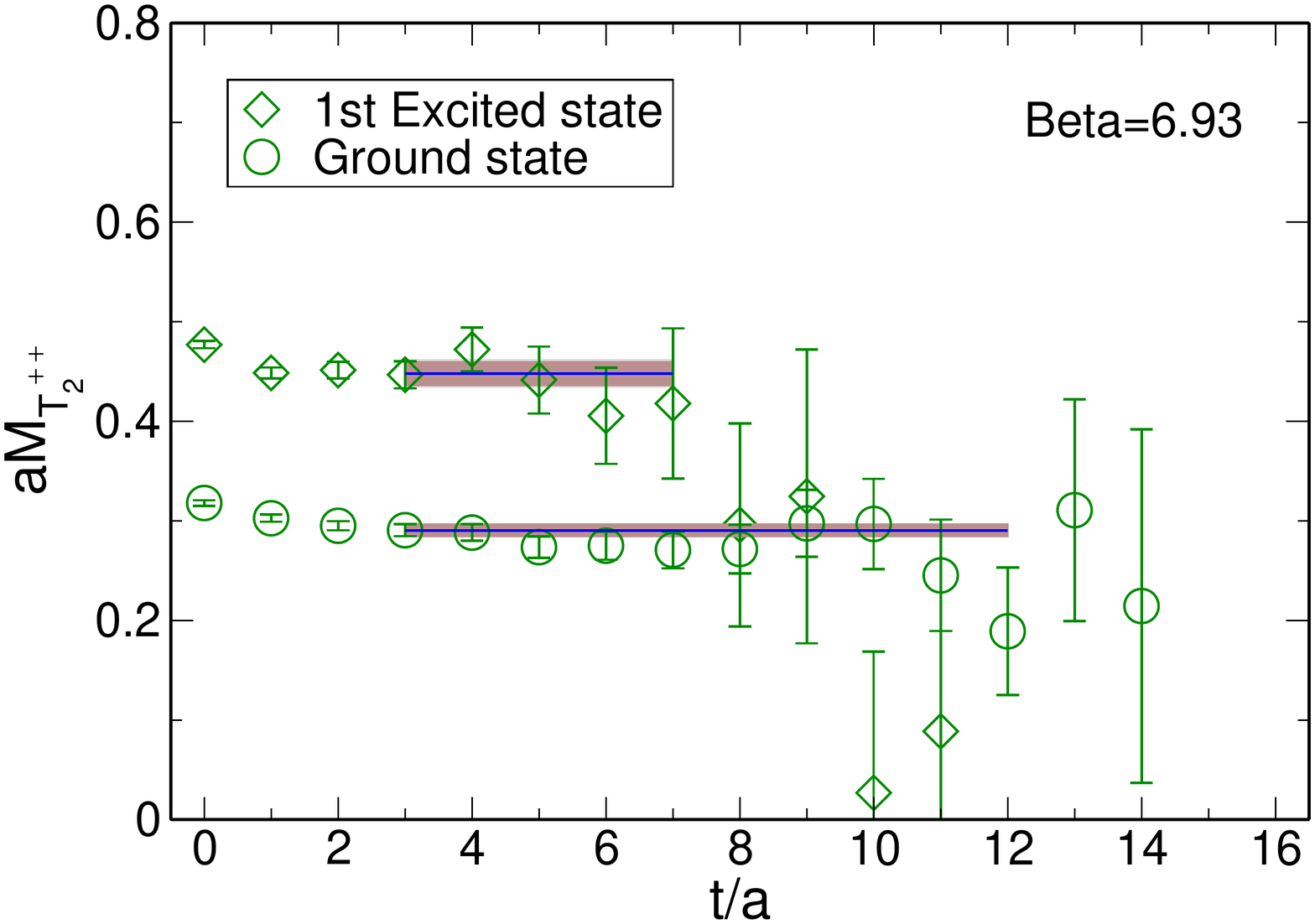}
	\caption{Effective mass plots for the ground state and the first excited state in the
	 $A_1^{++}$ (top-left), $A_1^{-+}$ (top-right), $E^{++}$ (bottom-left), and $T_{2}^{++}$ (bottom-right) channels 
	 at $\beta=6.93$.
	\label{fig:VM_B693}}
\end{figure*}

Let us first present the effective masses of glueballs obtained 
from the variational method using the $6\times 6$ correlation 
matrix constructed by the ${\cal O}_{\rm fish}$ operator with six different flow iterations. 
Figure~\ref{fig:VM_B620} show the effective mass plots of the 
first two eigenvalues in the $A_1^{++}$ (top-left), $A_1^{-+}$ (top-right), $E^{++}$ (bottom-left), and $T_{2}^{++}$ (bottom-right) representations at $\beta=6.20$.
Figures~\ref{fig:VM_B640},~\ref{fig:VM_B671}, and~\ref{fig:VM_B693} are also plotted for 
the results obtained at $\beta=6.40$, 6.71, and 6.93, respectively. In each panel of these figures, the horizontal solid lines represent each fit result obtained by
a correlated fit using a single-exponential functional form, and shaded bands display the fit range and one standard deviation. 
As can be seen, the variational analysis with the correlation matrix constructed in our chosen basis successfully separates
the first excited state from the ground state in each channel.
In Table~\ref{tab:GBmass}, we summarize the results of masses of the two lowest-lying glueball states
in all four channels, together with fit ranges used in the fits and value of $\chi^2$ per degrees of freedom (dof).

%
% FIG.9
%

%
% FIG.10
%

%
% FIG.11
%

%
% FIG.12
%

%
% TABLE.7
%
\begin{table*}[ht]                                                                                       
  \caption{Masses of the ground state and the first excited state of glueballs 
  obtained from the variational method using the $6\times 6$ correlation matrix
  constructed by the ${\cal O}_{\rm fish}$ operator with six different flow iterations. 
  \label{tab:GBmass}
  }
      \begin{ruledtabular}                                                                              
\begin{tabular}{| c    c    l c c    l c c  |}
\hline
&  & \multicolumn{3}{@{}c@{}}{Ground state} & \multicolumn{3}{@{}c@{}|}{1st excited-state} \cr
\cline{3-5}\cline{6-8}
State (Irreps) & $\beta$ & $aM_{G}$ & fit range & $\chi^2/{\rm dof}$  & $aM_{G}$ & fit range & $\chi^2/{\rm dof}$ \cr
\hline
$A_1^{++}$ & 6.2 & 0.5198(67) & [2, 8] & 0.82 & 0.928(30) & [2, 6] & 0.42 \cr
           & 6.4 & 0.4025(62) & [3, 8] & 0.61 & 0.714(32) & [3, 6] & 0.89 \cr
           & 6.71& 0.2664(45) & [3,10] & 0.85 & 0.518(10) & [2, 6] & 0.85\cr
           & 6.93& 0.1970(36) & [3,12] & 0.98 & 0.374(10) & [3, 9] & 0.30\cr
\hline
$E^{++}$   & 6.2 & 0.8032(74) & [1, 6] & 0.84 & 1.109(16) & [1, 4] & 0.66 \cr
           & 6.4 & 0.6035(85) & [2, 6] & 1.47 & 0.858(21) & [2, 5] & 0.83 \cr
           & 6.71& 0.3884(83) & [3,10] & 0.12 & 0.556(19) & [3, 7] & 0.66 \cr
           & 6.93& 0.2849(61) & [3,12] & 1.16 & 0.411(12) & [3, 6] & 0.25 \cr
\hline
$T_2^{++}$ & 6.2 & 0.7989(74) & [1, 6] & 0.31 & 1.117(16) & [1, 4] & 0.47 \cr
           & 6.4 & 0.5878(84) & [2, 6] & 1.58 & 0.843(21) & [2, 5] & 0.90 \cr
           & 6.71& 0.3871(85) & [3,10] & 0.35 & 0.544(20) & [3, 7] & 1.24\cr
     	   & 6.93& 0.2904(62) & [3,12] & 0.81 & 0.448(14) & [3, 7] & 1.36 \cr
\hline
$A_1^{-+}$ & 6.2 & 0.859(20)  & [2, 5] & 0.37 & 1.225(21) & [1, 3] & 2.13 \cr
           & 6.4 & 0.618(18)  & [3, 8] & 1.04 & 0.938(29) & [2, 4] & 1.14 \cr
           & 6.71& 0.415(10)  & [3, 9] & 0.87 & 0.629(14) & [2, 6] & 1.12\cr          		
     	   & 6.93& 0.313(10)  & [4,12] & 0.54 & 0.441(14) & [3, 8] & 0.55 \cr
\hline
\end{tabular}
\end{ruledtabular}                                                                                
\end{table*}

\clearpage
\subsection{Continuum extrapolation and comparison with previous results}
\label{sec:COMP_GB}

It is worth comparing our result obtained from the spatial gradient flow with previous results obtained from 
both the original gradient flow and a conventional approach. For this purpose, we choose 
the results obtained in the simulations performed on isotropic lattice with the standard Wilson plaquette action. 
The previous attempt to apply the gradient flow to the $A_1^{++}$ glueball spectroscopy was done 
by Chowdhury-Harindranath-Maiti~\cite{Chowdhury:2015hta} (denoted as CHJ result). 
Meyer~\cite{Meyer:2004gx} and Athenodorou-Teper~\cite{Athenodorou:2020ani} (denoted as AT result)
performed comprehensive studies of the low-lying glueball states using the conventional approach
at several lattice spacings.  

\begin{figure*}
	\includegraphics[width=0.48\textwidth, bb=0 0 792 612]{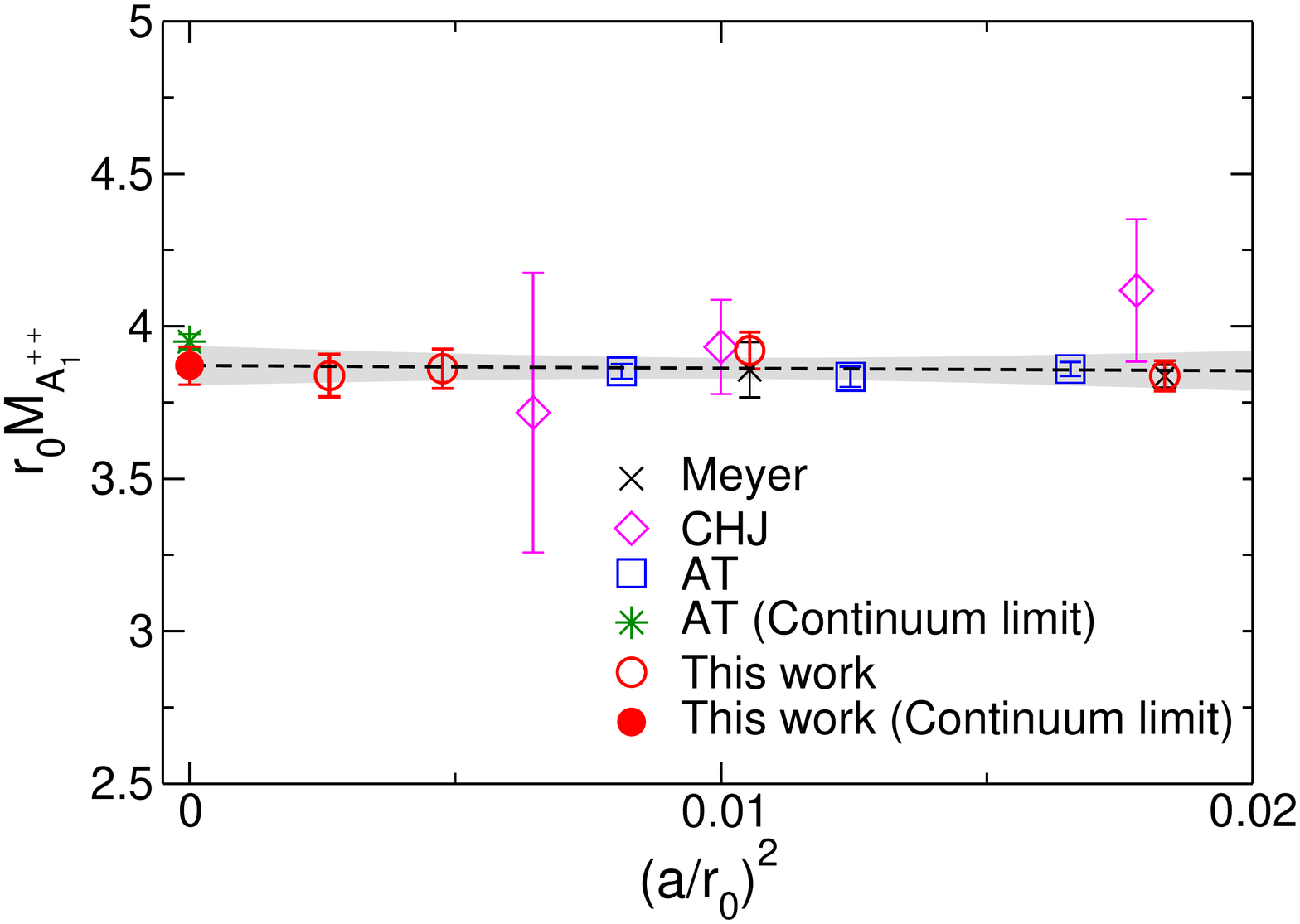}
	\includegraphics[width=0.48\textwidth, bb=0 0 792 612]{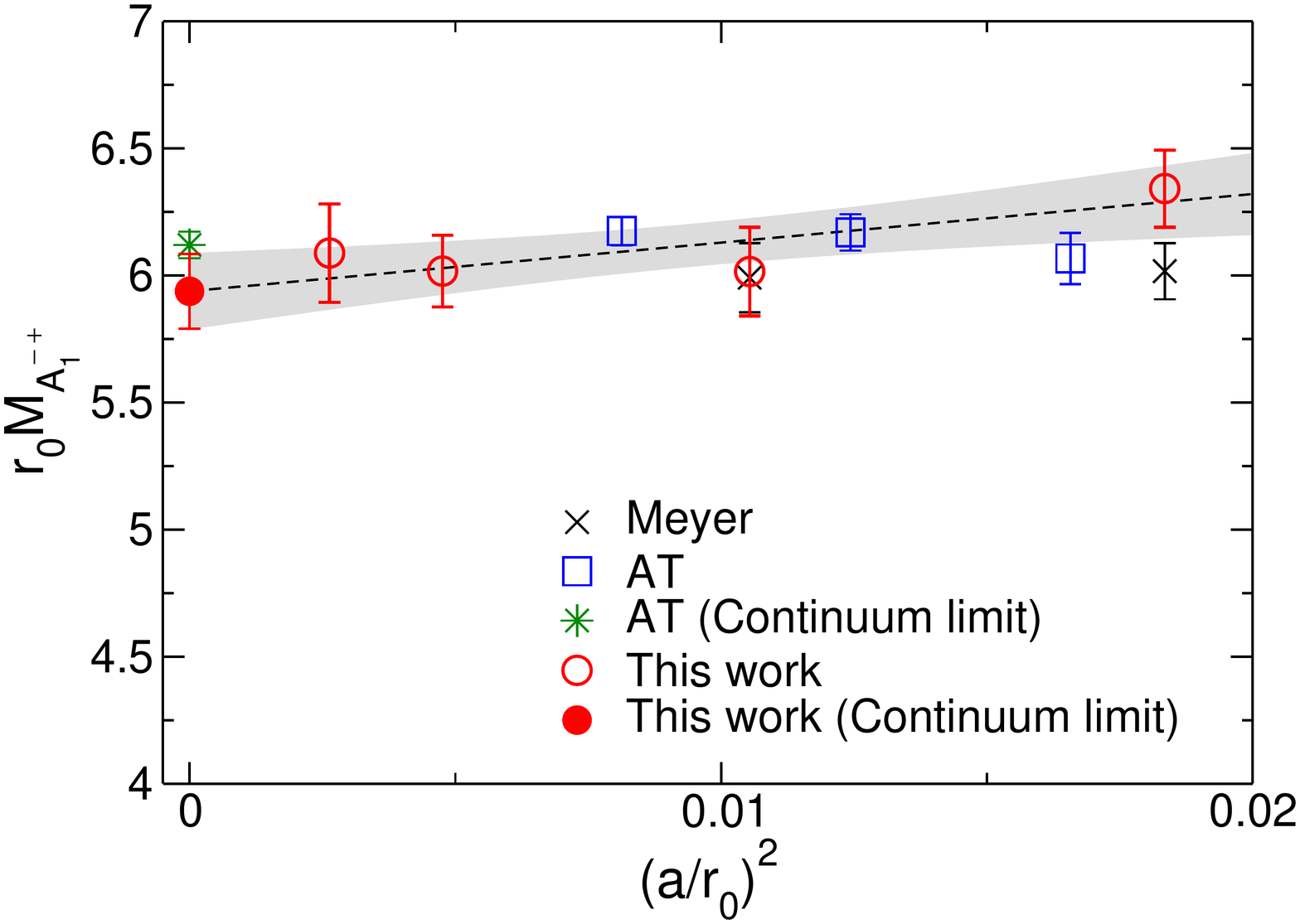}
	\includegraphics[width=0.48\textwidth, bb=0 0 792 612]{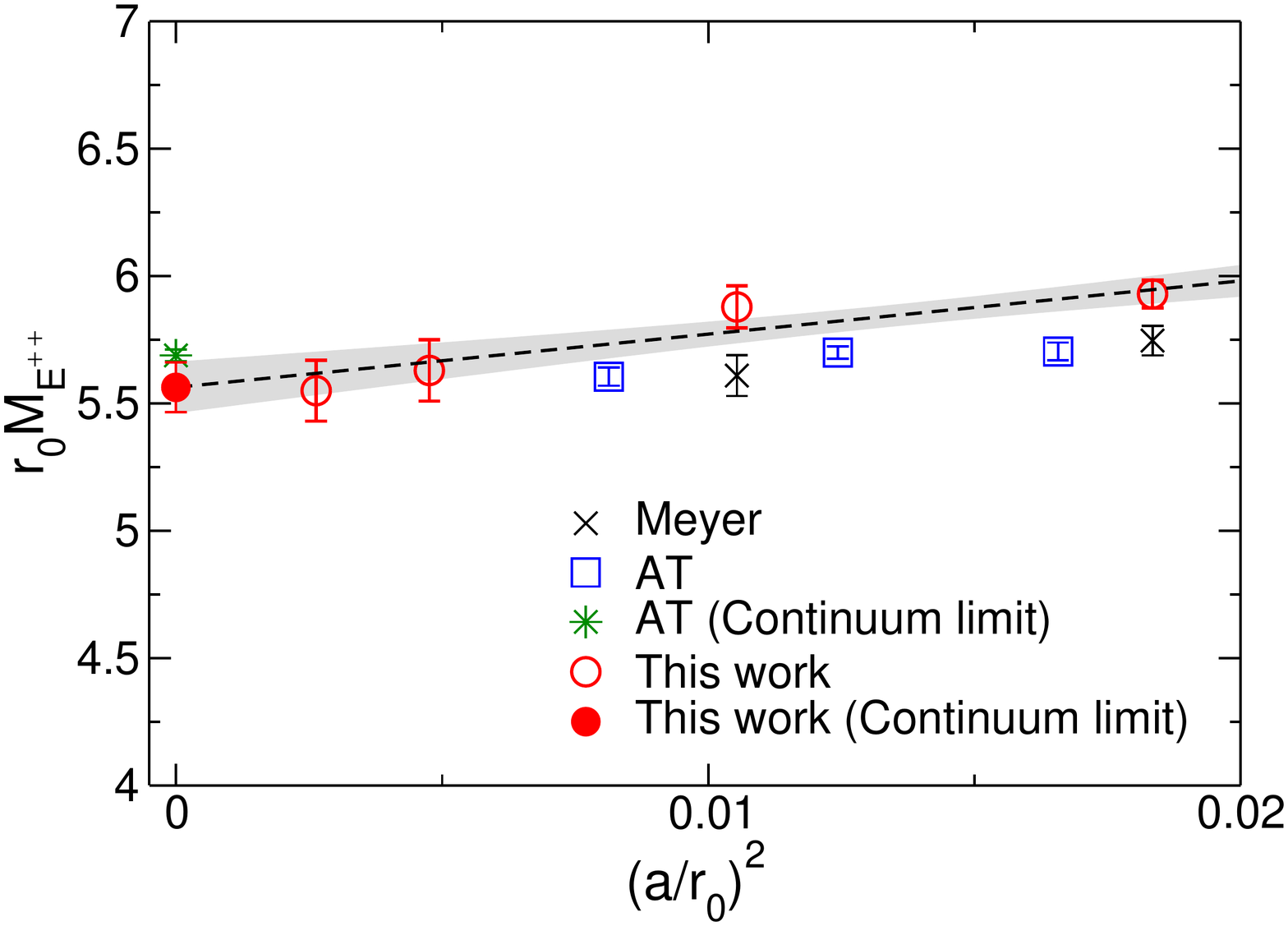}
	\includegraphics[width=0.48\textwidth, bb=0 0 792 612]{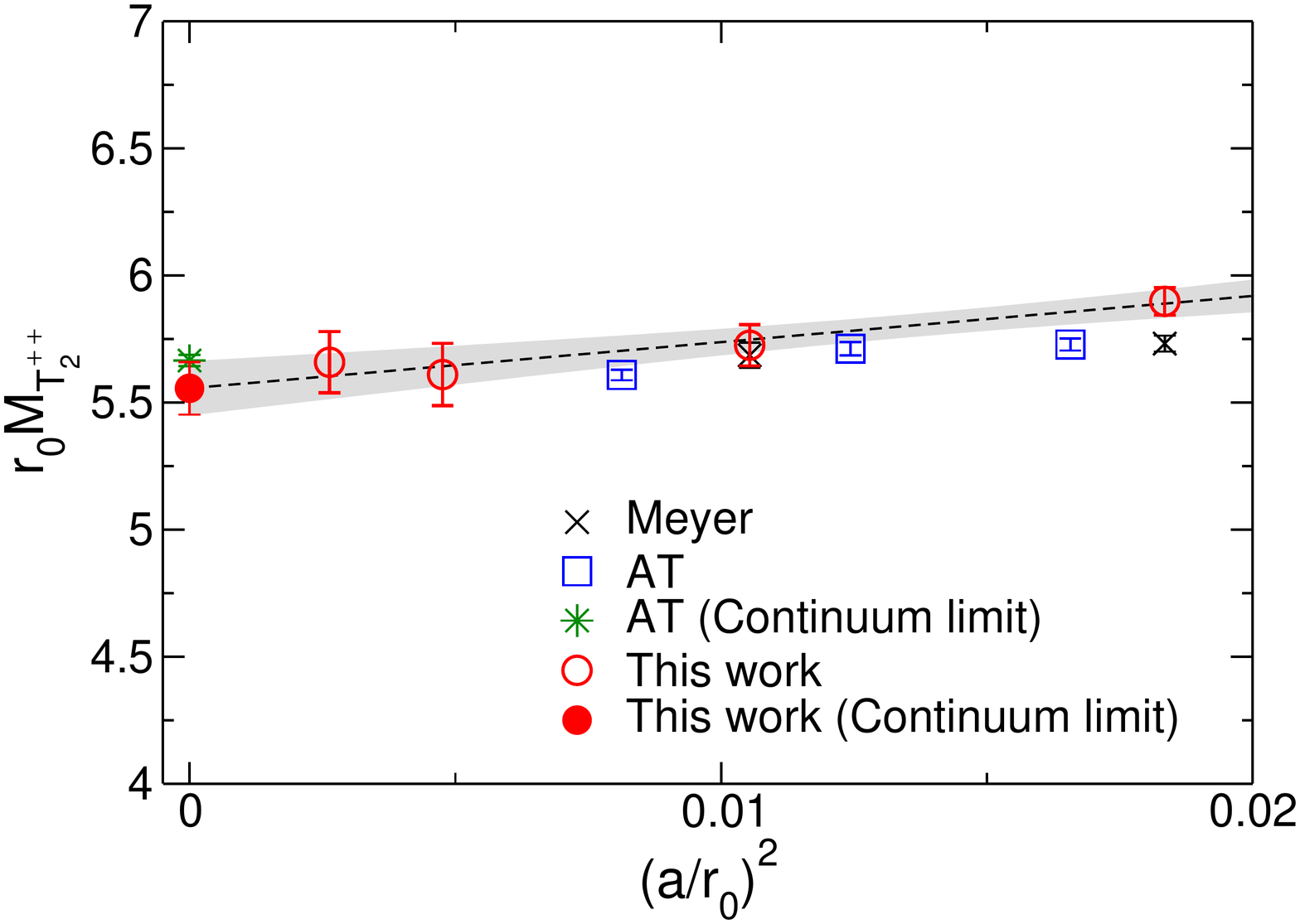}
	\caption{
	Comparisons of the glueball mass results for the ground states 
	in the $A_1^{++}$ (top-left), $A_1^{-+}$ (top-right), $E^{++}$ (bottom-left), and $T_{2}^{++}$ (bottom-right) 
	channels from
	this work, Meyer~\cite{Meyer:2004gx}, CHJ~\cite{Chowdhury:2015hta}, and AT~\cite{Athenodorou:2020ani}.
	Our results are calculated by the spatial gradient flow method. 
	On the other hand, both the Meyer and AT results are given by the conventional approach, while
	the CHJ result are given by the ordinary gradient flow.
	\label{fig:COMP_GBM_ALL}}
\end{figure*}

In Fig.~\ref{fig:COMP_GBM_ALL}, we show our results of the ground-state glueball masses calculated
in the $A_1^{++}$ (top-left), $A_1^{-+}$ (top-right), $E^{++}$ (bottom-left), and $T_{2}^{++}$ (bottom-right) 
channels. Our results are calculated by the spatial gradient flow method.
In each panel of Fig.~\ref{fig:COMP_GBM_ALL}, the dimensionless products of 
the glueball masses $M_G$ and the Sommer scale $r_0$ are shown as
functions of $(a/r_0)^2$ for a comparison with the previous works.

In the $A_1^{++}$ channel, our results are fairly consistent with the previous works. 
On the other hand, although overall agreements with Meyer and AT results are observed in the other three channels,
a more detailed comparison reveals a slight difference between these results and our results at the coarser lattice spacing. 
Especially, our data points calculated at $\beta=6.2$ are slightly overestimated except for the $A_1^{++}$ state. 
Nevertheless, our results obtained at the finer lattice spacings are consistent with the continuum-extrapolated AT result 
(marked as asterisk symbols) in all four channels. 

We next perform the continuum extrapolation of the glueball masses $M_G(0)$
from the glueball masses $M_G(a)$ measured at the finite lattice spacing $a$. 
To remove the lattice discretization corrections on the measured glueball masses $M_G(a)$, 
we use a linear fit with respect to $(a/r_0)^2$ as 
\begin{equation}
M_G(a)r_0=c_0+c_2\left(\frac{a}{r_0}\right)^2 
\end{equation}
with $c_0$ being the continuum-extrapolated glueball masses $M_G(0)$ in units of $r_0$.
The fit results using all four data points are shown in Fig.~\ref{fig:COMP_GBM_ALL}
as solid lines. The statistical uncertainty which is estimated by the jackknife analysis is 
indicated as the gray-shaded band in each panel. The data points are well described by 
the fitted curves. As mentioned above, our data points calculated at $\beta=6.2$ are slightly overestimated 
in comparison to the previous works except for the $A_1^{++}$ state, though our continuum results 
(marked as filled circles) are consistent with the continuum-extrapolated AT results obtained from their high-precision data
in all four channels. 

Our fit results are compiled in Table~\ref{tab:Cont_results}. 
The quoted values of $r_0M_G$ include a systematic error stemming from the continuum-extrapolation fits
as the second error, which are evaluated from a difference between the fit results obtained by all four data points
and three data points closest to the continuum. Table~\ref{tab:Cont_results} also includes the previous 
continuum-limit results for comparison.

In Fig.~\ref{fig:COMP_GBM_ALL}, we finally plot all of data included in Table~\ref{tab:Cont_results}.
The masses are given in terms of the Sommer scale $r_0$ along the left vertical axis, while the right vertical
axis is converted to physical units by $r_0 = 0.472(5)$ fm,~\footnote{
The quoted value is determined from $N_f>2$ lattice QCD simulations in Ref.~\cite{Sommer:2014mea}}
which was adopted in Ref.~\cite{Athenodorou:2020ani}.
For the $2^{++}$ results, we use a weighted average of $M_{2^{++}}=(2M_{E^{++}}+3M_{T_2^{++}})/5$
for the final estimation. The inner and outer error bars on our results represent their statistical and 
total (adding statistical and systematic errors in quadrature) uncertainties, respectively. 
Our final results of the ground-states masses of the $0^{++}$, $2^{++}$, and $0^{-+}$ glueballs
are obtained in physical units as follows:
%
% Eq.20-22
%
\begin{eqnarray}
M_{0^{++}}&=&1.618(26)(25)\;{\rm GeV} \\
M_{2^{++}}&=&2.324(42)(32)\;{\rm GeV}  \\
M_{0^{-+}}&=&2.483(61)(55)\;{\rm GeV},
\end{eqnarray}
where the first error is statistical, while the second one represents a systematic error in the continuum extrapolation
as explained above.  

It is worth recalling that our results and the results of Meyer~\cite{Meyer:2004gx} and AT~\cite{Athenodorou:2020ani} 
are obtained from the isotropic lattice simulations, while the results given by 
Morningstar-Peardon~\cite{Morningstar:1999rf} (denoted as MP result) 
and Chen {\it et al.}~\cite{Chen:2005mg} are obtained from the anisotropic lattice simulations.
The results from the isotropic lattice simulations are systematically underestimated in comparison to
those of the anisotropic lattice simulations. 
This may suggest that there remains some subtlety in taking the continuum limit for the results obtained from 
the anisotropic lattice simulations. It is beyond the scope of this study, while our aim is rather to demonstrate 
the feasibility of our proposed approach. 
Furthermore, as discussed in Appendix~\ref{app:effectiveness}, 
we found that the spatial gradient flow is a few times more effective than the original gradient 
flow and the conventional approach. 
We therefore stress that the spatial gradient flow method can efficiently reproduce the recent high precision 
results of the glueball masses~\cite{Athenodorou:2020ani} within the statistical uncertainties.

%
% FIG.13
%
%

%
% TABLE.8
%
\begin{table*}[ht]                                                                                       
  \caption{Comparison of the continuum-limit results of $r_0M_G$ 
  for the ground states of the $A_1^{++}$, $E^{++}$, $T_2^{++}$
  and $A_1^{-+}$ irreps. For comparison, the previous continuum-limit results are also included.
  \label{tab:Cont_results}
  }
\begin{ruledtabular}                                                                              
\begin{tabular}{| c l l l l l |}
\hline
State (Irreps) & This work  & 
Meyer~\cite{Meyer:2004gx} & AT~\cite{Athenodorou:2020ani} 
& MP~\cite{Morningstar:1999rf} & Chen~\cite{Chen:2005mg} \cr
\hline
$A_1^{++}$  & 3.871(62)(61) & 3.883(79) & 3.950(24) & 4.21(11) &  4.16(11)  \cr
$E^{++}$     & 5.563(98)(129) & 5.703(106) & 5.689(23)  & 5.85(2)   & 5.82(5)     \cr
$T_2^{++}$  & 5.556(104)(40) & 5.658(75) & 5.667(22) & 5.85(2)    & 5.83(4)     \cr
$A_1^{-+}$  & 5.938(147)(132) & 5.93(16) & 6.120(52)& 6.33(7)    & 6.25(6)     \cr
\hline
\end{tabular}
\end{ruledtabular}                                                                                
\end{table*}
%

%
% FIG.14
%
\begin{figure}
	\includegraphics[width=0.5\textwidth, bb=0 0 612 792]{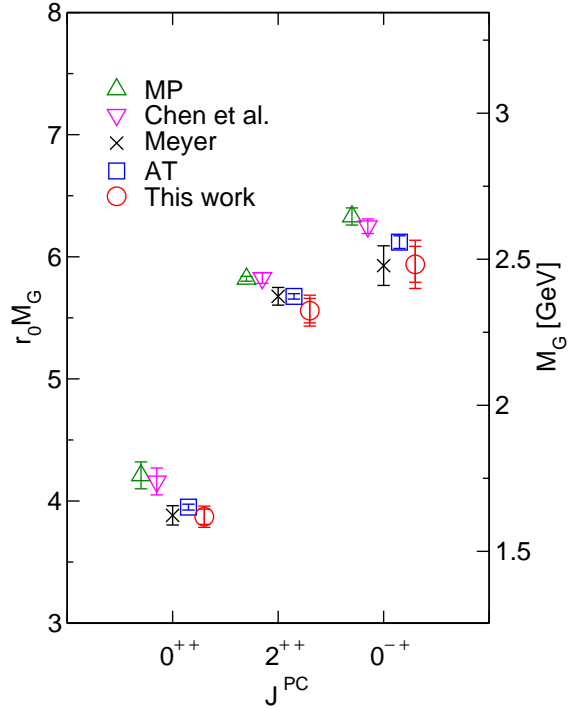}
	\caption{Comparison of our continuum-limit results (circles) with the previous continuum-limit results:
	MP (upper triangles), Chen et al. (lower triangles), Meyer (cross symbols), 
	and AT (squares) results, respectively.
	The masses are given in terms of the Sommer scale $r_0$ along the left vertical axis, while the right vertical
	axis is converted to physical units by $r_0 = 0.472$ fm, which is adopted in Ref.~\cite{Athenodorou:2020ani}.
	The inner and outer error bars on our results represent their statistical and total uncertainties, respectively. 
	}
        \label{fig:CONT_GBM_ALL}
\end{figure}

%\clearpage
%%%%%%%%%%%%%%  SEC7  %%%%%%%%%%%%%%%%%%%%%%%%
\section{Summary} 
\label{sec:SUMMARY}

We have studied the glueball two-point function with two types of the gradient flow method.
The original gradient flow, which makes the Wilson flow diffused in the four-dimensional space-time,
has some problem in measuring the glueball mass from the two-point function. It is known to be 
{\it over smearing} due to the overlap of two glueball operators extended in both space-time as reported in the previous study~\cite{Chowdhury:2015hta}. This particular issue makes the plateau behavior 
uncertain in the effective mass plot, so that it is difficult to extract the ground-state mass of the 
glueball with high accuracy. 

To avoid over smearing, we propose the spatial gradient flow approach
and also apply it to the glueball calculations. Our numerical simulations show that the spatial gradient flow
method works well as a noise-reduction technique, meanwhile it has a good property that the plateau behavior 
in the effective mass plot does not change with variation in flow time for sufficiently large flow time. 
The latter point gives an advantage for extracting the ground-state mass of the glueball with high accuracy without over smearing. 

It is also observed that the spatial gradient flow eliminates 
dependence of the Wilson loop shapes in the glueball two-point function
due to a strong isotropic nature. Therefore, the variational method based on the different shapes is not applicable.
Instead, the different diffuseness of the extended operator, which is given at the different flow time, 
are used for the variational analysis to separate the ground-state contribution from the excited-state
contributions.

To demonstrate the feasibility of our proposed method, we have determined the masses of 
the three lowest-lying glueball states, corresponding to the $0^{++}$, $2^{++}$, and $0^{-+}$ glueballs,
in the continuum limit by using four lattice QCD simulations 
for the lattice spacings ranging from 0.026 to 0.068 fm.  
Our results of the $0^{++}$, $2^{++}$ and $0^{-+}$ glueball 
masses are consistent with the previous works~\cite{{Meyer:2004gx},{Athenodorou:2020ani}}. 
Especially, it is worth emphasizing that the spatial gradient flow method can efficiently reproduce 
the recent high-precision results~\cite{Athenodorou:2020ani}, which are slightly underestimated
in comparison to the results given by the anisotropic lattice simulations~\cite{{Morningstar:1999rf},{Chen:2005mg}},
within the statistical uncertainties. 

We have also showed numerical equivalence between the spatial gradient flow and the stout smearing 
in the glueball calculations at the relatively fine lattice spacing of 0.0513(3) fm. 
This observation can be understood through the analytical derivation that demonstrates 
the equivalence between the gradient flow and stout smearing methods in the vicinity of the continuum limit 
as described in Appendix~\ref{sec:FLOW_STOUT}.
This fact can help to reflect actual efficiency for the glueball spectroscopy.
%since at the finer lattice spacings 
%the spatial gradient flow method can be replaced by the stout smearing method that roughly %leads to a factor of {\color{red}${\cal O}(10)$} improvement in the computational cost. 

As discussed in Appendix~\ref{app:effectiveness}, although the spatial gradient flow requires several times fewer statistics to achieve the same statistical accuracy than the conventional method, its computational cost is roughly a factor of ${\cal O}(10)$ higher than the conventional one. Thus, by replacing the spatial gradient flow method with the stout smearing method, which is almost as computationally inexpensive as the conventional method, the gradient flow approach becomes really an efficient scheme for the glueball spectroscopy.

We plan to extend our research to calculate the glueball three-point function
with the renormalized energy-momentum tensor formulated in the gradient flow method~\cite{Suzuki:2013gza}
in order to investigate the origin of the glueball masses. Such study is now in progress~\cite{future_work}. 

%--- acknowledgments ------------------------------------------------  
\begin{acknowledgments}
K. S. is supported by Graduate Program on Physics for the Universe (GP-PU)
of Tohoku University. 
Numerical calculations in this work were partially performed using Yukawa-21 
at the Yukawa Institute Computer Facility,
and also using Cygnus at Center for Computational Sciences (CCS), University of Tsukuba
under Multidisciplinary Cooperative Research Program of CCS (No. MCRP-2021-54 and No. MCRP-2022-42). 
This work was also supported in part by Grants-in-Aid for Scientific Research form the Ministry 
of Education, Culture, Sports, Science and Technology (No. 18K03605 and No. 22K03612).

\end{acknowledgments}
%\clearpage
%----------------------------------- appendix -------------------------------------------------------  
\appendix

%%%%%%%%%%%%%%%%%%%%%%%%%%%%%%%%%%%%%%%
\section{Equivalence between gradient flow and stout smearing}
\label{sec:FLOW_STOUT}
This section is devoted to a discussion of the equivalence between gradient flow and stout smearing referred 
in Sec.~\ref{sec:EQUIVE}. For this purpose, we will first provide the explicit form of $\partial_{x,\mu} S_W[U]$ 
appeared in the left-hand side of Eq.~(\ref{eq:gradient_flow}). 

The link derivative operator $\partial_{x,\mu}$ is defined as follows. 
The operator $\partial_{\mu, x}$ stands for the Lie algebra valued differential operator with respect to the
link variable~\cite{Luscher:2010iy}. Let us introduce the anti-Hermitian traceless $N\times N$ matrices $T^a$ $(a=1, ..., N^2-1)$ 
as generators of $SU(N)$ group.~\footnote{In this paper, we use the notational conventions adopted in the original L\"uscher's paper~\cite{Luscher:2010iy} Namely, they are normalized by ${\rm Tr}\left(T^aT^b\right)=-\frac{1}{2}\delta^{ab}$ and also satisfy the commutation relations $[T^a, T^b]=f_{abc}T^c$ with the structure constants $f^{abc}$. }
In general, with respect to a basis $T^a$, the elements $M$ of the Lie algebra of $SU(N)$ 
are given by $M=M^aT^a$ with real components $M^a$.
Therefore, the operator $\partial_{x,\mu}$ can be expressed with respect to a basis $T^a$ as
%
% Eq.
%
\begin{equation}
\partial_{\mu, x}=T^a\partial^a_{\mu, x},
\end{equation}
where the operators $\partial^a_{\mu, x}$ are defined by
\begin{equation}
\partial^a_{\mu, x}f(U)=\frac{d}{ds}f(e^{sX^a}U)|_{s=0}
\end{equation}
with
\begin{equation}
X^a(y,\nu)=
\begin{cases}
T^a & {\rm if}\ (y,\nu)=(x,\mu), \\
0 & {\rm otherwise}
\end{cases}
\end{equation}
and act as differential operators on functions $f$ of the link variable $U$.

When the link derivative $\partial_{x,\mu}$ acts on the action, we may simply focus on the term 
that explicitly depends on $U_\mu(x)$ in the action as
\begin{widetext}
\begin{equation}
  S_{W}[U] = -\frac{2}{g_0^2}\sum_{x, \mu>\nu}\left[
   {\rm Re}{\rm Tr}\left\{U_\mu(x)X^\dagger_{\mu}(x)
   \right\}+\left\{\mbox{terms independent of $U_\mu(x)$}\right\}\right],
    \label{eq:WilsonAction2}
\end{equation}
\end{widetext}
where $X_{\mu}(x)$ is the sum of all the path ordered products of the link variables, called the ``staple".
The staple $X_{\mu}(x)$ is given by
\begin{widetext}
\begin{equation}
X_{\mu}(x)=\sum_{\mu>\nu}\left[U_{\nu}(x)U_{\mu}(x+\hat{\nu})U_{\nu}^\dagger(x+\hat{\mu})
+U_{\nu}^\dagger(x-\hat{\nu})U_{\mu}(x-\hat{\nu})U_{\nu}(x-\hat{\nu}+\hat{\mu})
\right].
\end{equation}
\end{widetext}
If we set $\Omega_{\mu}(x)=X_\mu(x)U^\dagger_\mu(x)$, each basis component is given as
\begin{widetext}
\begin{equation}
    g_0^2\partial^a_{\mu, x}S_{W}[U]=-2{\rm Re}{\rm Tr}\left\{T^a \Omega^\dagger_\mu(x)\right\}=-{\rm Tr}\left\{T^a\left(\Omega^\dagger_\mu(x)-\Omega_\mu(x)\right)\right\},
\end{equation}
\end{widetext}
where $\Omega^\dagger_\mu(x)$ denotes the sum of all plaquettes that include $U_{\mu}(x)$. 
Therefore, we finally get
\begin{equation}
g_0^2\partial_{\mu, x}S_{W}[U]=-iQ_{\mu}(x)
\label{eq:DeriveAction}
\end{equation}
with 
\begin{equation}
Q_{\mu}(x)=\frac{i}{2}\left(\Omega^\dagger_\mu(x)-\Omega_\mu(x)\right)
-\frac{i}{2N}{\rm Tr}\left(\Omega^\dagger_\mu(x)-\Omega_\mu(x)\right), 
\label{eq:StoutPart}
\end{equation}
which becomes a Lie algebra valued quantity.

In the stout smearing, the link smearing is defined as the following recursive procedure~\cite{Morningstar:2003gk}.
Here, for simplicity, the stout smearing parameters $\rho_{\mu\nu}$ are taken as $\rho_{\mu\nu}=\rho$.
The link variables $U^{(k)}_{\mu}(x)$ at step $k$ are mapped into the link variables 
$U^{(k+1)}_{\mu}(x)$ using
\begin{equation}
    U^{(k+1)}_{\mu}(x)=\exp\left(i \rho Q_{\mu}^{(k)}(x)\right)U^{(k)}_{\mu}(x),
    \label{eq:StoutSmear}    
\end{equation}
where $Q_\mu^{(k)}(x)$ is given by Eq.~(\ref{eq:StoutPart}) with the stout link $U^{(k)}_{\mu}(x)$~\cite{Morningstar:2003gk}. 
By taking the logarithm of both sides of Eq.~(\ref{eq:StoutSmear}), one can get
\begin{equation}
    \Delta_k\log U^{(k)}_{\mu}(x)=i \rho Q_{\mu}^{(k)}(x),
    \label{eq:log_stout_discrete}
\end{equation}
where $\Delta_k$ represents a forward difference with respect to $k$ as ${\Delta_k f(k) \equiv f(k+1)-f(k)}$. 

Next, let us introduce a continuous variable $s=k\rho$, and then reexpress the link variable $U^{(k)}_{\mu}(x)$
by writing a function of $s$ as $\tilde{U}_{\mu}(x, s)$. 
Since $\frac{\partial}{\partial s}f(s)=\lim_{\rho \rightarrow 0}\frac{f(s+\rho)-f(s)}{\rho}$, the above difference equation~(\ref{eq:log_stout_discrete}) becomes the differential equation
with respect to the variable $s$ in the limit of $\rho \rightarrow 0$. 
\begin{equation}
    \frac{\partial}{\partial s}\log \tilde{U}_{\mu}(x, s)= -g_0^2\partial_{\mu, x}S_W[\tilde{U}],
    \label{eq:log_stout_cont}
\end{equation}
where the left-hand side of Eq.~(\ref{eq:log_stout_discrete}) is 
rewritten by using the expression of Eq.~(\ref{eq:DeriveAction}). 
Finally, Eq.~(\ref{eq:log_stout_cont}) reduces to a gradient flow equation with respect to the link variable $\tilde{U}_{\mu}(x,s)$ as
\begin{equation}
    \frac{\partial}{\partial s}\tilde{U}_{\mu}(x,s)\cdot\left\{\tilde{U}_{\mu}(x,s)\right\}^{-1} 
    = -g_0^2\partial_{\mu, x}S_W[\tilde{U}] + {\cal O}(a)
    \label{eq:floweq_stout}
    \end{equation}
in the vicinity of the continuum limit.~\footnote{When $U(s)$ is a matrix Lie group, $\frac{\partial U}{\partial s}U^{-1}$ is given 
in terms of $\frac{\partial}{\partial s}\log U$ as
\begin{equation}
\frac{\partial U}{\partial s}U^{-1}=\frac{\partial}{\partial s}\log U +\frac{1}{2!}\left[\log U, \frac{\partial}{\partial s}\log U\right]+\frac{1}{3!}\left[\log U, \left[\log U, \frac{\partial}{\partial s}\log U\right]\right]+\cdot\cdot\cdot. 
\end{equation}
In the case when $U$ is the link variable, the power series of $\log U$ in the left-hand side can be neglected 
for the small lattice spacing $a$.}

It is clear that Eq.~(\ref{eq:floweq_stout}) is equivalent to Eq.~(\ref{eq:gradient_flow}) at the leading order. 
Since the variable $s=k\rho$ directly corresponds to the Wilson flow time $\tau$, 
the perturbative matching relation of $\tau$, $\rho$ and the number of stout smearing steps $n_{\rm st}$ 
as $\tau=\rho n_{\rm st}$ found in Ref.~\cite{Alexandrou:2017hqw} is also rigorously proved. 
Therefore, the gradient flow equation is certainly regarded 
as a continuous version of the recursive update procedure in the stout smearing {\it at the smaller lattice spacing}. 
When $S_W$ is replaced by $S_{sW}$ in the gradient flow equation, and $\rho_{\mu\nu}$
is set as $\rho_{ij}=\rho$ and $\rho_{4\mu}=\rho_{\mu4}=0$ in the stout smearing,
above consideration fully supports our finding that there is the numerical equivalence between
the spatial gradient flow and the spatial stout smearing in the glueball spectroscopy 
at the relatively fine lattice spacing of 0.0513(3) fm. 

%%%%%%%%%%%%%%%%%%%%%
\section{Effectiveness of the spatial gradient flow}
\label{app:effectiveness}
In this appendix, we aim to assess effectiveness of our proposed method in comparison to the conventional approach. A simple indicator of effectiveness or efficiency of a given method to calculate the glueball mass is defined as the following index:
\begin{widetext}
\begin{equation}
  \mbox{Effectiveness index (EI)} = \left[\frac{(\rm Error)}{(\rm Central\ value)}\right]^2\times (\rm No.\ of\ measurements),
\end{equation}
\end{widetext}
which is inversely proportional to 
the relative size of the square of the signal-to-noise ratio with respect to the statistics.
When the EI index gets smaller, efficiency of the method becomes better with fixed statistics. Table~\ref{tab:effectiveness} compiles the values of effectiveness of respective smearing methods among three simulations (CHJ, AT, and this work) performed at the similar lattice spacing (${\beta \approx 6.4}$). 
According to the EI value, the spatial gradient flow or the stout smearing with the high value of $n_{\rm st}$ is several times more effective than the original gradient flow and the conventional approach (see Ref.~\cite{Lucini:2004my} for details of the smearing and fuzzing methods used in Ref.~\cite{Athenodorou:2020ani}).

It should be noted that the EI value does not reflect actual efficiency since the computational cost for the gradient flow method is relatively higher than the conventional approach. Indeed, in our actual numerical code, we find
that the single-link smearing including APE smearing and stout smearing
are a factor of ${\cal O}(10)$ faster than the gradient flow method even with the same numbers 
of flow iterations $n_{\rm flow}$ and smearing steps $n_{\rm st}$.
Moreover, the required number of flow iterations increases quadratically as 
the lattice spacing decreases.

Nevertheless, as numerically found in Sec.~\ref{sec:EQUIVE} and analytically proven in Appendix~\ref{app:effectiveness}, {\it the gradient flow method can 
be replaced by the stout smearing at the finer lattice spacing}
with keeping the same value of EI as shown in Table~\ref{tab:effectiveness}. Since the stout smearing is comparable to the conventional approach regarding the computational cost, the gradient flow approach is really an efficient scheme for the glueball spectroscopy and would have an advantage in dynamical lattice QCD simulations for glueball observables.
\clearpage
%
% TABLE.9
%
\begin{table*}[ht]
  \caption{Comparison of effectiveness of respective smearing methods among three 
  simulations (CHJ, AT, and this work) performed at the similar lattice spacing (${\beta \approx 6.4}$).
  $N_{\rm total}$ denotes the number of total measurements in each simulation.
  \label{tab:effectiveness}
  }
\begin{ruledtabular}    
\begin{tabular}{|l c c c c c c |}
\hline
Label  & $\beta$ & $L^3\times T$& $N_{\rm total}$ & Method & $aM_{A_1^{++}}$  & {\rm EI}  \cr
\hline
This work 
& 6.40  & $32^3\times 32$ & 3000 
& Spatial gradient flow & 0.404(9)  & 1.5\cr
& & & & Stout smearing (high $n_{\rm st}$)& 0.403(9)   &1.5\cr
& & & & Stout smearing (low $n_{\rm st}$)& 0.442(31)  &14.8\cr
& & & & Gradient flow ($R_d=0.23$ fm)& 0.446(14) &3.0\cr
CHJ~\cite{Chowdhury:2015hta}  & 6.42 & $32^3\times 64$ &1958 
& Gradient flow ($R_d=0.3$ fm)& 0.393(15) &2.9 \cr
& & & & Gradient flow ($R_d=0.35$ fm)&0.387(18) &4.3\cr
AT~\cite{Athenodorou:2020ani} & 6.338 & $30^3\times 30$ & 80000 
& Conventional approach & 0.4276(37) &6.0 \cr
\hline
\end{tabular}
\end{ruledtabular}  
\end{table*}

%--- bibliography ---------------------------------------------------  

\end{document}